\documentclass[aps,prd,preprint,superscriptaddress,byrevtex,showpacs,tightenlines]{revtex4}

\usepackage{graphicx}
\usepackage{color}
\usepackage{amsmath}

\begin{document}

\preprint{\vbox{ \hbox{   }
                 \hbox{Belle Preprint 2008-11}
                 \hbox{KEK Preprint 2008-5}
                 \hbox{arXiv:0803.2158[hep-ex]}
}}

\title{\quad \\[0.5cm] \boldmath Measurement of the Moments of the\\
  Photon Energy Spectrum in $B\to X_s\gamma$~Decays and\\
  Determination of $|V_{cb}|$ and $m_b$ at Belle}



\affiliation{Budker Institute of Nuclear Physics, Novosibirsk}
\affiliation{University of Cincinnati, Cincinnati, Ohio 45221}
\affiliation{Department of Physics, Fu Jen Catholic University, Taipei}
\affiliation{The Graduate University for Advanced Studies, Hayama}
\affiliation{Hanyang University, Seoul}
\affiliation{University of Hawaii, Honolulu, Hawaii 96822}
\affiliation{High Energy Accelerator Research Organization (KEK), Tsukuba}
\affiliation{University of Illinois at Urbana-Champaign, Urbana, Illinois 61801}
\affiliation{Institute of High Energy Physics, Chinese Academy of Sciences, Beijing}
\affiliation{Institute of High Energy Physics, Vienna}
\affiliation{Institute of High Energy Physics, Protvino}
\affiliation{Institute for Theoretical and Experimental Physics, Moscow}
\affiliation{J. Stefan Institute, Ljubljana}
\affiliation{Kanagawa University, Yokohama}
\affiliation{Korea University, Seoul}
\affiliation{Kyungpook National University, Taegu}
\affiliation{\'Ecole Polytechnique F\'ed\'erale de Lausanne (EPFL), Lausanne}
\affiliation{Faculty of Mathematics and Physics, University of Ljubljana, Ljubljana}
\affiliation{University of Maribor, Maribor}
\affiliation{University of Melbourne, School of Physics, Victoria 3010}
\affiliation{Nagoya University, Nagoya}
\affiliation{Nara Women's University, Nara}
\affiliation{National Central University, Chung-li}
\affiliation{National United University, Miao Li}
\affiliation{Department of Physics, National Taiwan University, Taipei}
\affiliation{H. Niewodniczanski Institute of Nuclear Physics, Krakow}
\affiliation{Nippon Dental University, Niigata}
\affiliation{Niigata University, Niigata}
\affiliation{University of Nova Gorica, Nova Gorica}
\affiliation{Osaka City University, Osaka}
\affiliation{Osaka University, Osaka}
\affiliation{Panjab University, Chandigarh}
\affiliation{RIKEN BNL Research Center, Upton, New York 11973}
\affiliation{Saga University, Saga}
\affiliation{University of Science and Technology of China, Hefei}
\affiliation{Seoul National University, Seoul}
\affiliation{Sungkyunkwan University, Suwon}
\affiliation{University of Sydney, Sydney, New South Wales}
\affiliation{Toho University, Funabashi}
\affiliation{Tohoku Gakuin University, Tagajo}
\affiliation{Tohoku University, Sendai}
\affiliation{Department of Physics, University of Tokyo, Tokyo}
\affiliation{Tokyo Institute of Technology, Tokyo}
\affiliation{Tokyo Metropolitan University, Tokyo}
\affiliation{Tokyo University of Agriculture and Technology, Tokyo}
\affiliation{Virginia Polytechnic Institute and State University, Blacksburg, Virginia 24061}
\affiliation{Yonsei University, Seoul}
   \author{C.~Schwanda}\affiliation{Institute of High Energy Physics, Vienna} 
   \author{P.~Urquijo}\affiliation{University of Melbourne, School of Physics, Victoria 3010} 
   \author{E.~Barberio}\affiliation{University of Melbourne, School of Physics, Victoria 3010} 
   \author{A.~Limosani}\affiliation{University of Melbourne, School of Physics, Victoria 3010} 
   \author{I.~Adachi}\affiliation{High Energy Accelerator Research Organization (KEK), Tsukuba} 
   \author{H.~Aihara}\affiliation{Department of Physics, University of Tokyo, Tokyo} 
   \author{K.~Arinstein}\affiliation{Budker Institute of Nuclear Physics, Novosibirsk} 
   \author{T.~Aushev}\affiliation{\'Ecole Polytechnique F\'ed\'erale de Lausanne (EPFL), Lausanne}\affiliation{Institute for Theoretical and Experimental Physics, Moscow} 
   \author{S.~Bahinipati}\affiliation{University of Cincinnati, Cincinnati, Ohio 45221} 
   \author{A.~M.~Bakich}\affiliation{University of Sydney, Sydney, New South Wales} 
% ...
\collaboration{The Belle Collaboration}

\begin{abstract}
  Using the previous Belle measurement of the inclusive photon energy in
  $B\to X_s\gamma$~decays, we determine the first and second moments of
  this spectrum for minimum photon energies in the $B$~meson rest frame
  ranging from 1.8 to 2.3~GeV. Combining these measurements with recent
  Belle data on the lepton energy and hadronic mass moments in $B\to
  X_c\ell\nu$~decays, we perform fits to theoretical expressions derived
  in the 1S and kinetic mass schemes and extract the magnitude of the
  Cabibbo-Kobayashi-Maskawa (CKM) matrix element $V_{cb}$, the $b$-quark
  mass and other non-perturbative parameters. In the 1S scheme analysis
  we find $|V_{cb}|=(41.56\pm 0.68(\mathrm{fit})\pm 0.08(\tau_B))\times
  10^{-3}$ and $m_b^\mathrm{1S}=(4.723\pm 0.055)$~GeV. In the kinetic
  scheme, we obtain $|V_{cb}|=(41.58\pm 0.69(\mathrm{fit})\pm
  0.08(\tau_B)\pm 0.58(\mathrm{th}))\times 10^{-3}$ and
  $m_b^\mathrm{kin}=(4.543\pm 0.075)$~GeV.
\end{abstract}


\maketitle

\section{Introduction}

The most precise determinations of the Cabibbo-Kobayashi-Maskawa (CKM)
matrix element~$|V_{cb}|$~\cite{Kobayashi:1973fv} are obtained using
combined fits to inclusive $B$~decay
distributions~\cite{Bauer:2004ve,Buchmuller:2005zv,Abdallah:2005cx,:2007yaa}.
These analyses are based on calculations of the semileptonic decay
rate and spectral moments in $B\to X_c\ell\nu$ and $B\to
X_s\gamma$~decays in the frameworks of the Operator Product Expansion
(OPE) and the Heavy Quark Effective Theory
(HQET)~\cite{Bauer:2004ve,Benson:2003kp,Gambino:2004qm,Benson:2004sg},
which predict these quantities in terms of $|V_{cb}|$ and a number of
non-perturbative heavy quark (HQ) parameters including the $b$-quark
mass~$m_b$.

Analyses combining measurements from different
experiments~\cite{Bauer:2004ve,Buchmuller:2005zv} quote the most
precise numbers for $|V_{cb}|$ and $m_b$. However, as the correlated
systematic uncertainties are not precisely known, there is some
concern that uncertainties are underestimated. In this analysis, we
have chosen the opposite approach and perform fits to the data from
the Belle experiment only. In addition, we use two independent sets of
theoretical expressions, derived in the 1S~\cite{Bauer:2004ve} and
kinetic mass~\cite{Gambino:2004qm,Benson:2004sg} schemes respectively,
to test the compatibility of these two frameworks.

The present document is organized as follows: Sect.~\ref{sect:1}
describes the measurement of the first and second moment of the
inclusive photon energy spectrum in $B\to X_s\gamma$, $\langle
E_\gamma\rangle$ and $\langle(E_\gamma-\langle
E_\gamma\rangle)^2\rangle$, using the Belle measurement of this decay
in Ref.~\cite{Koppenburg:2004fz}. In the previously published analysis
the first and second moments were obtained for one value of the
minimum energy threshold, namely $E_\mathrm{min}=1.8$~GeV. Here we
report additional measurements with
$E_\mathrm{min}=1.9,2.0,2.1,2.2,2.3$~GeV, and perform a re-evaluation
of the systematic error. In Sect.~\ref{sect:2} we use these data
together with the recent Belle measurements of the lepton energy and
hadronic mass moments in $B\to
X_c\ell\nu$~decays~\cite{Urquijo:2006wd,Schwanda:2006nf} to extract
$|V_{cb}|$ and $m_b$ using theoretical expressions derived in the 1S
and kinetic mass schemes.
\section{\boldmath Moments of the $B\to X_s\gamma$ photon energy
  spectrum} \label{sect:1}

\subsection{\boldmath Review of the Belle $B\to X_s\gamma$
  Measurement}

The analysis described in Ref.~\cite{Koppenburg:2004fz} uses
$e^+e^-\to\Upsilon(4S)\to B\bar B$~events equivalent to 140~fb$^{-1}$
of integrated luminosity (ON sample) and 15~fb$^{-1}$ taken 60~MeV
below the $\Upsilon(4S)$~resonance energy (OFF sample). Photon
candidates with energy greater than 1.5~GeV as measured in the
$\Upsilon(4S)$~rest frame are reconstructed. Vetoes are applied to
photon candidates with high likelihood of originating from $\pi^0$ or
$\eta$~decays to two photons.

In general, the background of photons from the $e^+e^-\to q\bar
q$~continuum is dominant. It is suppressed with event shape variables
used as the inputs to two Fisher
discriminants~\cite{Fisher:1936et}. The first discriminant
distinguishes spherically-shaped $B\bar B$ from jet-like continuum
events and includes the Fox-Wolfram moments~\cite{Fox:1978vu}, the
thrust calculated using all particles detected in the event including
and excluding the candidate photon, and the angles of the
corresponding thrust axes with respect to the beam and candidate
photon directions, respectively. The second discriminant is designed
to exploit the topology of $B\to X_s\gamma$~events by utilizing the
energy sum of detected particles, which is measured in three angular
regions bounded by cones that are subtended from the direction of the
candidate photon in the $\Upsilon(4S)$~frame; defined as
$0^\circ-30^\circ$ (forward), $30^\circ-140^\circ$ (middle), and
$140^\circ-180^\circ$ (backward).

\begin{figure}
  \begin{center}
    \includegraphics[width=0.49\columnwidth]{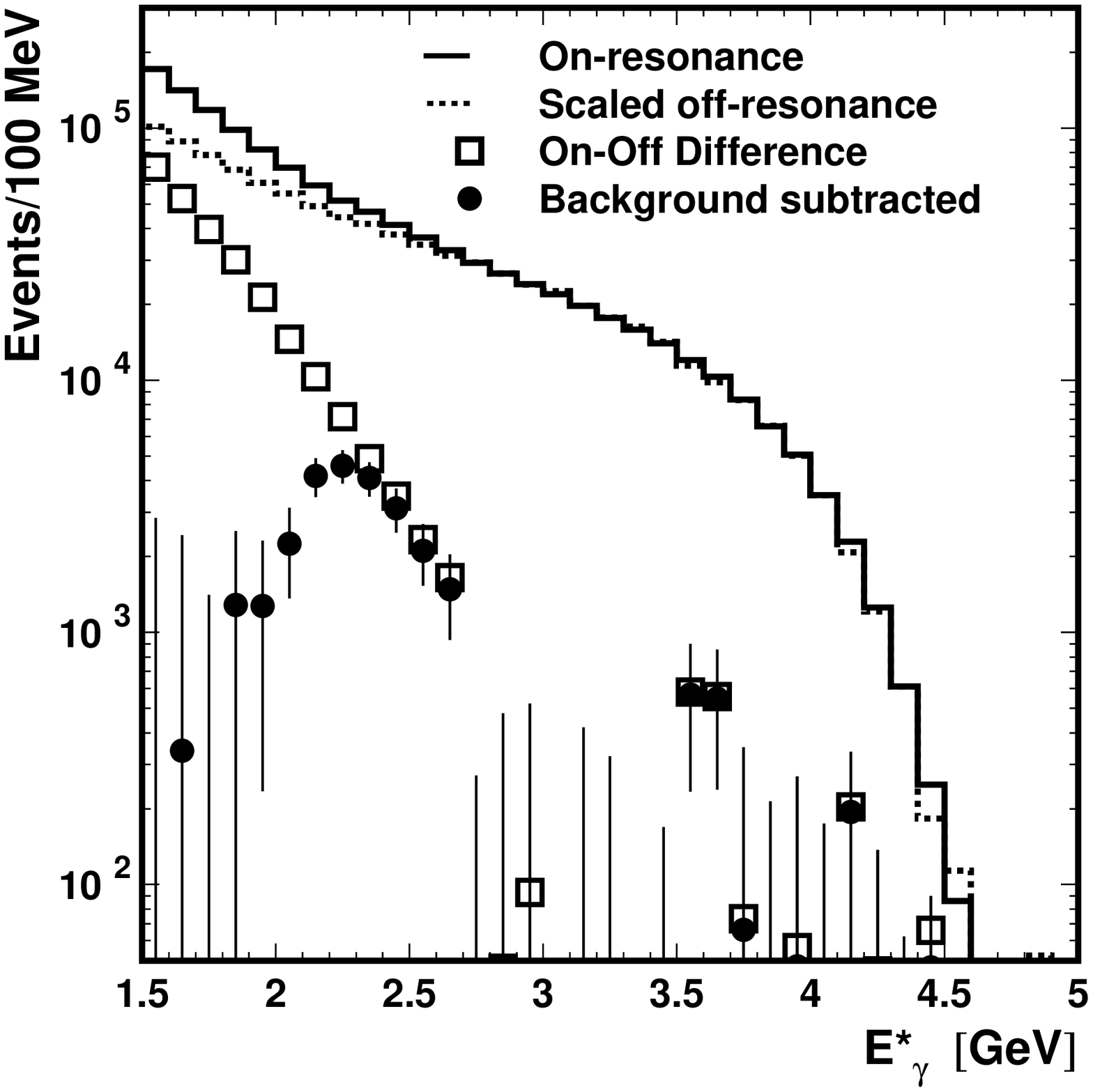}
    \includegraphics[width=0.49\columnwidth]{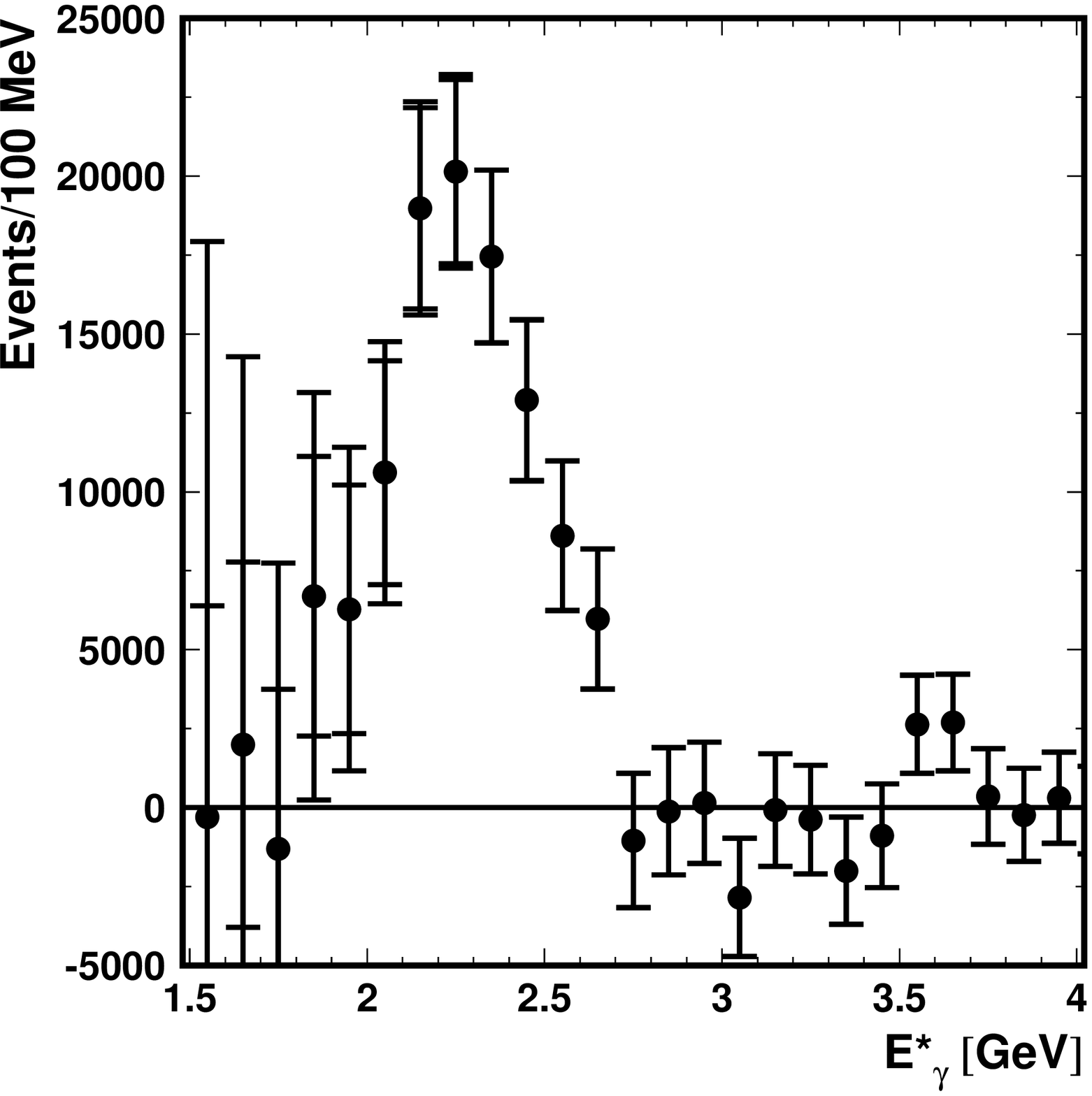}
  \end{center}
  \caption{Left: raw photon energy spectrum in the
  $\Upsilon(4S)$~frame; right: photon energy spectrum after background
  subtraction and efficiency correction where the inner error bars are
  the statistical uncertainties and the outer error bars show the total
  errors, which include the systematic uncertainties. These plots are
  reproduced from Ref.~\cite{Koppenburg:2004fz}.} \label{fig:1_1}
\end{figure}

After these selections are applied, the remaining continuum background
is removed by subtracting scaled OFF data from the ON data
set. Backgrounds in $B\bar B$~events, including photons from $\pi^0$
and $\eta$ (veto leakage), other real photons (mainly from  $\omega$,
$\eta'$, and $J/\psi$), clusters in the calorimeter not due to single
photons (mainly electrons interacting with matter, $K^0_L$ and $\bar
n$) and beam background, are estimated from Monte Carlo (MC)
simulation (Fig.~\ref{fig:1_1}).

\subsection{Moment Measurements}

We calculate the truncated first and second moments, $\langle
E_\gamma\rangle$ and $\langle(E_\gamma-\langle
E_\gamma\rangle)^2\rangle$, of the efficiency corrected spectrum in
Fig.~\ref{fig:1_1} for minimum photon energies ranging from 1.8 to
2.3~GeV. The following corrections are applied to these moments: The
non-zero $B$~meson momentum in the $\Upsilon(4S)$~rest frame changes
the first moment of the photon energy by 0.2\% and adds a Doppler
broadening of 0.006~GeV$^2$ to the second moment; the finite energy
resolution, uncorrected in Fig.~\ref{fig:1_1}, causes a broadening of
the spectrum and increases the second moment by 0.004~GeV$^2$; the
100~MeV binning in Fig.~\ref{fig:1_1} increases the second moment by
0.0008~GeV$^2$.

The above corrections assume a symmetric photon energy distribution,
and do not account for expected and known asymmetries in the true
spectrum and detector response, respectively. To account for these effects
an additional bias correction, derived from a MC simulation, is
implemented. The $B\to X_s\gamma$~model contains decays
of the form $B\to K^\ast\gamma$, where $K^\ast$ is any known spin-$1$
resonance with strangeness $S=1$. The relative amounts of these decays
are adjusted by matching the total photon spectrum to the theoretical
model of Ref.~\cite{Kagan:1998ym}. The bias correction, calculated
as the difference of the true moment and the moment measured in the
$B\to X_s\gamma$~MC simulation once all aforementioned corrections are
applied, is listed in Table~\ref{tab:1_1}.
\begin{table}
  \caption{Residual bias correction to $\langle E_\gamma\rangle$ and
    $\langle(E_\gamma-\langle E_\gamma\rangle)^2\rangle$ as a function
    of $E_\mathrm{min}$.}
  \label{tab:1_1}
    \begin{center}
      \begin{tabular}{c|@{\extracolsep{.3cm}}cc}
	\hline\hline
	$E_\mathrm{min}$ (GeV) & $\Delta\langle E_\gamma\rangle$ &
	$\Delta\langle(E_\gamma-\langle E_\gamma\rangle)^2\rangle$\\
	\hline
	1.8 & +2.0\% &  0.0\% \\
	1.9 & +1.6\%   & -0.4\% \\
	2.0 & +1.2\%   & -7.1\% \\
	2.1 & +0.8\%   & -17.4\% \\
	2.2 & +0.2\%   & -35.3\% \\
	2.3 & -0.3\%   & -57.9\% \\
	\hline \hline
      \end{tabular}
    \end{center}
\end{table}

\subsection{Systematic Uncertainties}

The error bars of the efficiency corrected spectrum of
Fig.~\ref{fig:1_1} show the total error including the systematic
uncertainty related to the scaling of the MC background samples
(sizable in the first energy bins). In the calculation of the
moments, we consider also the following sources of systematic
uncertainty: uncertainty in the OFF data scaling factor; possible
difference in ON and OFF data selection efficiencies; uncertainty in the
$B\bar B$~data/MC correction; we vary by $\pm 20\%$ the background
from $\eta^\prime$, $\omega$ and bremsstrahlung; uncertainty of the
$\eta$~veto efficiency; we consider an alternate signal MC that favors
high-mass resonances decaying into high-multiplicity final states,
where the fraction of $\gamma K\pi$~final states, somewhat
overestimated in our default sample, matches our previous
measurement~\cite{Nishida:2003yw}; and we vary the photon detection
efficiency in both signal and background samples by its measured
uncertainty ($\pm 2.3\%$).

We also assign systematic uncertainties to the corrections applied to
the moments:
an alternate energy resolution correction
that neglects the lower energy tail in the resolution is
implemented and the difference is assigned as systematic uncertainty;
a $\pm 100\%$ uncertainty on the binning correction for the second
moment is assigned; we also implement a $\pm 50\%$ variation on the
bias correction for the first moment while for the second moment the
correction is re-calculated using the alternate signal MC sample.

The total systematic uncertainty on each moment measurement is
obtained by summing the aforementioned contributions in quadrature
(Tables~\ref{tab:1_2} and \ref{tab:1_3}).
\begin{table}
  \caption{Systematic uncertainties contributing to the first
    moment~$\langle E_\gamma\rangle$ as a function of the lower energy
    threshold $E_\mathrm{min}$ in GeV.}
  \label{tab:1_2}
  \begin{center}
    \begin{tabular}{l|@{\extracolsep{.3cm}}cccccc}
      \hline \hline
      $E_\mathrm{min}$ (GeV) & 1.8 & 1.9 & 2.0 & 2.1 & 2.2 & 2.3\\
      \hline
      MC scaling & 0.021 & 0.012 & 0.006 & 0.003 & 0.002 & 0.001\\
      OFF scaling & 0.004 & 0.001 & 0.001 & 0.002 & 0.002 & 0.002\\
      ON/OFF efficiency & 0.000  & 0.000  & 0.000  & 0.000  & 0.000 & 0.000\\
      $B\bar B$ data/MC correction & 0.005 & 0.003 & 0.002 & 0.001 &
      0.000 & 0.000\\
      other $\gamma$s in $B\bar B$ & 0.010 & 0.004 & 0.002 & 0.000 &
      0.000 & 0.000\\
      $\eta$~veto efficiency & 0.001 & 0.001 & 0.000 & 0.000 & 0.000 &
      0.000\\
      signal MC & 0.004 & 0.004 & 0.004 & 0.004 & 0.004 & 0.003\\
      $\gamma$ efficiency & 0.001 & 0.001 & 0.000 & 0.000 & 0.000 & 0.000\\
      bias correction & 0.022 & 0.018 & 0.014 & 0.009 & 0.002 & 0.003\\
      \hline
      total systematic & 0.033 & 0.023 & 0.016 & 0.010 & 0.005 &
      0.004\\
      \hline \hline
    \end{tabular}
  \end{center}
\end{table}
\begin{table}
  \caption{Systematic uncertainties contributing to the second
    moment~$\langle(E_\gamma-\langle E_\gamma\rangle)^2\rangle$ as a
    function of the lower energy threshold $E_\mathrm{min}$ in GeV$^2$.}
  \label{tab:1_3}
  \begin{center}
    \begin{tabular}{l|@{\extracolsep{.3cm}}cccccc}
      \hline \hline
      $E_\mathrm{min}$ (GeV) & 1.8 & 1.9 & 2.0 & 2.1 & 2.2 & 2.3\\
      \hline
      MC scaling & 0.0060 & 0.0027 & 0.0009 & 0.0003 & 0.0001 & 0.0001\\
      OFF scaling & 0.0018 & 0.0010 & 0.0006 & 0.0005 & 0.0005 & 0.0004\\
      ON/OFF efficiency & 0.0000 & 0.0000 & 0.0000 & 0.0000 & 0.0000 &
      0.0000\\
      $B\bar B$ data/MC correction & 0.0010 & 0.0004 & 0.0001 & 0.0000
      & 0.0000 & 0.0000\\
      other $\gamma$s in $B\bar B$ & 0.0024 & 0.0008 & 0.0002 & 0.0000
      & 0.0000 & 0.0000\\
      $\eta$~veto efficiency & 0.0003 & 0.0001 & 0.0000 & 0.0000 &
      0.0000 & 0.0000\\
      signal MC & 0.0007 & 0.0005 & 0.0004 & 0.0003 & 0.0002 & 0.0000\\
      $\gamma$ efficiency & 0.0003 & 0.0001 & 0.0000 & 0.0000 & 0.0000
      & 0.0000\\
      energy resolution & 0.0020 & 0.0020 & 0.0021 & 0.0022 & 0.0023 & 0.0024\\
      binning & 0.0008 & 0.0008 & 0.0008 & 0.0008 & 0.0008 & 0.0008\\
      bias correction & 0.0068 & 0.0040 & 0.0026 & 0.0005 & 0.0004 & 0.0011\\
      \hline
      total systematic & 0.0099 & 0.0055 & 0.0036 & 0.0024 & 0.0025 &
      0.0028\\
      \hline \hline
    \end{tabular}
  \end{center}
\end{table}

\subsection{Results}

\begin{figure}
  \begin{center}
    \includegraphics[width=0.49\columnwidth]{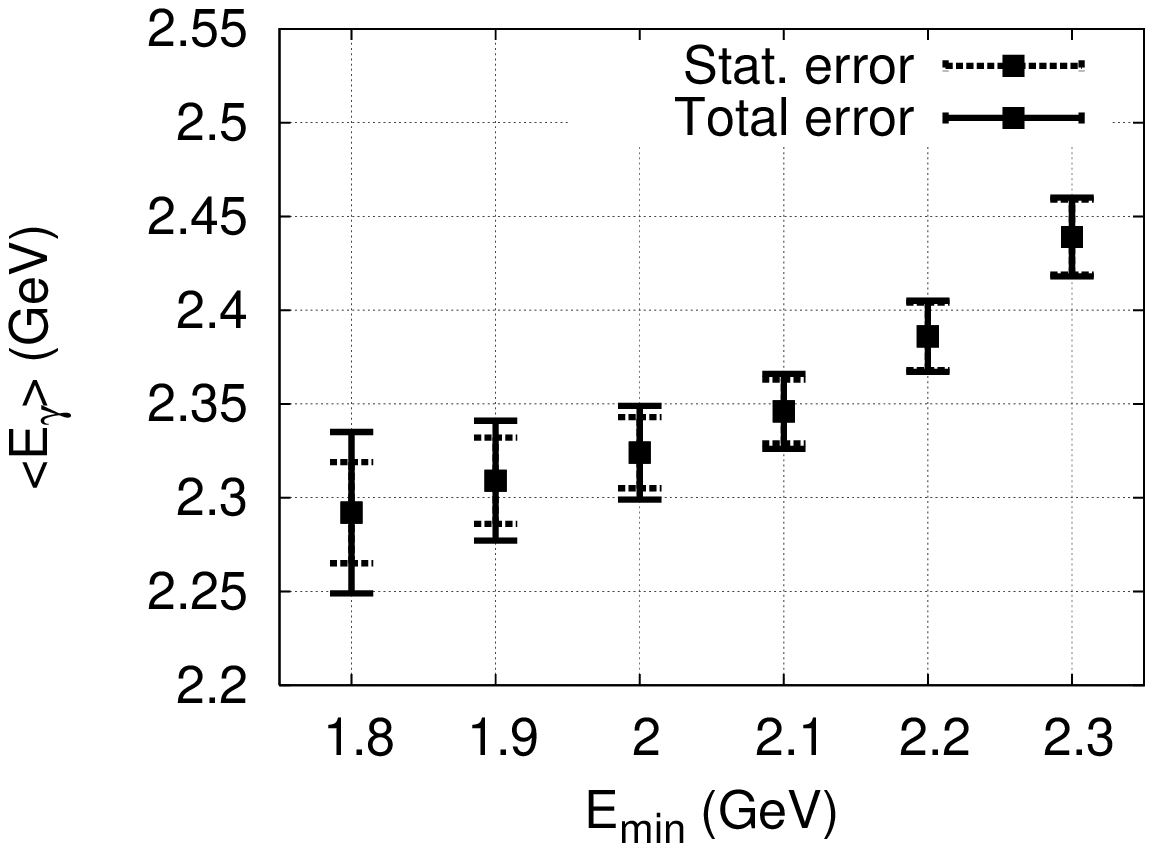}
    \includegraphics[width=0.49\columnwidth]{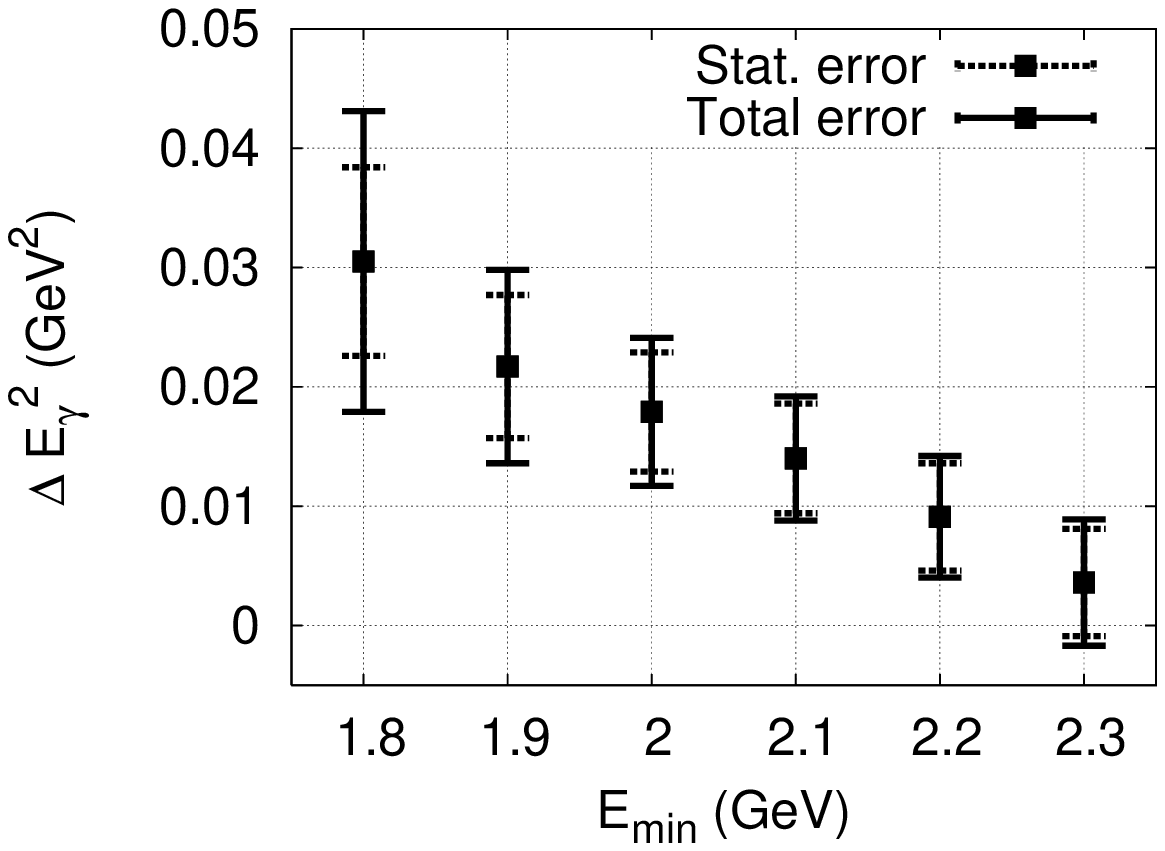}
  \end{center}
  \caption{Measurements of $\langle E_\gamma\rangle$ and
    $\langle(E_\gamma-\langle E_\gamma\rangle)^2\rangle$ as a function
    of the minimum photon energy $E_\mathrm{min}$ ($\Delta
    E^2_\gamma=\langle(E_\gamma-\langle E_\gamma\rangle)^2\rangle$).}
    \label{fig:1_2}
\end{figure}

The measurements of the first and second moments of the photon energy
spectrum in $B\to X_s\gamma$ for minimum photon energies ranging from
1.8~GeV to 2.3~GeV are shown in Table~\ref{tab:1_4} and
Fig.~\ref{fig:1_2}. Our results agree with the data from
CLEO~\cite{Chen:2001fja} and BaBar~\cite{Aubert:2005cua}.
\begin{table}
  \caption{Measurements of $\langle E_\gamma\rangle$ and
    $\langle(E_\gamma-\langle E_\gamma\rangle)^2\rangle$ as a function
    of the minimum photon energy $E_\mathrm{min}$. The first error is
    statistical and the second is systematic.} \label{tab:1_4}
  \begin{center}
    \begin{tabular}{c|@{\extracolsep{.3cm}}cc}
      \hline \hline
      $E_\mathrm{min}$ (GeV) & $\langle E_\gamma\rangle$ (GeV) &
      $\langle(E_\gamma-\langle E_\gamma\rangle)^2\rangle$\\
      \hline
      1.8 & $2.292\pm 0.027\pm 0.033$ & $0.0305\pm 0.0079\pm 0.0099$\\
      1.9 & $2.309\pm 0.023\pm 0.023$ & $0.0217\pm 0.0060\pm 0.0055$\\
      2.0 & $2.324\pm 0.019\pm 0.016$ & $0.0179\pm 0.0050\pm 0.0036$\\
      2.1 & $2.346\pm 0.017\pm 0.010$ & $0.0140\pm 0.0046\pm 0.0024$\\
      2.2 & $2.386\pm 0.018\pm 0.005$ & $0.0091\pm 0.0045\pm 0.0025$\\
      2.3 & $2.439\pm 0.020\pm 0.004$ & $0.0036\pm 0.0045\pm 0.0028$\\
      \hline \hline
    \end{tabular}
  \end{center}
\end{table}

The statistical and systematic errors on the first and second moments
at $E_\mathrm{min}=1.8$~GeV are slightly different from the values
quoted in Ref.~\cite{Koppenburg:2004fz} and supersede our previously
published values. The change in the uncertainties is due to the use of
the toy MC approach and to the additional contribution from the
uncertainty in the bias correction.

The correlations between the different moment measurements are
estimated using a toy MC approach: Starting from the efficiency
corrected spectrum in Fig.~\ref{fig:1_1} we create new spectra by
generating values of a Gaussian random variable for the contents of
each bin, where the mean and standard deviation of the Gaussian
correspond to the bin yield and its uncertainty in the original
spectrum. The moments and their fluctuations with respect to each
other were measured for each generated spectrum, and finally averaged
to yield the covariance matrix, from which the uncertainties due to
statistics and systematics scaling were obtained. The covariance
matrix was also obtained from systematic variations due to the
aforementioned corrections to the moments. The method assumes 100\%
correlation of any two truncated moments due to any single systematic
variation. The covariance matrices that are derived from statistics
and systematics are added to yield the overall covariance matrix, from
which the correlations between any of the truncated moments are
deduced. Tables~\ref{tab:1_5}$-$\ref{tab:1_7} show the correlation coefficients derived from this study.
\begin{table}
  \caption{Correlation coefficients between the $\langle
    E_\gamma\rangle$~measurements.} \label{tab:1_5}
  \begin{center}
    \begin{tabular}{cc|cccccc}
      \hline \hline
      \multicolumn{2}{c|}{$E_\mathrm{min}$} &
      \multicolumn{6}{c}{$\langle E_\gamma\rangle$}\\
      \multicolumn{2}{c|}{(GeV)} & 1.8  & 1.9  & 2.0  & 2.1  & 2.2  &
      2.3\\
      \hline
      & 1.8 & $\phantom{-}1.00$ & $\phantom{-}0.79$ &
      $\phantom{-}0.68$ & $\phantom{-}0.56$ & $\phantom{-}0.38$ &
      $\phantom{-}0.22$\\
      & 1.9 & & $\phantom{-}1.00$ & $\phantom{-}0.82$ &
      $\phantom{-}0.70$ & $\phantom{-}0.52$ & $\phantom{-}0.33$\\
      $\langle E_\gamma\rangle$ & 2.0 & & & $\phantom{-}1.00$
      & $\phantom{-}0.86$ & $\phantom{-}0.67$ & $\phantom{-}0.47$\\
      & 2.1 & & & & $\phantom{-}1.00$ & $\phantom{-}0.84$ &
      $\phantom{-}0.65$\\
      & 2.2 & & & & & $\phantom{-}1.00$ & $\phantom{-}0.86$\\
      & 2.3 & & & & & & $\phantom{-}1.00$\\
      \hline \hline
    \end{tabular}
  \end{center}
\end{table}
\begin{table}
  \caption{Correlation coefficients between the $\langle
    E_\gamma\rangle$ and $\langle(E_\gamma-\langle
    E_\gamma\rangle)^2\rangle$~measurements.} \label{tab:1_6}
  \begin{center}
    \begin{tabular}{cc|cccccc}
      \hline \hline
      \multicolumn{2}{c|}{$E_\mathrm{min}$} &
      \multicolumn{6}{c}{$\langle(E_\gamma-\langle
      E_\gamma\rangle)^2\rangle$}\\
      \multicolumn{2}{c|}{(GeV)} & 1.8  & 1.9  & 2.0  & 2.1  & 2.2  &
      2.3\\
      \hline
      & 1.8 & $-0.46$ & $-0.18$ & $-0.01$ & $\phantom{-}0.04$ &
      $\phantom{-}0.01$ & $-0.01$\\
      & 1.9 & $-0.06$ & $-0.21$ & $\phantom{-}0.05$ &
      $\phantom{-}0.12$ & $\phantom{-}0.10$ & $\phantom{-}0.07$\\
      $\langle E_\gamma\rangle$ & 2.0 & $-0.14$ & $\phantom{-}0.15$ &
      $\phantom{-}0.12$ & $\phantom{-}0.23$ & $\phantom{-}0.20$ &
      $\phantom{-}0.17$\\
      & 2.1 & $\phantom{-}0.27$ & $\phantom{-}0.37$ &
      $\phantom{-}0.43$ & $\phantom{-}0.42$ & $\phantom{-}0.39$ &
      $\phantom{-}0.34$\\
      & 2.2 & $\phantom{-}0.38$ & $\phantom{-}0.55$ &
      $\phantom{-}0.67$ & $\phantom{-}0.75$ & $\phantom{-}0.66$ &
      $\phantom{-}0.61$\\
      & 2.3 & $\phantom{-}0.43$ & $\phantom{-}0.63$ &
      $\phantom{-}0.79$ & $\phantom{-}0.91$ & $\phantom{-}0.88$ &
      $\phantom{-}0.79$\\
      \hline \hline
    \end{tabular}
  \end{center}
\end{table}
\begin{table}
  \caption{Correlation coefficients between the $\langle(E_\gamma-\langle
    E_\gamma\rangle)^2\rangle$~measurements.} \label{tab:1_7}
  \begin{center}
    \begin{tabular}{cc|cccccc}
      \hline \hline
      \multicolumn{2}{c|}{$E_\mathrm{min}$} &
      \multicolumn{6}{c}{$\langle(E_\gamma-\langle
      E_\gamma\rangle)^2\rangle$}\\
      \multicolumn{2}{c|}{(GeV)} & 1.8  & 1.9  & 2.0  & 2.1  & 2.2  &
      2.3\\
      \hline
      & 1.8 & $\phantom{-}1.00$ & $\phantom{-}0.72$ &
      $\phantom{-}0.63$ & $\phantom{-}0.49$ & $\phantom{-}0.39$ &
      $\phantom{-}0.30$\\
      & 1.9 & & $\phantom{-}1.00$ & $\phantom{-}0.83$ &
      $\phantom{-}0.71$ & $\phantom{-}0.61$ & $\phantom{-}0.52$\\
      $\langle(E_\gamma-$ & 2.0 & & & $\phantom{-}1.00$ &
      $\phantom{-}0.89$ & $\phantom{-}0.80$ & $\phantom{-}0.71$\\
      $-\langle E_\gamma\rangle)^2\rangle$ & 2.1 & & & &
      $\phantom{-}1.00$ & $\phantom{-}0.96$ & $\phantom{-}0.91$\\
      & 2.2 & & & & & $\phantom{-}1.00$ & $\phantom{-}0.97$\\
      & 2.3 & & & & & &  $\phantom{-}1.00$\\
      \hline\hline
    \end{tabular}
  \end{center}
\end{table}

\begin{figure}
  \begin{center}
    \includegraphics[width=0.49\columnwidth]{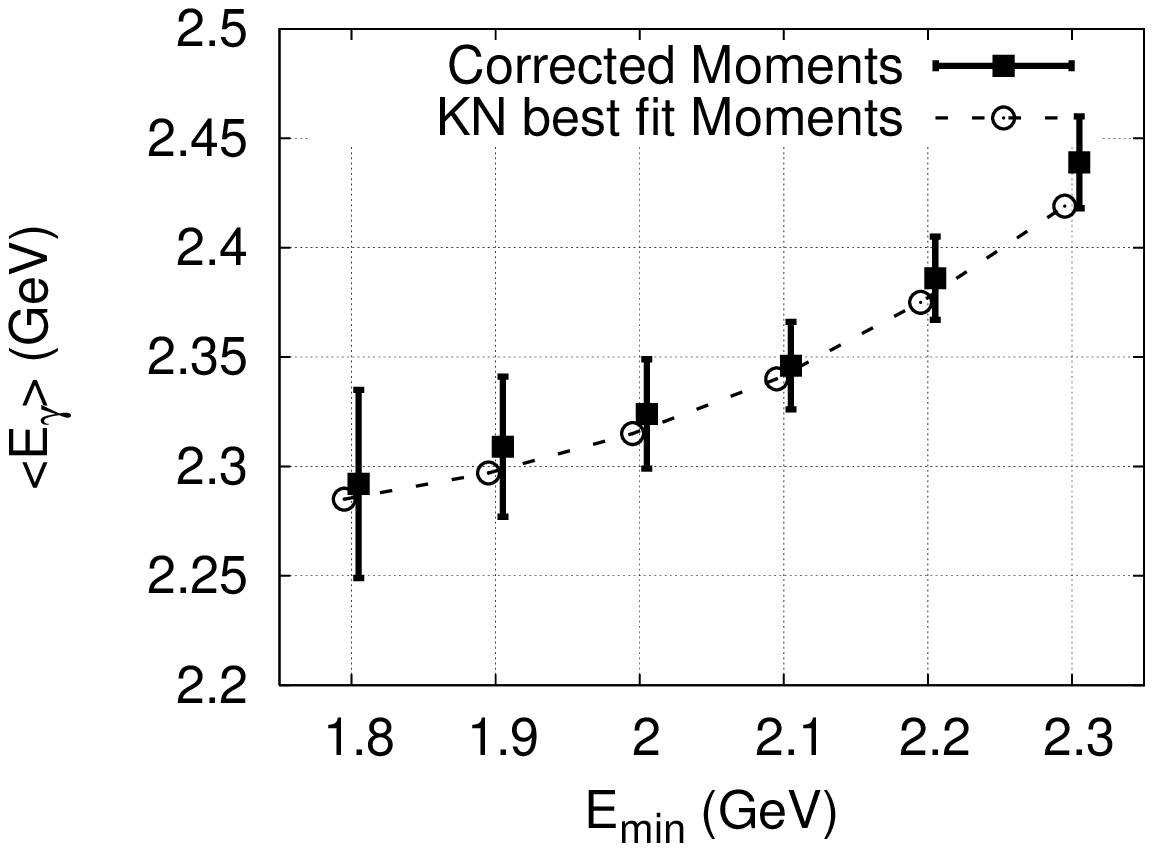}
    \includegraphics[width=0.49\columnwidth]{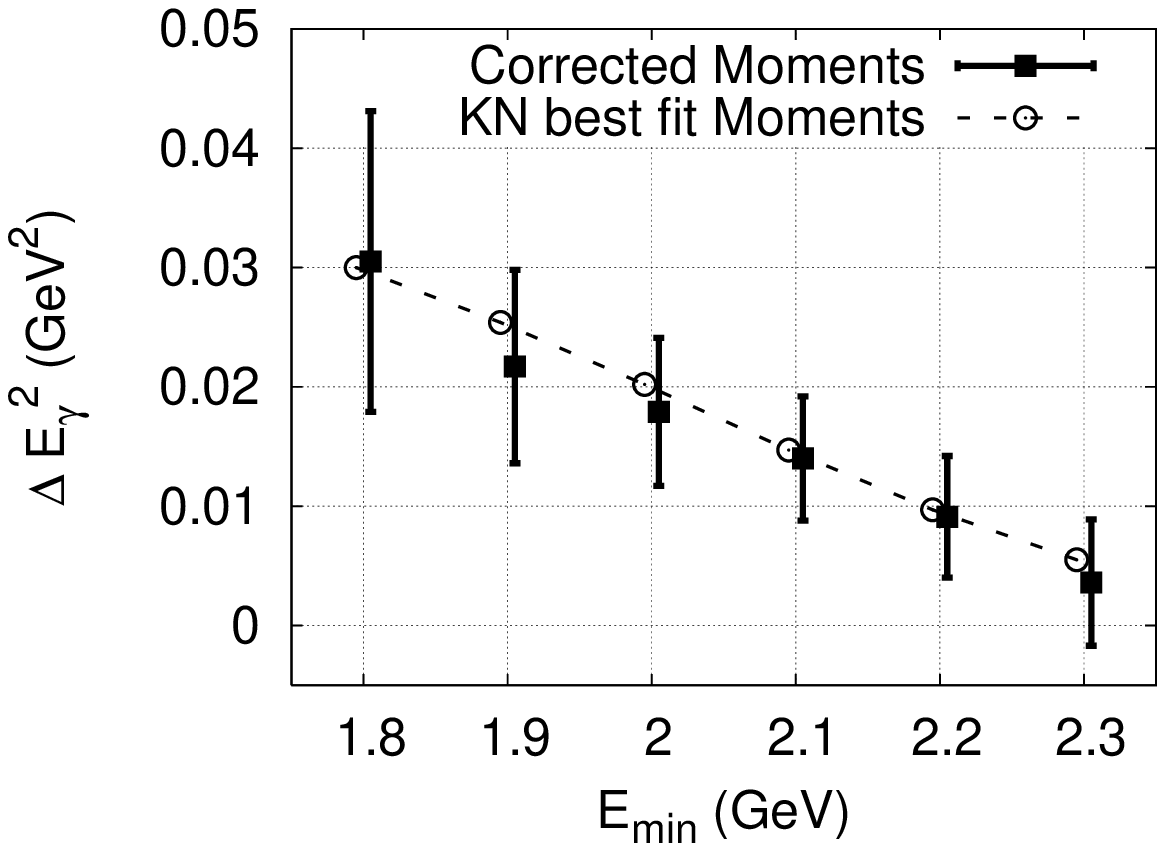}
  \end{center}
  \caption{Cross-check of the moment measurements. The moment
  measurements presented here are compared to the moments predicted in
  the Kagan-Neubert prescription, tuned to fit our data ($\Delta
  E^2_\gamma=\langle(E_\gamma-\langle E_\gamma\rangle)^2\rangle$).}
  \label{fig:1_3}
\end{figure}

To cross-check these moment measurements, we extract the moments from
the Kagan-Neubert (KN) photon spectrum~\cite{Kagan:1998ym} tuned to fit our
data~\cite{Limosani:2004jk} ($m_b(\mathrm{KN})=4.62$~GeV,
$\mu_\pi^2(\mathrm{KN})=0.40$~GeV$^2$). We generate the photon
spectrum in the rest frame of the $B$~meson with these parameters and
extract the moments in the range $E_\mathrm{min}=1.8, \dots,
2.3$~GeV. The results are plotted in Fig.~\ref{fig:1_3} along with the
moment measurements presented here. We find very good agreement
between these independent methods.

\section{\boldmath Extraction of $|V_{cb}|$ and $m_b$ from inclusive
  $B$~decays} \label{sect:2}

\subsection{Experimental Inputs}

Belle has measured the partial branching fractions~$\Delta\mathcal{B}$
and the first, second, third and fourth moments of the truncated
electron energy spectrum in $B\to X_ce\nu$, $\langle E_\ell\rangle$,
$\langle (E_\ell-\langle E_\ell\rangle)^2\rangle$, $\langle (E_\ell-\langle
E_\ell\rangle)^3\rangle$ and $\langle (E_\ell-\langle
E_\ell\rangle)^4\rangle$, for nine different electron energy
thresholds ($E_\mathrm{min}=0.4$, 0.6, 0.8, 1.0, 1.2, 1.4, 1.6, 1.8
and 2.0~GeV)~\cite{Urquijo:2006wd}. This analysis uses
$\Upsilon(4S)\to B\bar B$~events equivalent to 140~fb$^{-1}$ of
integrated luminosity. The hadronic decay of one $B$~meson is fully
reconstructed and $B\to X_ce\nu$~decays of the other $B$ are selected by
requiring an identified electron amongst the particles remaining in
the event.

In addition, Belle has measured the first, second central and second
non-central moments of the hadron invariant mass squared
($M^2_X$)~spectrum in $B\to X_c\ell\nu$, $\langle M^2_X\rangle$,
$\langle(M^2_X-\langle M^2_X\rangle)^2\rangle$ and $\langle
M^4_X\rangle$, for seven different lepton energy thresholds
($E_\mathrm{min}=0.7$, 0.9, 1.1, 1.3, 1.5, 1.7 and
1.9~GeV)~\cite{Schwanda:2006nf}. This analysis is also based on
140~fb$^{-1}$ of $\Upsilon(4S)$~data. Again, one $B$~meson is fully
reconstructed and a charged lepton (electron or muon) from the decay
of the other $B$ is required. The hadronic $X_c$~system is
reconstructed by summing the 4-momenta of the particles remaining in
the event.

The measurements of the first and second moments of the photon energy
spectrum in $B\to X_s\gamma$, $\langle E_\gamma\rangle$ and
$\langle(E_\gamma-\langle E_\gamma\rangle)^2\rangle$, have been
described previously in this document. They are available for six
different photon energy thresholds ($E_\mathrm{min}=1.8$, 1.9, 2.0,
2.1, 2.2 and 2.3~GeV).

Hence, there are a total of 71 Belle measurements of inclusive spectra
in $B$~decays available for the global analysis~\cite{ref:1}. The
measurements actually used in the 1S and kinetic mass scheme fit
analyses are shown in Table~\ref{tab:2_1}: We have excluded
measurements that do not have corresponding theoretical predictions;
measurements with high $E_\mathrm{min}$ cut-offs ({\it i.e.}, electron
energy and hadronic mass moments with $E_\mathrm{min}>1.5$~GeV and
photon energy moments with $E_\mathrm{min}>2$~GeV) are not used to
determine the HQ parameters, as theoretical expressions are not
considered reliable in this
region~\cite{Benson:2004sg,Uraltsev:2003bw}; finally, we have also
excluded measurements where correlations with neighboring points are too
high as these measurements do not add new information to the fit and
introduce numerical problems such as negative eigenvalues of the
covariance matrix.
\begin{table}
  \caption{Experimental inputs used in the 1S and kinetic mass scheme
    analyses. Both analyses use a total of 25
    measurements.}\label{tab:2_1}
  \begin{center}
    \begin{tabular}{c@{\extracolsep{.3cm}}|l}
      \hline \hline
      & Measurements used\\
      \hline
      & $n=0$: $E_\mathrm{min}=0.6$, 1.0, 1.4~GeV\\
      Lepton energy moments~$\langle E^n_\ell\rangle$ &
      $n=1$: $E_\mathrm{min}=0.6$, 0.8, 1.0, 1.2, 1.4~GeV\\
      & $n=2$: $E_\mathrm{min}=0.6$, 1.0, 1.4~GeV\\
      & $n=3$: $E_\mathrm{min}=0.8$, 1.0, 1.2~GeV\\
      \hline
      Hadronic mass moments~$\langle M^{2n}_X\rangle$ & $n=1$:
      $E_\mathrm{min}=0.7$, 1.1, 1.3, 1.5~GeV\\
      & $n=2$: $E_\mathrm{min}=0.7$, 0.9, 1.3~GeV\\
      \hline
      Photon energy moments~$\langle E^n_\gamma\rangle$ & $n=1$:
      $E_\mathrm{min}=1.8$, 2.0~GeV\\
      & $n=2$: $E_\mathrm{min}=1.8$, 2.0~GeV\\
      \hline \hline
    \end{tabular}
  \end{center}
\end{table}

The value of $|V_{cb}|$ is dependent on the average lifetime~$\tau_B$ of
neutral and charged $B$~mesons. In the following analyses we use
$\tau_B=(1.585\pm 0.006)$~ps based on Ref.~\cite{Barberio:2007cr} and
assume equal production of charged and neutral $B$~mesons.
\subsection{1S Mass Scheme Analysis}

\subsubsection{Theoretical Input}

The parameters appearing in the OPE depend on the choice of the mass
scheme, {\it i.e.}, the definition of $m_b$. The 1S scheme eliminates
the $b$-quark pole mass by relating it to the mass of the
$\Upsilon(1S)$. Truncated spectral moments in $B\to X_c\ell\nu$ have
been derived in this scheme up to
$\mathcal{O}(1/m^3_b)$~\cite{Bauer:2004ve}. The theoretical
expressions are of the form
\begin{equation}
  \begin{split}
    \langle X\rangle_{E_\mathrm{min}} =
    X^{(1)}+X^{(2)}\Lambda+X^{(3)}\Lambda^2+X^{(4)}\Lambda^3+X^{(5)}\lambda_1+X^{(6)}\Lambda\lambda_1+X^{(7)}\lambda_2+X^{(8)}\Lambda\lambda_2+X^{(9)}\rho_1\\+X^{(10)}\rho_2+X^{(11)}\tau_1+X^{(12)}\tau_2+X^{(13)}\tau_3+X^{(14)}\tau_4+X^{(15)}\epsilon+X^{(16)}\epsilon^2_{\rm{BLM}}+X^{(17)}\epsilon\Lambda~,
  \end{split} \label{eq:2_1}
\end{equation}
where $\langle X\rangle$ stands for any experimental observable in
Table~\ref{tab:2_1} and $X^{(i)}$, $i=1, \dots, 17$, are
perturbatively calculable coefficients that depend on
$E_\mathrm{min}$. The computations include radiative contributions of
$\mathcal{O}(\epsilon)$ and $\mathcal{O}(\epsilon^2_{\rm BLM})$, the
so-called BLM contribution at $\mathcal{O}(\epsilon^2)$. The HQ
parameters are $\Lambda$ at leading order, $\lambda_1$ and $\lambda_2$
at $\mathcal{O}(1/m_b^2)$, and $\tau_1$, $\tau_2$, $\tau_3$, $\tau_4$,
$\rho_1$ and $\rho_2$ at $\mathcal{O}(1/m_b^3)$. The CKM magnitude
$|V_{cb}|$ enters through the predictions of the partial semileptonic
branching fractions,
\begin{equation}
  \Delta\mathcal{B}_{E_\mathrm{min}}=\frac{G_F^2m^5}{192\pi^3}|V_{cb}|^2\eta_\mathrm{QED}\tau_B\langle
  X\rangle_{\Delta\mathcal{B},E_\mathrm{min}}~,
\end{equation}
where $m$ is the 1S reference mass, $m=m_{\Upsilon(\mathrm{1S})}/2$,
$G_F^2m^5/(192\pi^3)=5.4\times 10^{-11}$~ps$^{-1}$,
$\eta_\mathrm{QED}=1.007$ and $\langle
X\rangle_{\Delta\mathcal{B},E_\mathrm{min}}$ is an expression of the
form of Eq.~\ref{eq:2_1}.

The analysis in the 1S mass scheme determines a total of seven
parameters: $|V_{cb}|$, $\Lambda$, $\lambda_1$, $\tau_1$, $\tau_2$,
$\tau_3$ and $\rho_1$. Following the prescriptions in
Ref.~\cite{Bauer:2004ve}, $\tau_4$ is set to zero and the measured
$B^*-B$ and $D^*-D$~mass splittings allow us to constrain some of the
HQET matrix elements in Eq.~\ref{eq:2_1}:
$\lambda_2=0.1227-0.0145\lambda_1$ and $\rho_2=0.1361+\tau_2$. The
parameter~$\Lambda$ is the difference between the $b$-quark and the
reference mass,
$\Lambda=m_{\Upsilon(\mathrm{1S})}/2-m_b^\mathrm{1S}$. We will present
our results in terms of $m_b^\mathrm{1S}$.

\subsubsection{The Fit}

The expressions in the 1S scheme are fitted to the data using the
$\chi^2$~minimization technique and the MINUIT
program~\cite{James:1975dr}. The covariance matrix used in the fit
takes into account both experimental and theoretical
uncertainties. Following the approach in Ref.~\cite{Bauer:2004ve}, an
element of the combined experimental and theoretical error matrix is
given by
\begin{equation}
  \sigma_{ij}^2=\sigma_i\sigma_j c_{ij}~,
\end{equation}
where $i$ and $j$ denote the observables and $c_{ij}$ is the
experimental correlation matrix element. The total error on the
observable~$i$ is defined as
\begin{eqnarray}
  \sigma_i & = & \sqrt{(\sigma_i^\mathrm{exp})^2 + (Af_nm_B^{2n})^2 +
      (B_i/2)^2} \nonumber\\
      & & \mbox{for the $n$th hadron moment~,} \nonumber\\
  \sigma_i & = & \sqrt{(\sigma_i^\mathrm{exp})^2 + (Af_n(m_B/2)^n)^2 +
      (B_i/2)^2} \nonumber\\
      & & \mbox{for the $n$th lepton moment~,} \nonumber\\
  \sigma_i & = & \sqrt{(\sigma_i^\mathrm{exp})^2 + (Af_n(m_B/2)^n)^2 +
      (B_i/2)^2} \nonumber\\
      & & \mbox{for the $n$th photon moment~,}
\end{eqnarray}
and $f_0=f_1=1$, $f_2=1/4$ and $f_3=1/(6\sqrt{3})$. Here,
$\sigma_i^{\rm{exp}}$ are the experimental errors, $B_i=X^{(16)}$ are
the coefficients of the last computed terms in the perturbation series
(used to estimate the uncertainty on the uncalculated higher order
perturbative terms), and $A$ is a dimensionless parameter that
contains different theoretical uncertainties (uncalculated power
corrections, uncalculated effects of order
$(\alpha_s/4\pi)\Lambda_\mathrm{QCD}^2/m_b^2$, and effects not
included in the OPE, {\it i.e.}, duality violation). For lepton and
hadron moments, we fix $A=0.001$~\cite{Bauer:2004ve}. For photon
moments, the factor $A$ is 0.001 multiplied by the ratio of the energy
difference from the endpoint, relative to that for
$E_\mathrm{min}=1.8$~GeV, to account for the increase in shape
function effects as one limits the allowed region of the photon
spectrum.

As the fit does not provide strong constraints on the
$1/m^3_b$~parameters, we add the following extra terms to the
$\chi^2$~function,
\begin{equation}
  \chi^2_{\langle\mathcal{O}\rangle}=
    \begin{cases} 
      0 & |\langle\mathcal{O}\rangle|\le m_\chi^3~,\\
      \left(|\langle\mathcal{O}\rangle|-m_\chi^3\right)^2/M_\chi^6 &
      |\langle\mathcal{O}\rangle|>m_\chi^3~,
    \end{cases} 
\end{equation}
where ($m_\chi$,$M_\chi$) are both quantities of
$\mathcal{O}(\Lambda_\mathrm{QCD})$, and $\langle\mathcal{O}\rangle$
are the matrix elements of any of the $\mathcal{O}(1/m^3_b)$~operators
in the fit. For the central value of the fit, we take
$M_\chi=m_\chi=500$~MeV~\cite{Bauer:2004ve}.

\subsubsection{Results and Discussion}

\begin{figure}
  \begin{center}
    \includegraphics[width=0.24\columnwidth]{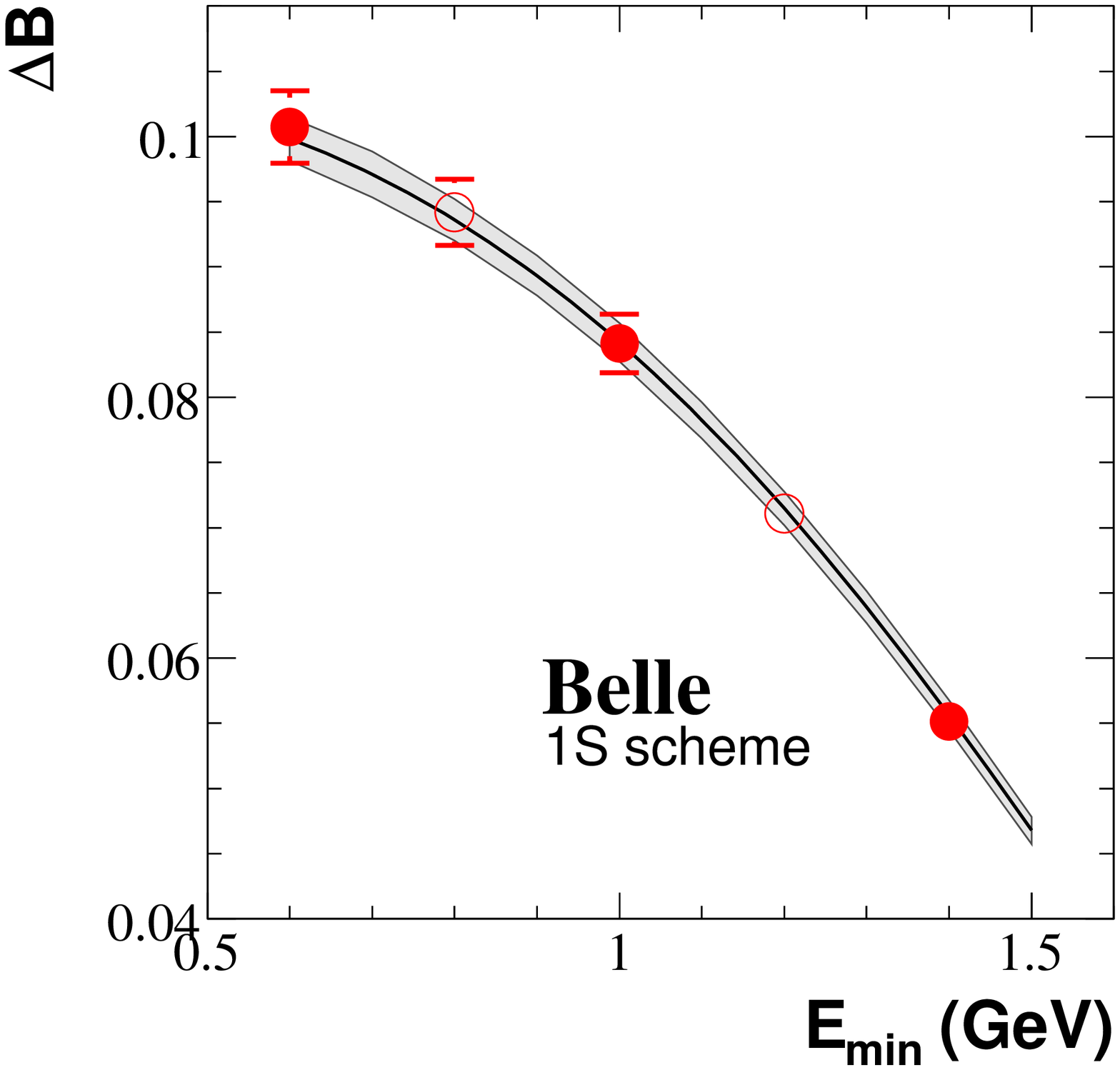}
    \includegraphics[width=0.24\columnwidth]{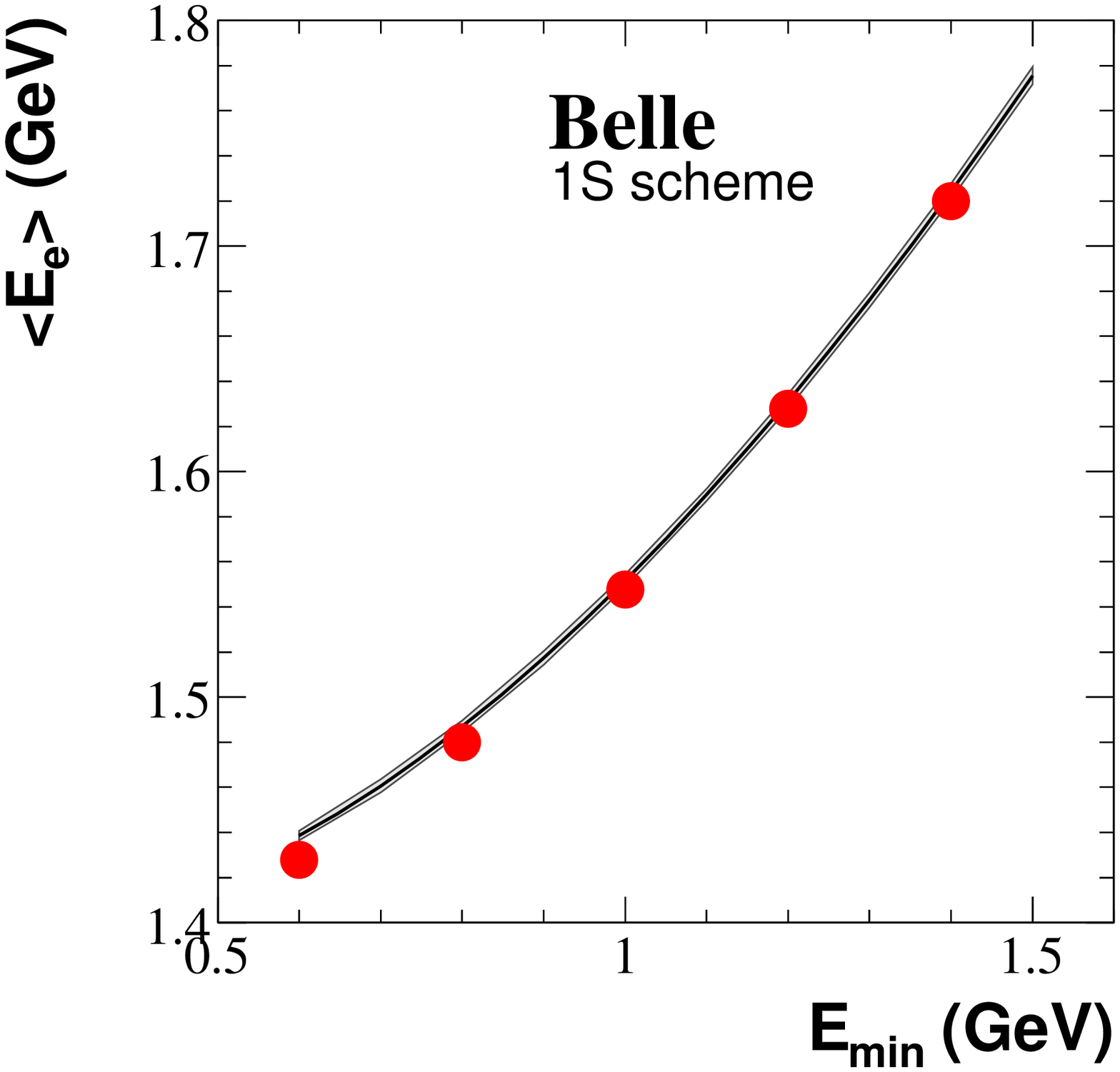}
    \includegraphics[width=0.24\columnwidth]{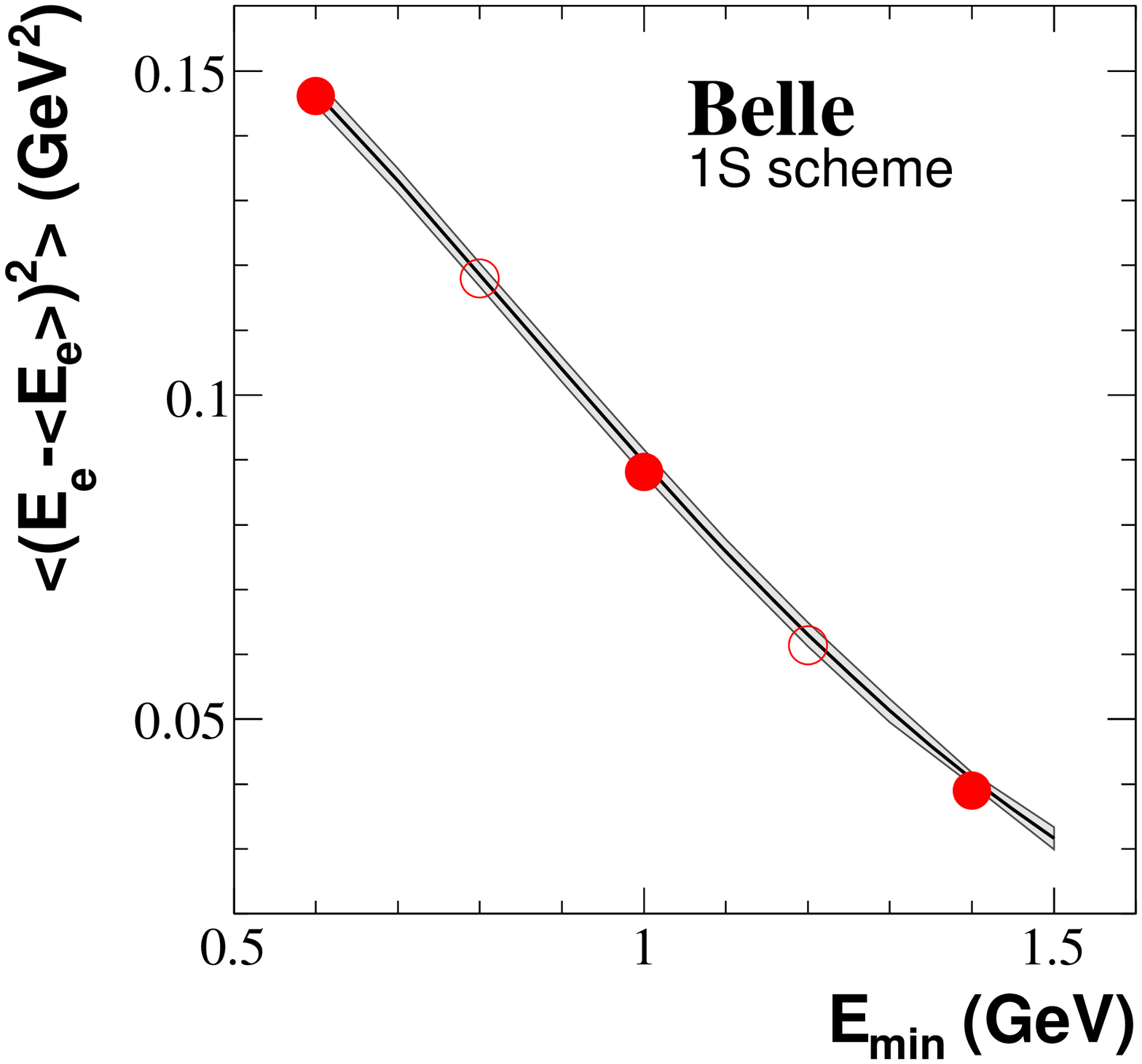}
    \includegraphics[width=0.24\columnwidth]{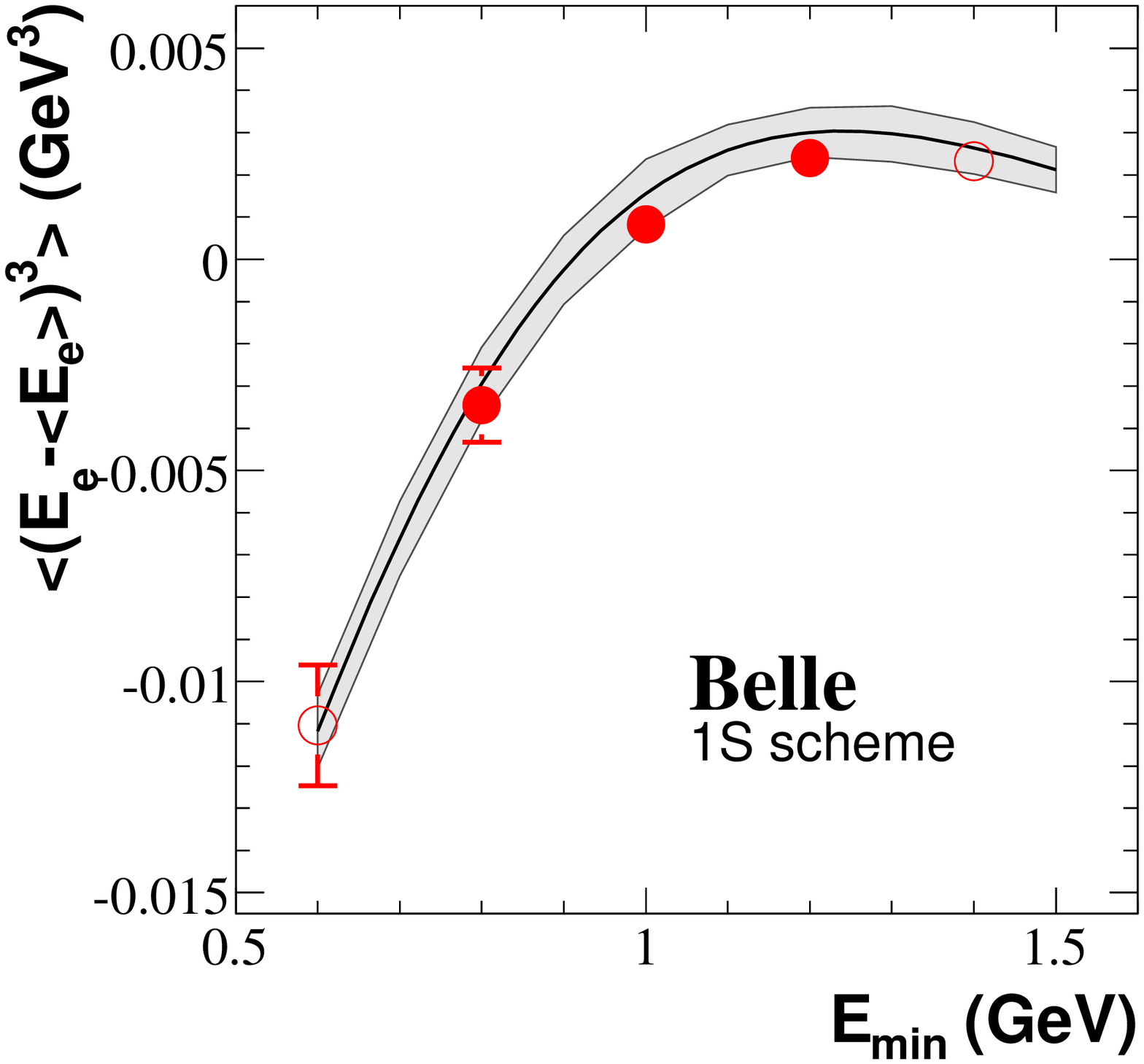}\\
    \includegraphics[width=0.24\columnwidth]{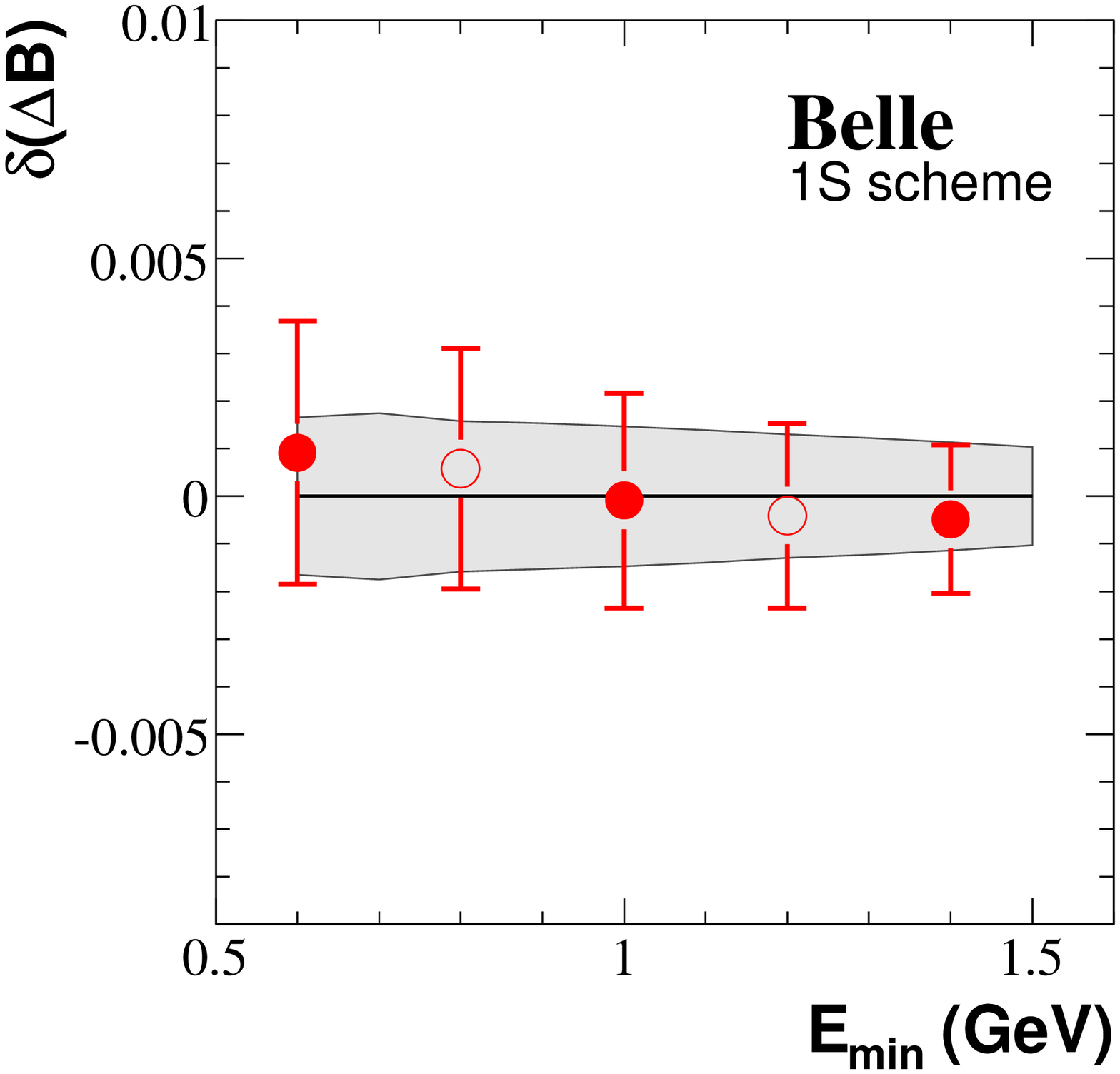}
    \includegraphics[width=0.24\columnwidth]{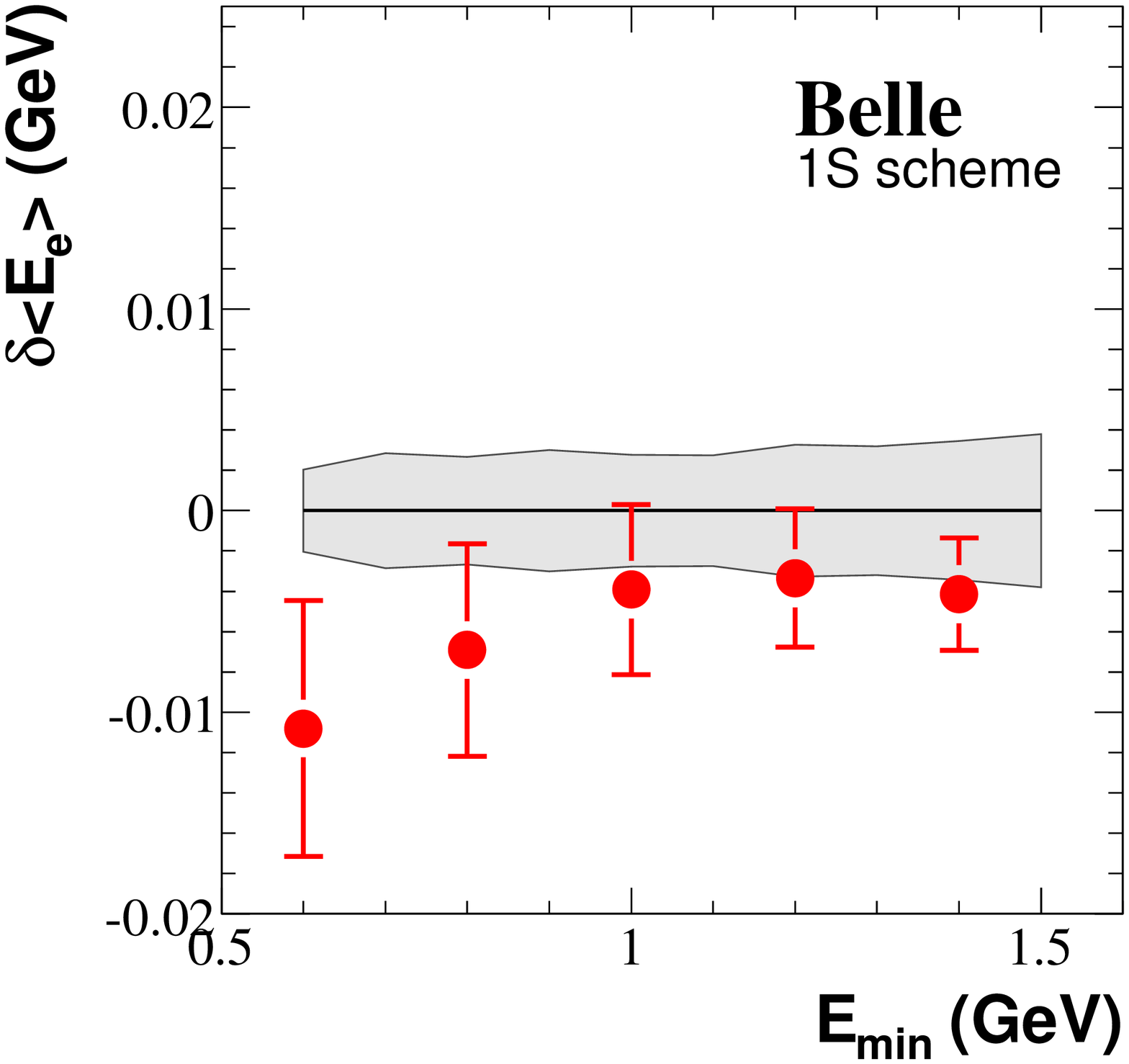}
    \includegraphics[width=0.24\columnwidth]{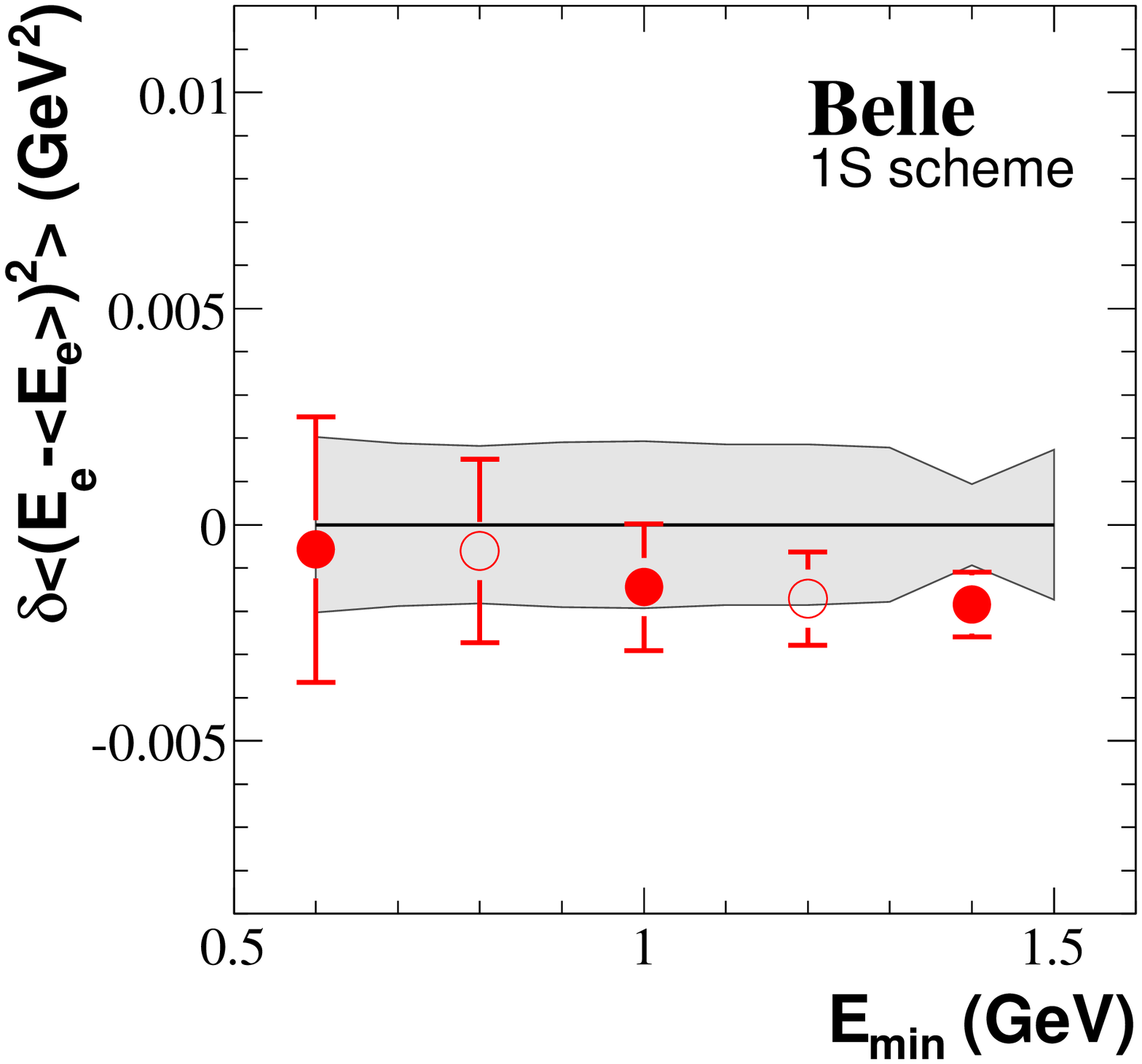}
    \includegraphics[width=0.24\columnwidth]{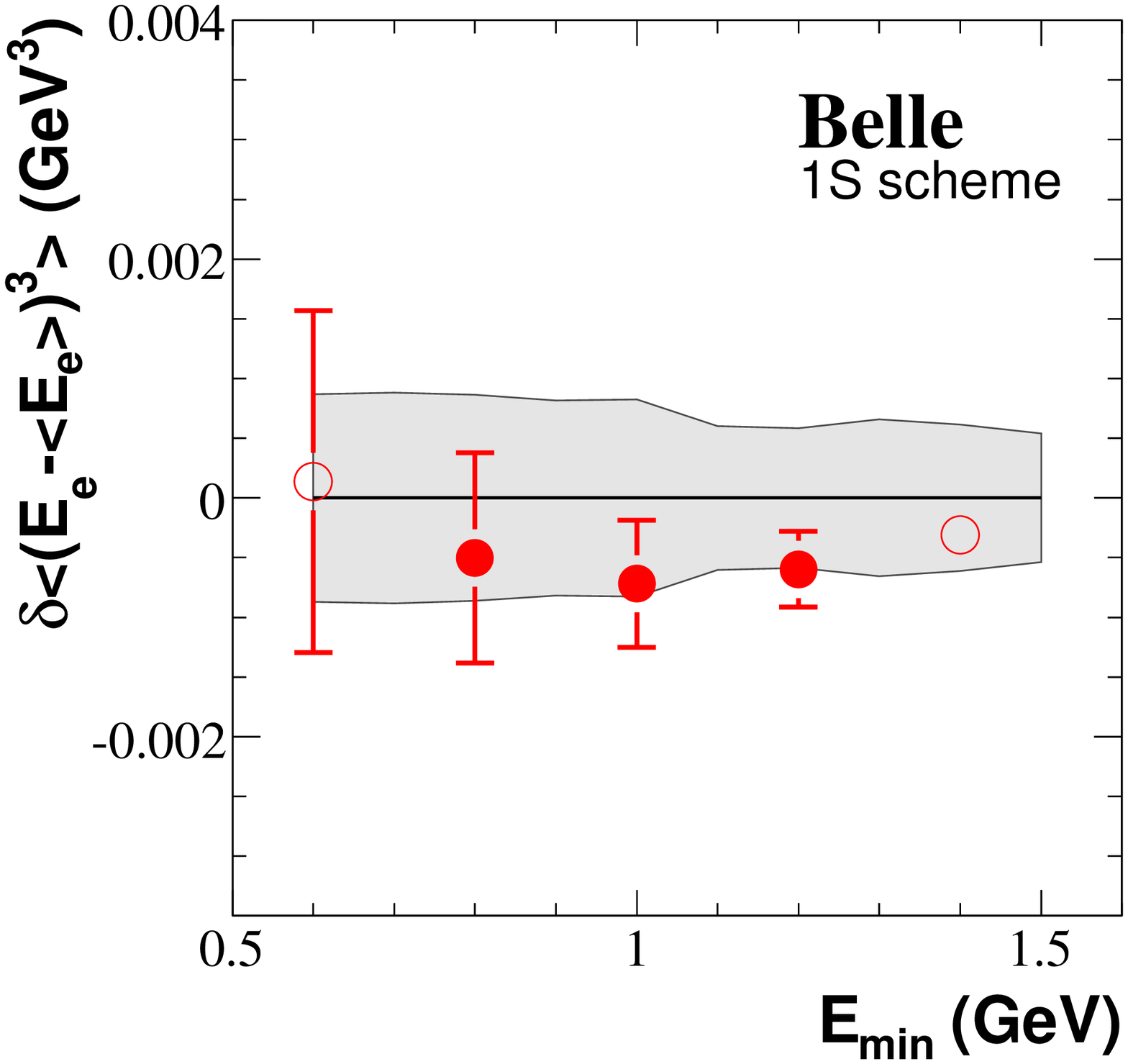}
  \end{center}
  \caption{Comparison of the measured electron energy moments and the
    1S~scheme predictions (upper row), and difference between the
    measurements and the predictions (lower row). The error bars show
    the experimental uncertainties. The error bands represent the theory
    error. Filled circles are data points used in the fit, and open
    circles are unused measurements.} \label{fig:2_1}
\end{figure}

\begin{figure}
  \begin{center}
    \includegraphics[width=0.24\columnwidth]{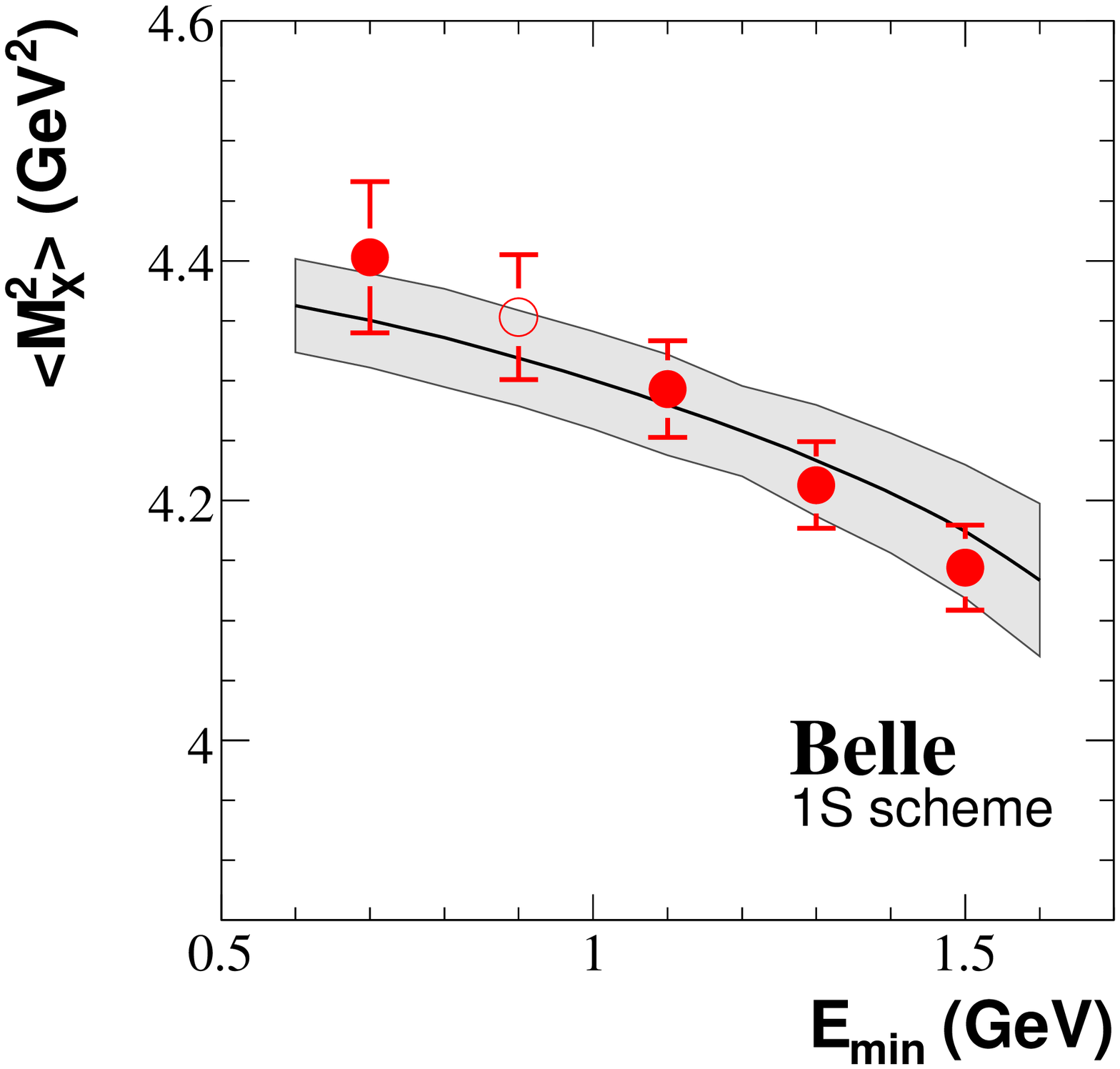}
    \includegraphics[width=0.24\columnwidth]{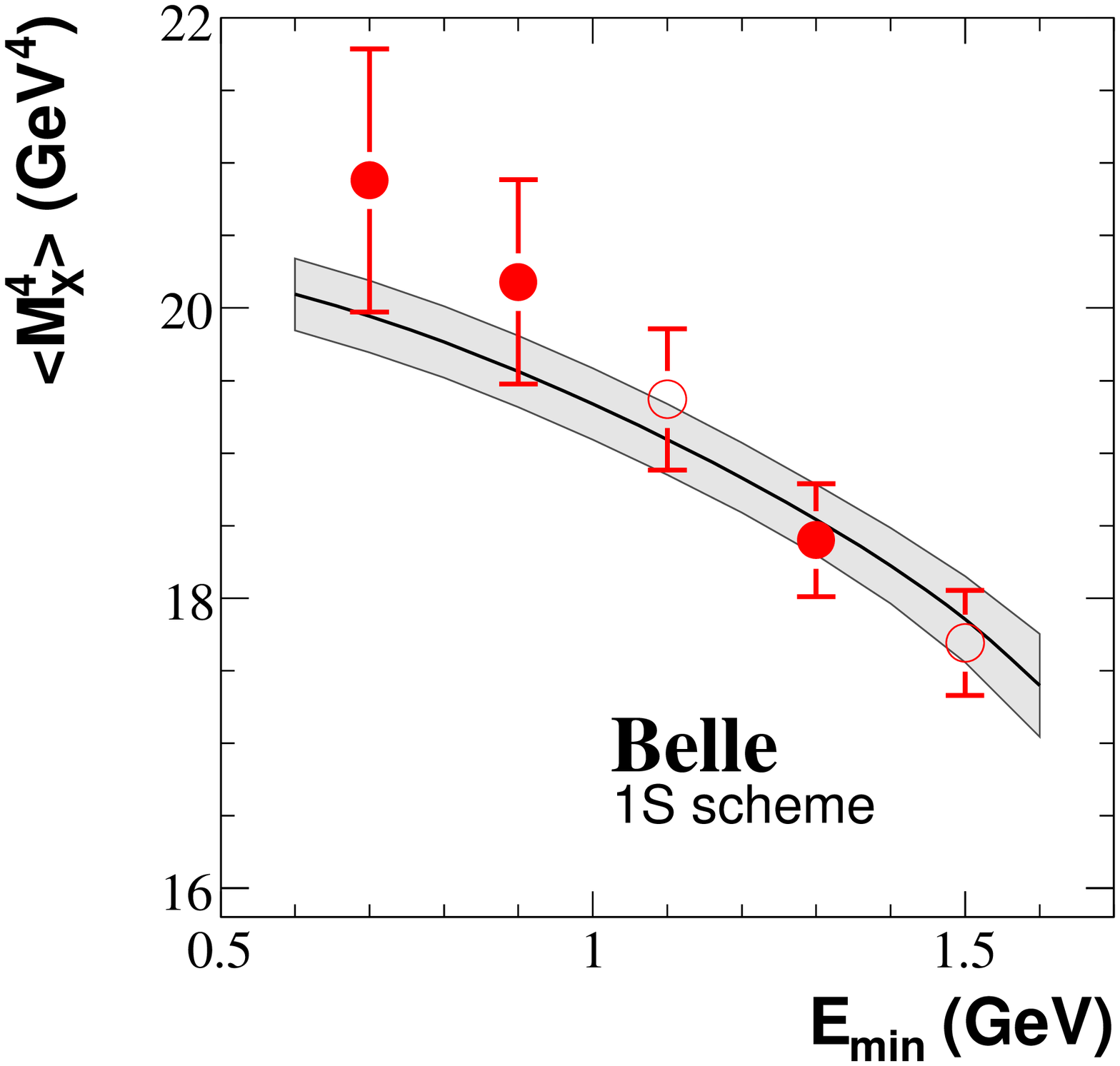}
    \includegraphics[width=0.24\columnwidth]{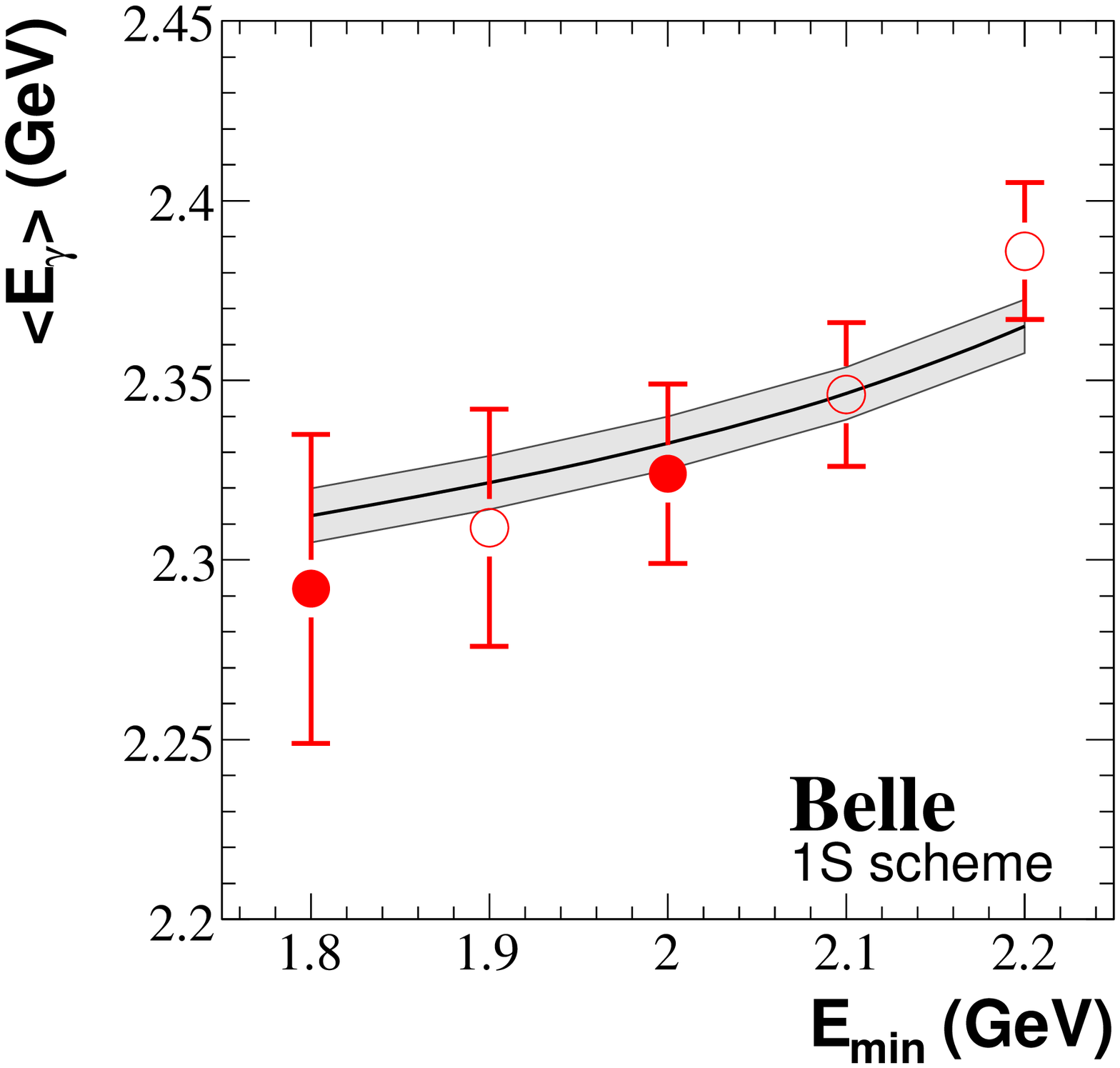}
    \includegraphics[width=0.24\columnwidth]{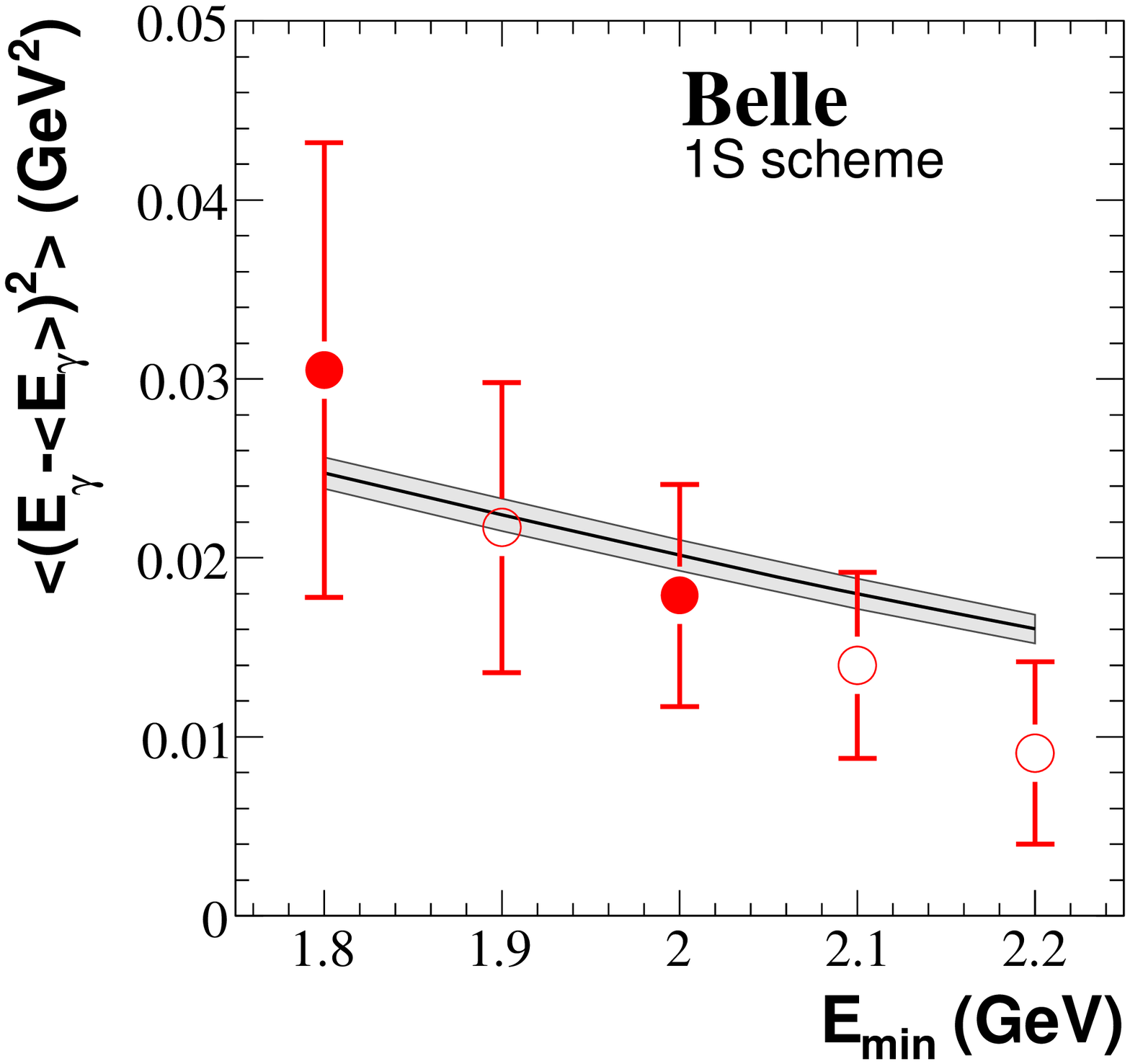}\\
    \includegraphics[width=0.24\columnwidth]{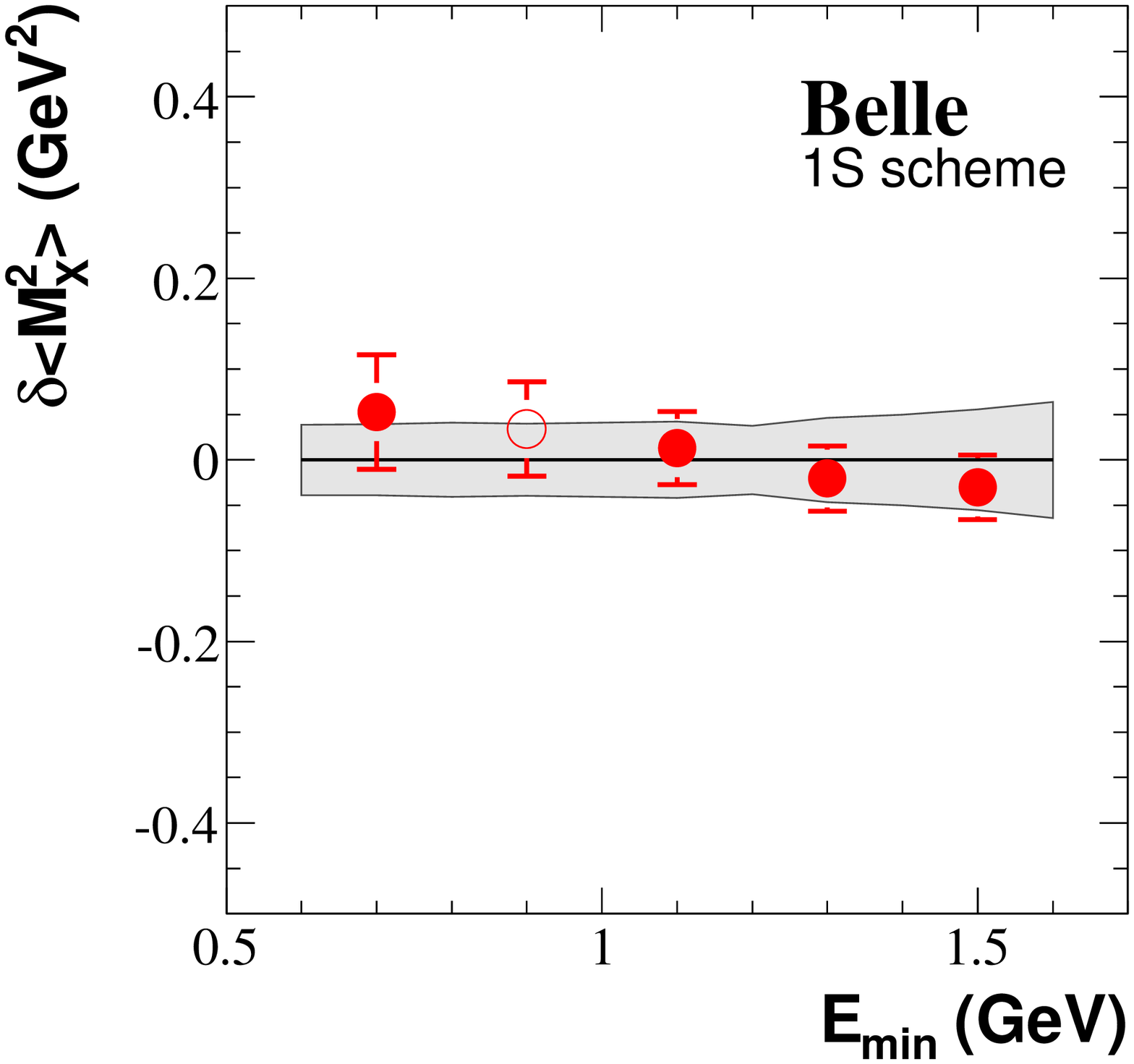}
    \includegraphics[width=0.24\columnwidth]{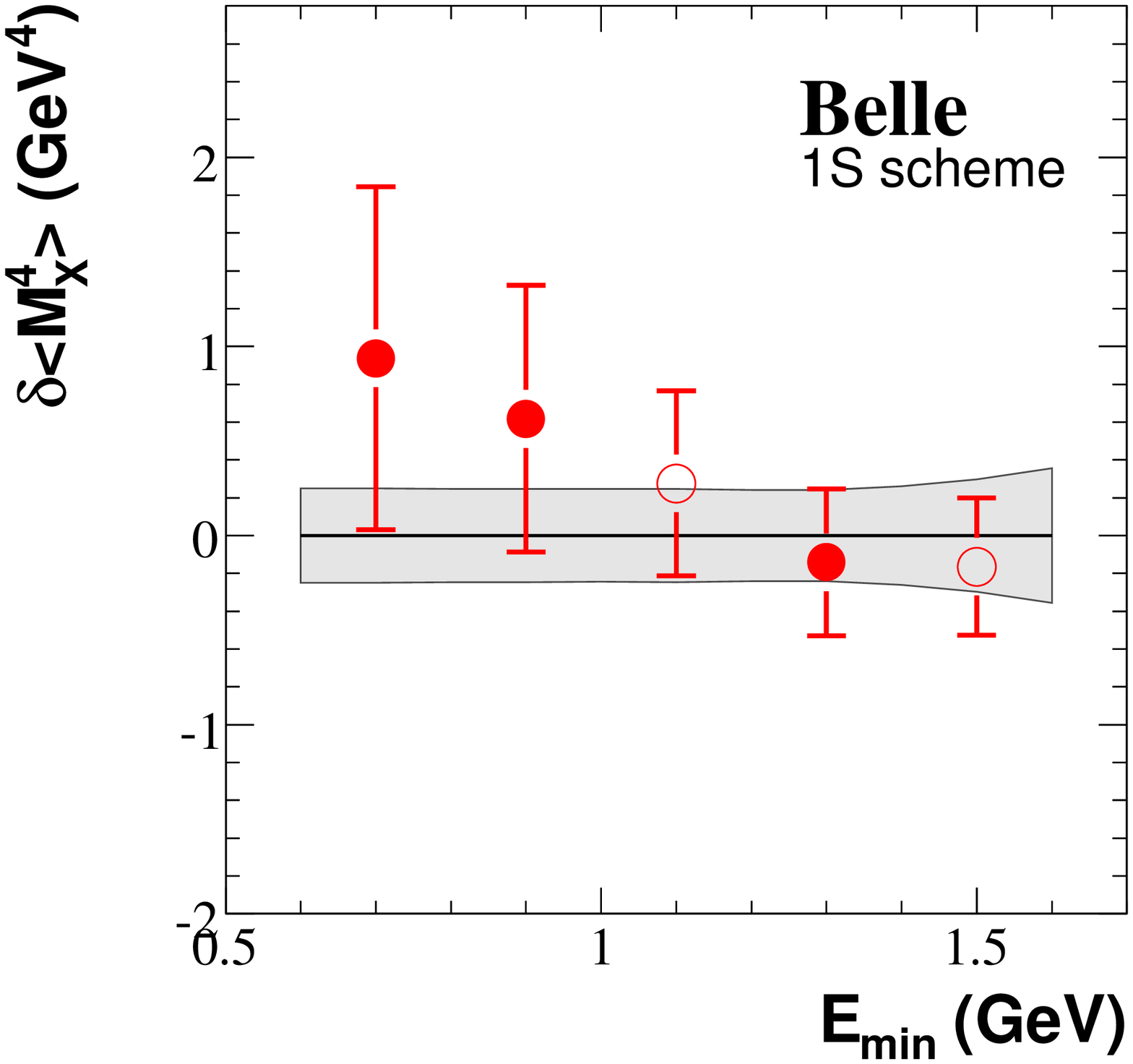}
    \includegraphics[width=0.24\columnwidth]{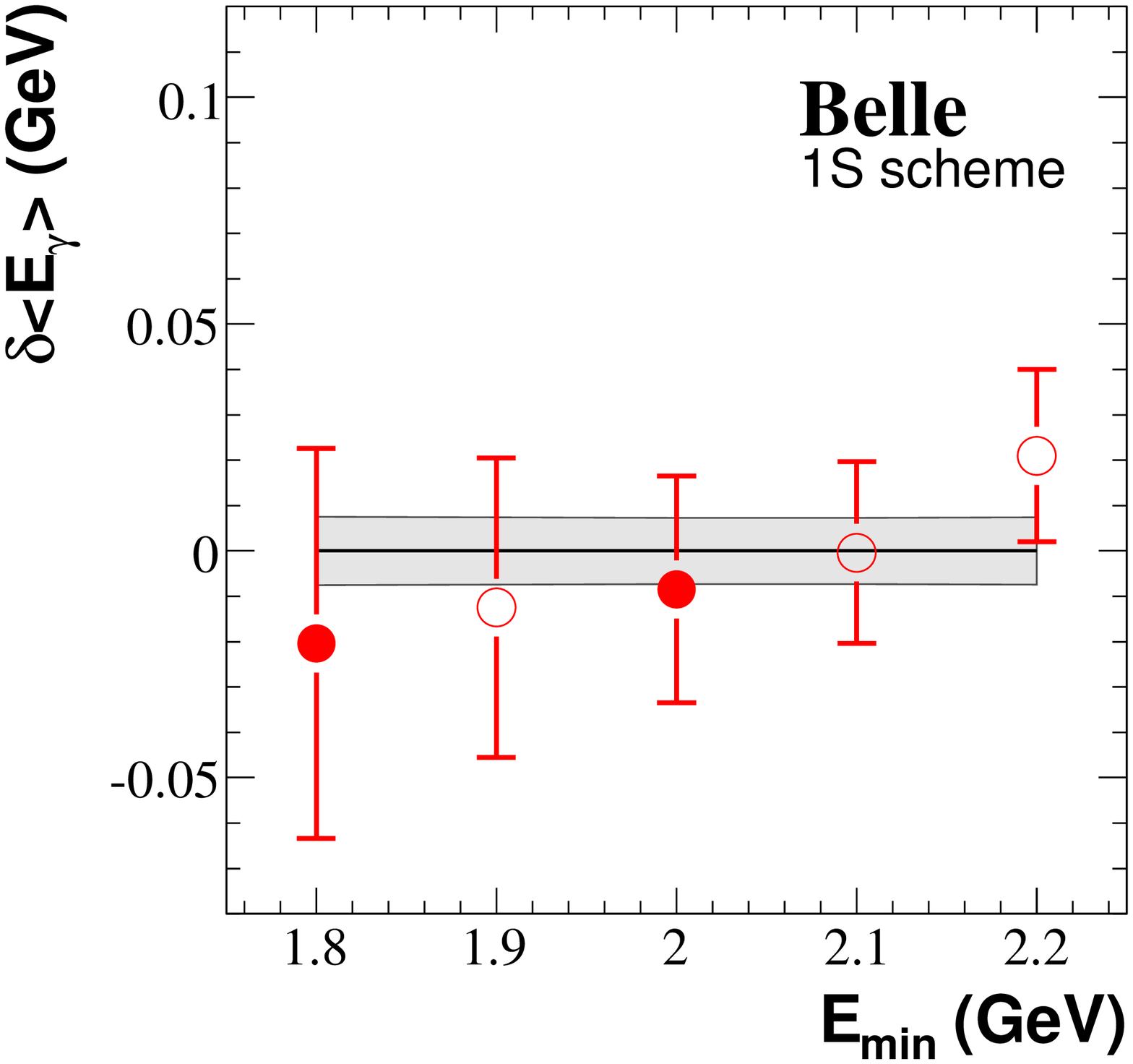}
    \includegraphics[width=0.24\columnwidth]{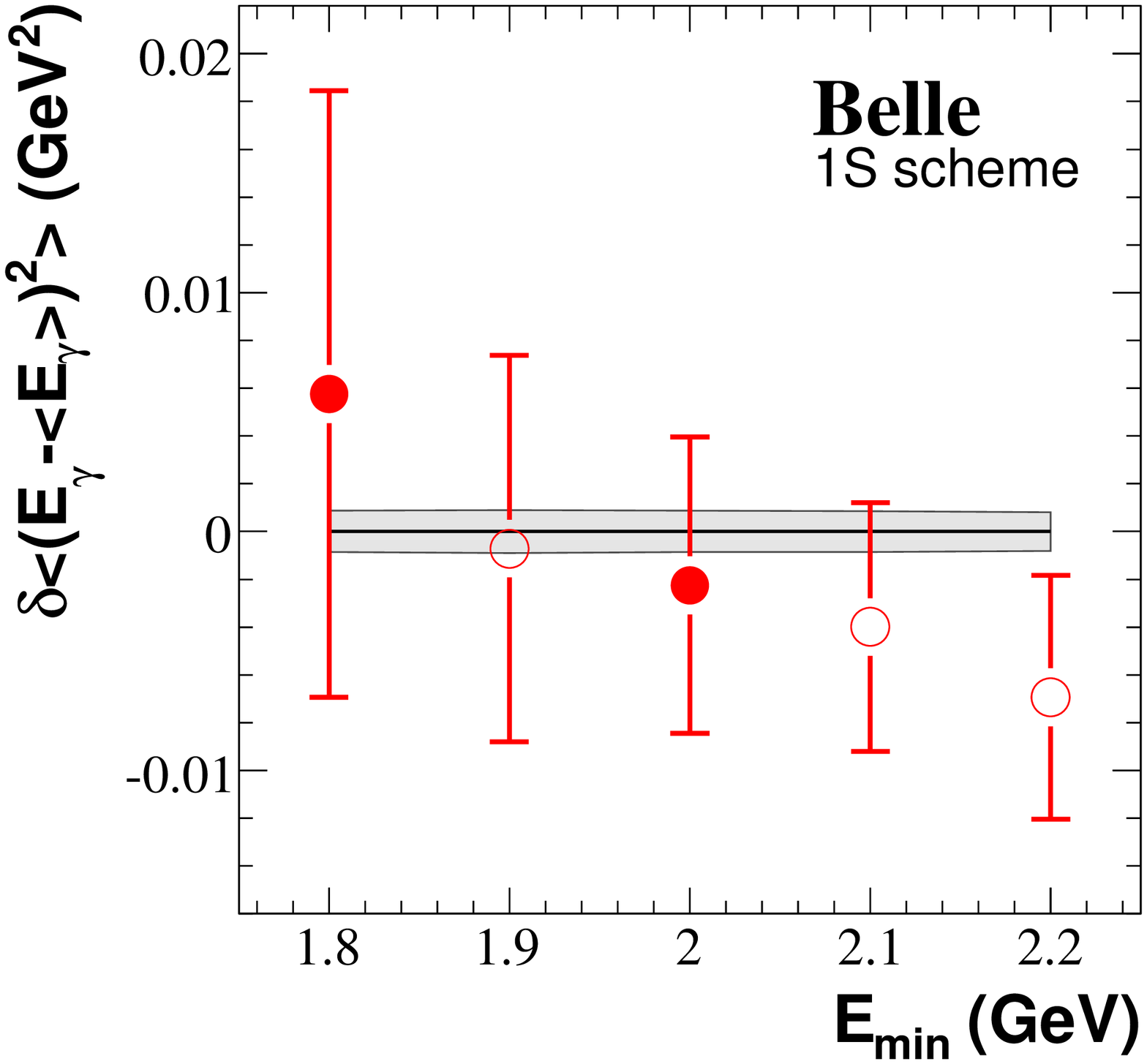}
  \end{center}
  \caption{Same as Fig.~\ref{fig:2_1} for the measured hadronic mass
    and photon energy moments and the 1S~scheme predictions.}
    \label{fig:2_2}
\end{figure}

The results for the fit parameters are given in
Table~\ref{tab:2_2}. Using the measurement of the partial branching
fraction at $E_\mathrm{min}=0.6$~GeV, we obtain for the semileptonic
branching fraction (over the full lepton energy range) ${\mathcal
B}_{X_c\ell\nu}=(10.60\pm 0.28)\%$. A comparison of the measured
moments and the 1S~scheme predictions is shown in Figs.~\ref{fig:2_1}
and \ref{fig:2_2}.
\begin{table}
  \caption{Result of fit in the 1S mass scheme. The
  $\sigma(\mathrm{fit})$~error contains the experimental and
  theoretical uncertainties in the moments. The $\sigma(\tau_B)$~error
  on $|V_{cb}|$ is due to the uncertainty in the average $B$~meson
  lifetime. In the lower part of the table, the correlation matrix of
  the parameters is given.} \label{tab:2_2}
  \begin{center}
    \begin{tabular}{l|@{\extracolsep{.1cm}}ccccccc}
      \hline \hline
      & $|V_{cb}|$ (10$^{-3}$) & $m_b$ (GeV) & $\lambda_1$ (GeV$^2$) &
      $\rho_1$ (GeV$^3$) & $\tau_1$ (GeV$^3$) & $\tau_2$ (GeV$^3$) &
      $\tau_3$ (GeV$^3$)\\
      \hline
      value & 41.56 & \phantom{$-$}4.723 & $-$0.303 &
      \phantom{$-$}0.067 & \phantom{$-$}0.125 & $-$0.101 &
      \phantom{$-$}0.125\\
      $\sigma$(fit) & 0.68 & \phantom{$-$}0.055 & \phantom{$-$}0.046 &
      \phantom{$-$}0.030 & \phantom{$-$}0.005 & \phantom{$-$}0.056 &
      \phantom{$-$}0.005\\
      $\sigma$($\tau_B$) & 0.08 & & & & & & \\
      \hline
      $|V_{cb}|$ & 1.000 & $-$0.121 & \phantom{$-$}0.003 &
      \phantom{$-$}0.195 & \phantom{$-$}0.008 & $-$0.432 &
      \phantom{$-$}0.021\\
      $m_b$ & & \phantom{$-$}1.000 & \phantom{$-$}0.893 & $-$0.137 &
      $-$0.002 & $-$0.509 & $-$0.006\\
      $\lambda_1$ & & & \phantom{$-$}1.000 & $-$0.410 & $-$0.041 &
      $-$0.429 & $-$0.045\\
      $\rho_1$ & & & & \phantom{$-$}1.000 & \phantom{$-$}0.009 &
      $-$0.533 & \phantom{$-$}0.028\\
      $\tau_1$ & & & & & \phantom{$-$}1.000 & \phantom{$-$}0.005 &
      \phantom{$-$}0.000\\
      $\tau_2$ & & & & & & \phantom{$-$}1.000 & \phantom{$-$}0.007\\
      $\tau_3$ & & & & & & & \phantom{$-$}1.000\\
      \hline \hline
    \end{tabular}
  \end{center}
\end{table}

\begin{figure}
  \begin{center}
    \includegraphics[width=0.49\columnwidth]{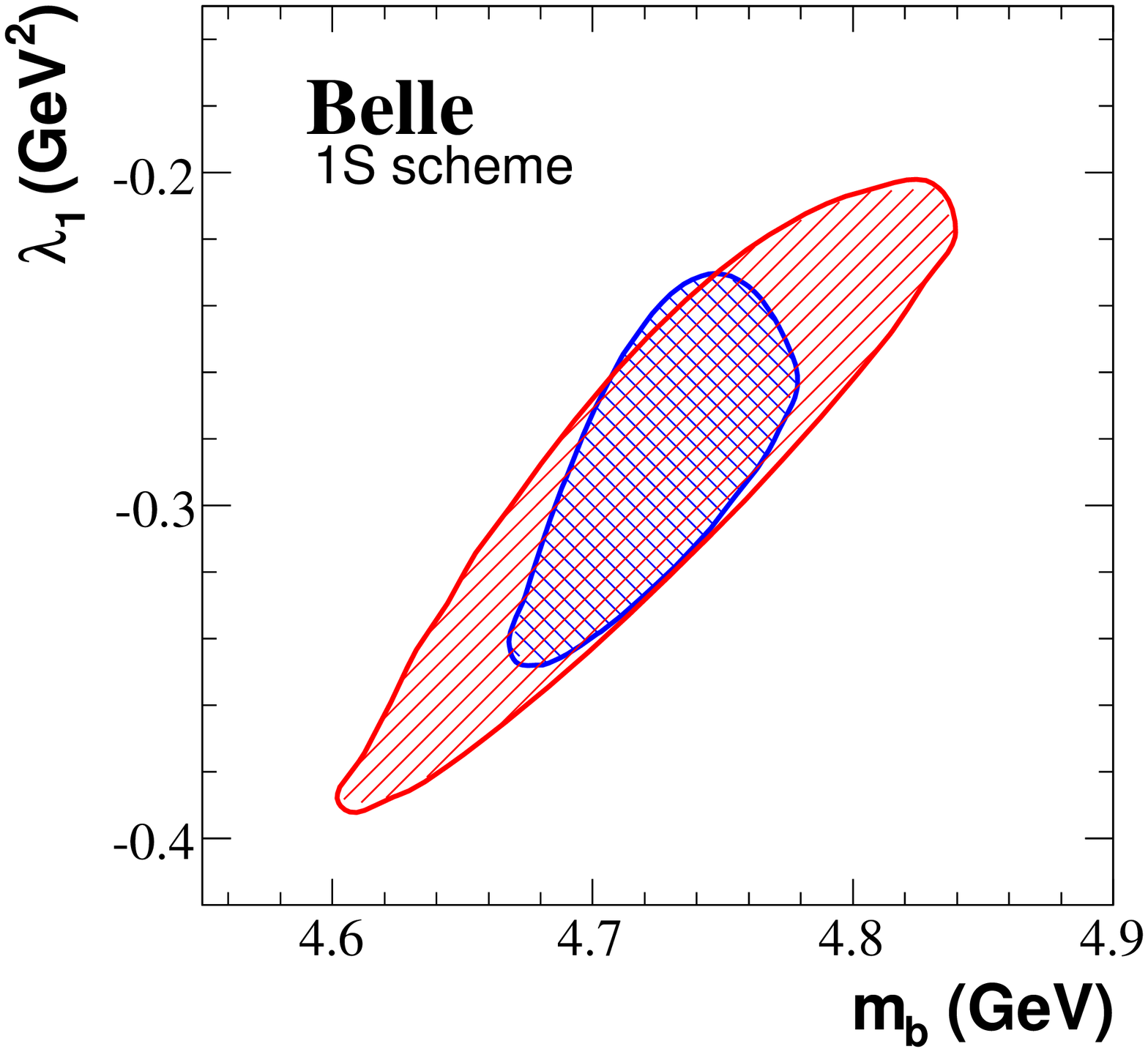}
    \includegraphics[width=0.49\columnwidth]{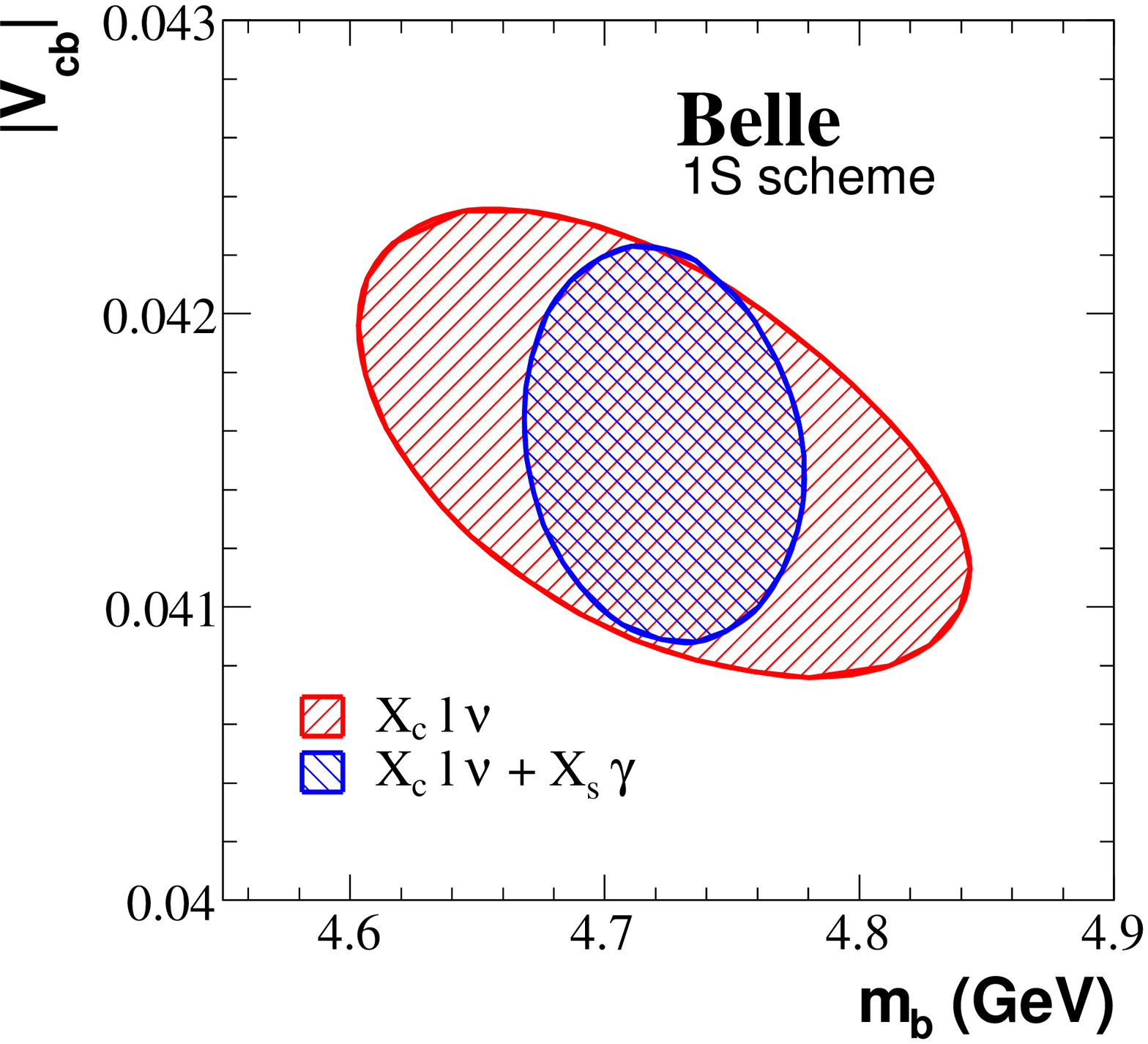}
  \end{center}
  \caption{$\Delta\chi^2=1$~contours for the fit to all moments and
  the fit to the $B\to X_c\ell\nu$~data only.} \label{fig:2_3}
\end{figure}

We have verified the stability of the fit by considering the following
variations (Table~\ref{tab:2_3}): (a)~by repeating the fit only for the
$B\to X_c\ell\nu$~data (21 measurements); (b)~by releasing the
$m_\chi$~constraint on the higher order parameters; (c)~by repeating
the fit with all theoretical uncertainties set to zero. In
Table~\ref{tab:2_3} the default fit corresponds to setup
(d). Figure~\ref{fig:2_3} shows the $\Delta\chi^2=1$~contour plots for
the fits corresponding to setups (a) and (d) in Table~\ref{tab:2_3}.
\begin{table}
  \caption{Stability of the fit in the 1S mass scheme. The different
    setups are explained in the text. Setup (d) corresponds to the
    default fit.} \label{tab:2_3}
  \begin{center}
    \begin{tabular}{c|@{\extracolsep{.3cm}}cccc}
      \hline \hline
      Setup & $\chi^2/\mathrm{ndf.}$ & $|V_{cb}|$ (10$^{-3}$) &
      $m_b$ (GeV) & $\lambda_1$ (GeV$^2$)\\
      \hline
      (a) & 6.4/14 & $41.55\pm 0.80$ & $4.718\pm 0.119$ & $-0.308\pm 0.092$\\
      (b) & 5.6/18 & $41.28\pm 0.86$ & $4.699\pm 0.060$ & $-0.491\pm 0.084$\\
      (c) & 16.6/18 & $41.10\pm 0.54$ & $4.666\pm 0.046$ & $-0.341\pm 0.031$\\
      (d) & 7.3/18 & $41.56\pm 0.68$ & $4.723\pm 0.055$ & $-0.303\pm 0.046$\\
      \hline\hline
    \end{tabular}
  \end{center}
\end{table}
\subsection{Kinetic Mass Scheme Analysis}

\subsubsection{Theoretical Input}

Spectral moments of the lepton energy and hadronic mass in $B\to
X_c\ell\nu$~decays have been derived in the kinetic mass scheme up to
$\mathcal{O}(1/m^3_b)$~\cite{Gambino:2004qm}. Compared to the
original publication, the theoretical expressions in our fit
contain an improved calculation of the perturbative corrections to the
lepton energy moments~\cite{ref:2} and account for the
$E_\mathrm{min}$~dependence of the perturbative corrections to the
hadronic mass moments~\cite{Uraltsev:2004in}. For the photon energy
moments in $B\to X_s\gamma$, the (biased) OPE prediction and the bias
correction have been calculated~\cite{Benson:2004sg}.

These expressions depend on the following set of non-perturbative
parameters: the $b$- and $c$-quark masses $m_b$ and $m_c$, $\mu^2_\pi$
and $\mu^2_G$ at $\mathcal{O}(1/m^2_b)$ and $\tilde\rho^3_D$ and
$\rho^3_{LS}$ at $\mathcal{O}(1/m^3_b)$~\cite{ref:3}. In our analysis,
we determine these six parameters together with the semileptonic
branching fraction (over the full lepton energy range) ${\mathcal
B}_{X_c\ell\nu}$. The total number of parameters in the fit is thus
seven.

The CKM magnitude~$|V_{cb}|$ is calculated using the following
expression~\cite{Benson:2003kp},
\begin{eqnarray}
  \frac{|V_{cb}|}{0.0417} & = & \left(\Gamma(B\to
  X\ell\nu)\frac{1.55~\mathrm{ps}}{0.105}\right)^{1/2}\times
  [1+0.30(\alpha_s-0.22)] \nonumber \\
  \lefteqn{\times
  [1-0.66(m_b-4.6~\mathrm{GeV})+0.39(m_c-1.15~\mathrm{GeV})} \nonumber\\
  \lefteqn{+0.013(\mu^2_\pi-0.4~\mathrm{GeV}^2)+0.09(\tilde\rho^3_D-0.1~\mathrm{GeV}^3)}
  \nonumber \\
  \lefteqn{+0.05(\mu^2_G-0.35~\mathrm{GeV}^2)-0.01(\rho^3_{LS}+0.15~\mathrm{GeV}^3)]~,} \label{eq:2_2}
\end{eqnarray}
where $\Gamma(B\to X\ell\nu)$ is the semileptonic width of the $B$~meson.

\subsubsection{The Fit}

As in the 1S scheme case, the fit is performed using the
$\chi^2$~minimization technique and the MINUIT
program~\cite{James:1975dr}. The covariance matrix used is the sum of
matrices corresponding to the experimental and theoretical
uncertainties. The theoretical covariance matrix is constructed
following the recipe in Ref.~\cite{Gambino:2004qm}:

The non-perturbative uncertainties (\textit{i.e.}, the uncertainties
related to the $1/m_b$~expansion) are evaluated by varying $\mu^2_\pi$
and $\mu^2_G$ ($\tilde\rho^3_D$ and $\rho^3_{LS}$) by $\pm 20\%$ ($\pm
30\%$) around their ``nominal'' values of $\mu^2_\pi=0.4$~GeV$^2$,
$\tilde\rho^3_D=0.1$~GeV$^3$, $\mu^2_G=0.35$~GeV$^2$ and
$\rho^3_{LS}=-0.15$~GeV$^3$. The perturbative uncertainties
(\textit{i.e.}, the uncertainties related to the expansion in
$\alpha_s$) are estimated by varying $\alpha_s$ within $\pm 0.04$
($\pm 0.1$) around the central value of 0.22 (0.3) for the lepton and
photon energy (hadronic mass) moments. The difference in the treatment
of $\alpha_s$ for the hadronic mass moments is due to the fact that
the calculation of the perturbative corrections to these moments is
less complete.

The theoretical uncertainty in the moment predictions is the quadratic
sum of these different contributions. The theoretical covariance
matrix is then constructed by treating these errors as fully
correlated for a given moment with different $E_\mathrm{min}$ while
they are treated as uncorrelated between moments of different order.

For the moments of the photon energy spectrum, we take 30\% of the
absolute value of the bias correction as its uncertainty. This
additional theoretical error is considered uncorrelated for moments
with different $E_\mathrm{min}$ and different order.

The experimental data from $B^*-B$~mass splitting and heavy quark sum
rules constrain the parameters $\mu^2_G$ and $\rho^3_{LS}$ to
$0.35\pm 0.07$~GeV$^2$ and $-0.15\pm 0.1$~GeV$^3$, respectively. We
account for these constraints by adding the following additional terms
to the $\chi^2$~function,
\begin{equation}
  (\mu^2_G-0.35~\mathrm{GeV}^2)^2/(0.07~\mathrm{GeV}^2)^2+(\rho^3_{LS}+0.15~\mathrm{GeV}^3)^2/(0.1~\mathrm{GeV}^3)^2~.
\end{equation}
To calculate $|V_{cb}|$ using Eq.~\ref{eq:2_2} and properly account
for the correlations of the HQ~parameters, we make $|V_{cb}|$ a free
parameter of the fit, calculate $\Gamma(B\to X_c\ell\nu)$ with
Eq.~\ref{eq:2_2} and add the following term to the $\chi^2$~function,
\begin{equation}
  (\frac{\mathcal{B}_{X_c\ell\nu}}{\Gamma(B\to
  X_c\ell\nu)}-\tau_B)^2/\sigma^2~.
\end{equation}
The uncertainty~$\sigma$ accounts for the experimental uncertainty in
$\tau_B$ and an additional 1.4\% theoretical uncertainty in extracting
$|V_{cb}|$ using Eq.~\ref{eq:2_2}~\cite{Benson:2003kp}. We have
verified that this method of calculating $|V_{cb}|$ does not change
the fit result for the other parameters.

\subsubsection{Results and Discussion}

\begin{figure}
  \begin{center}
    \includegraphics[width=0.24\columnwidth]{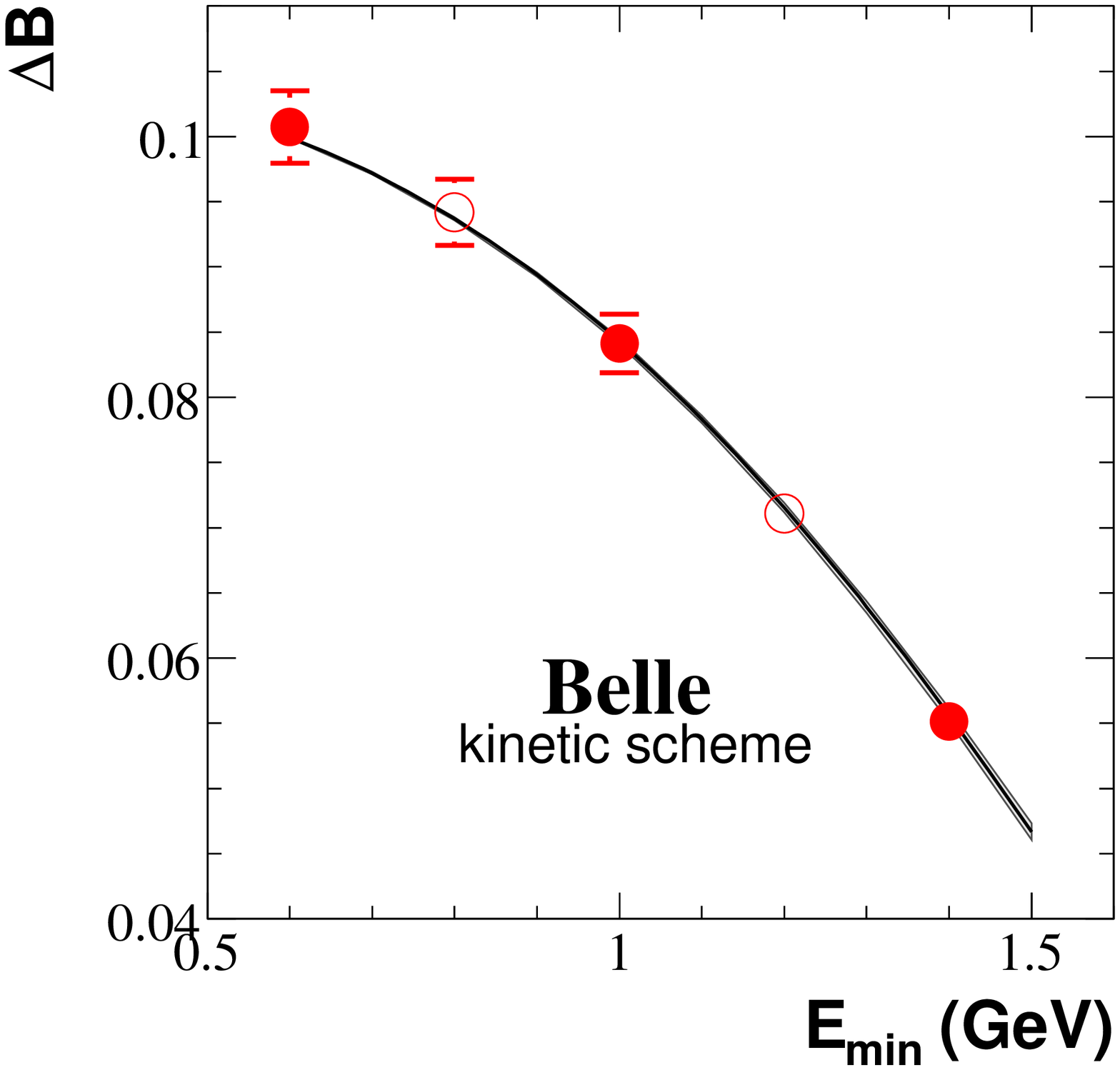}
    \includegraphics[width=0.24\columnwidth]{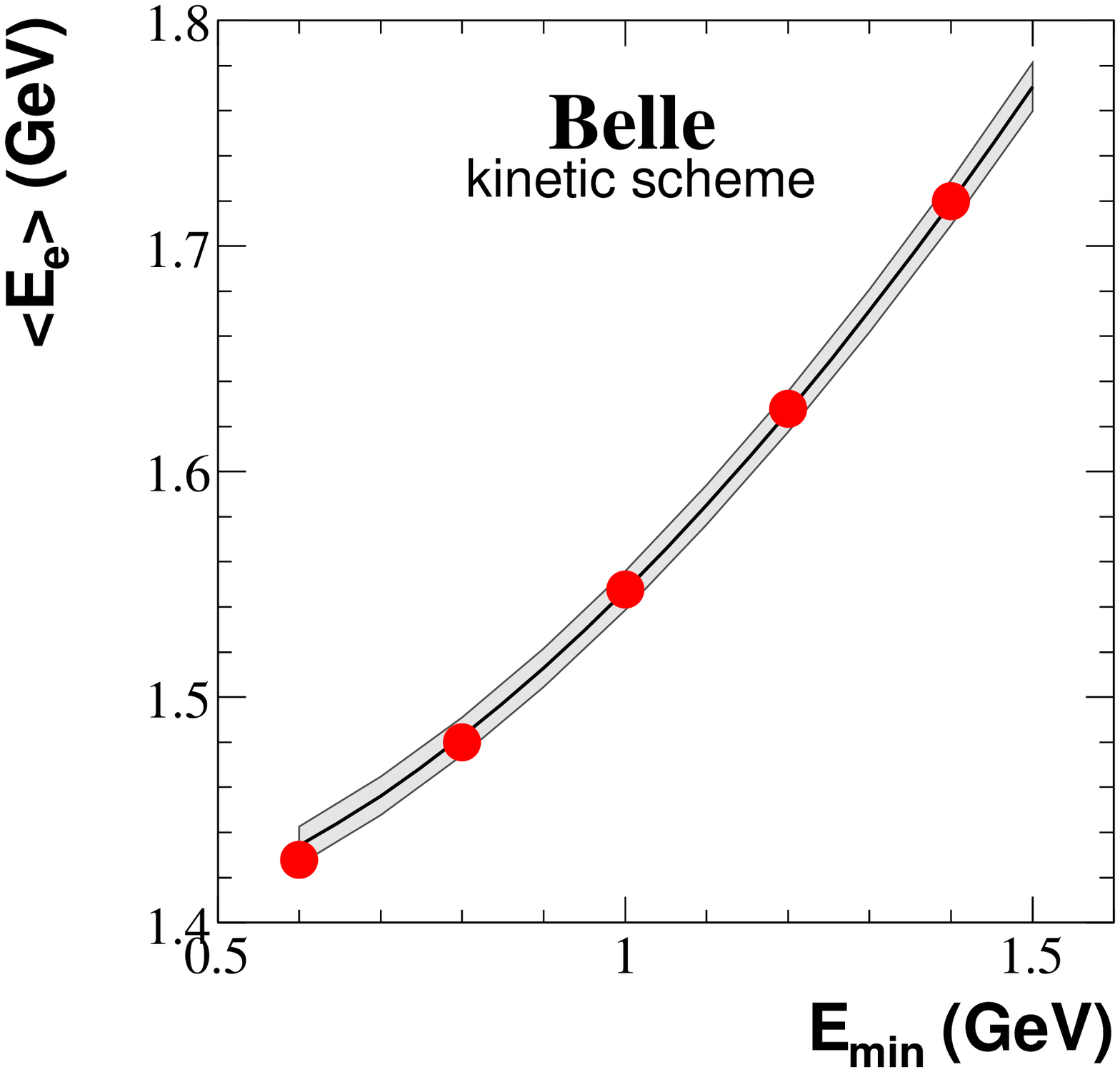}
    \includegraphics[width=0.24\columnwidth]{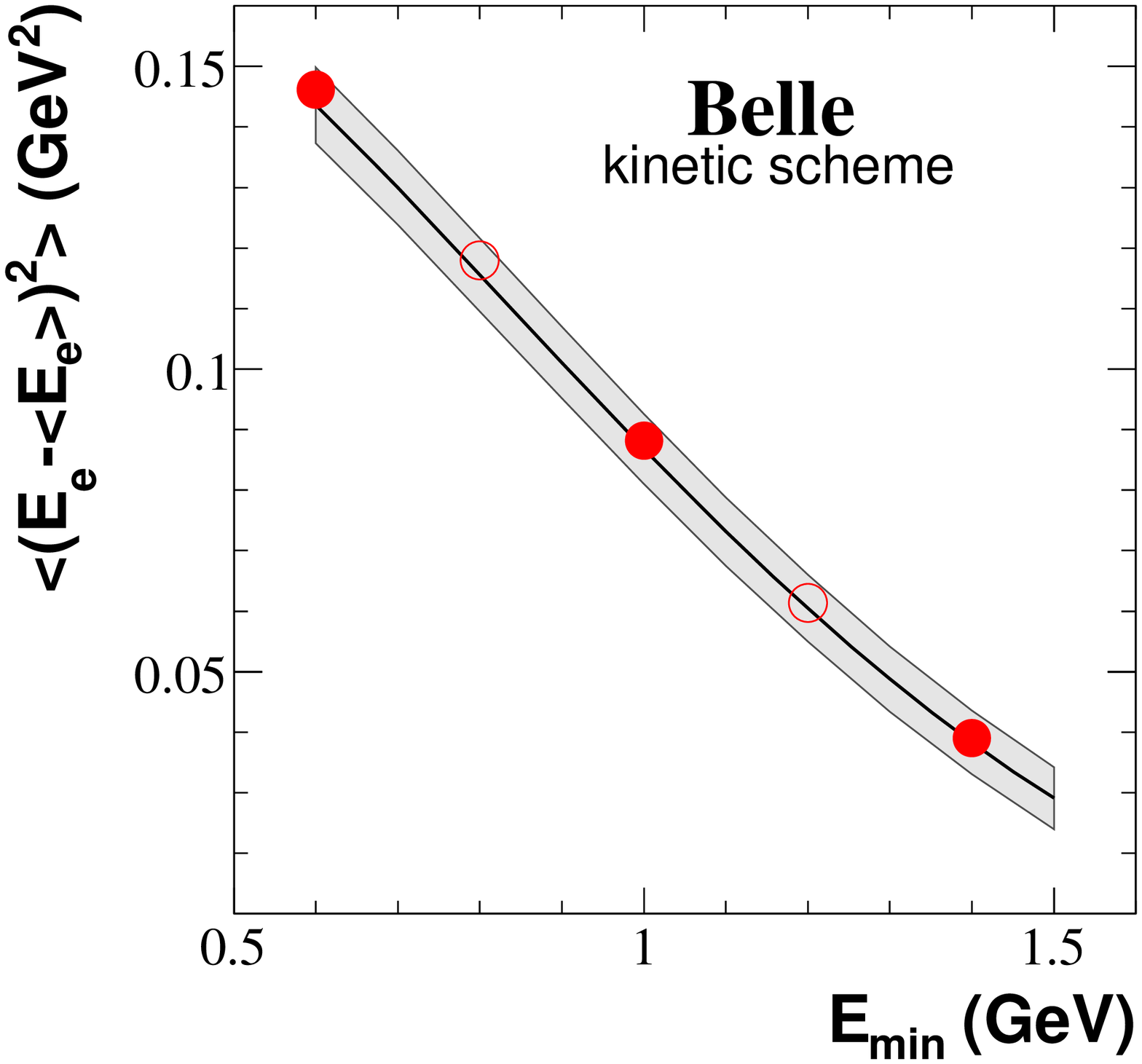}
    \includegraphics[width=0.24\columnwidth]{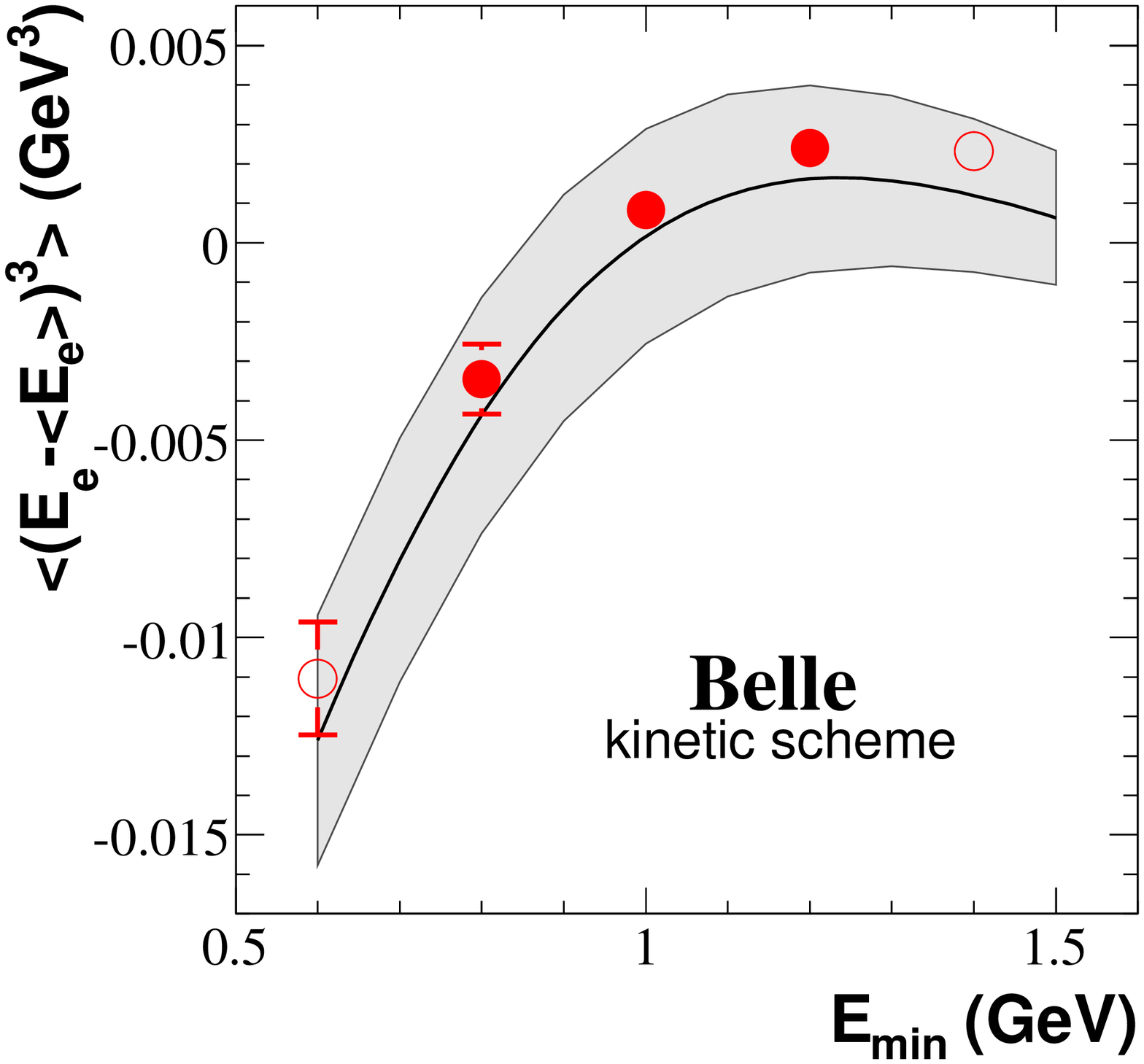}\\
    \includegraphics[width=0.24\columnwidth]{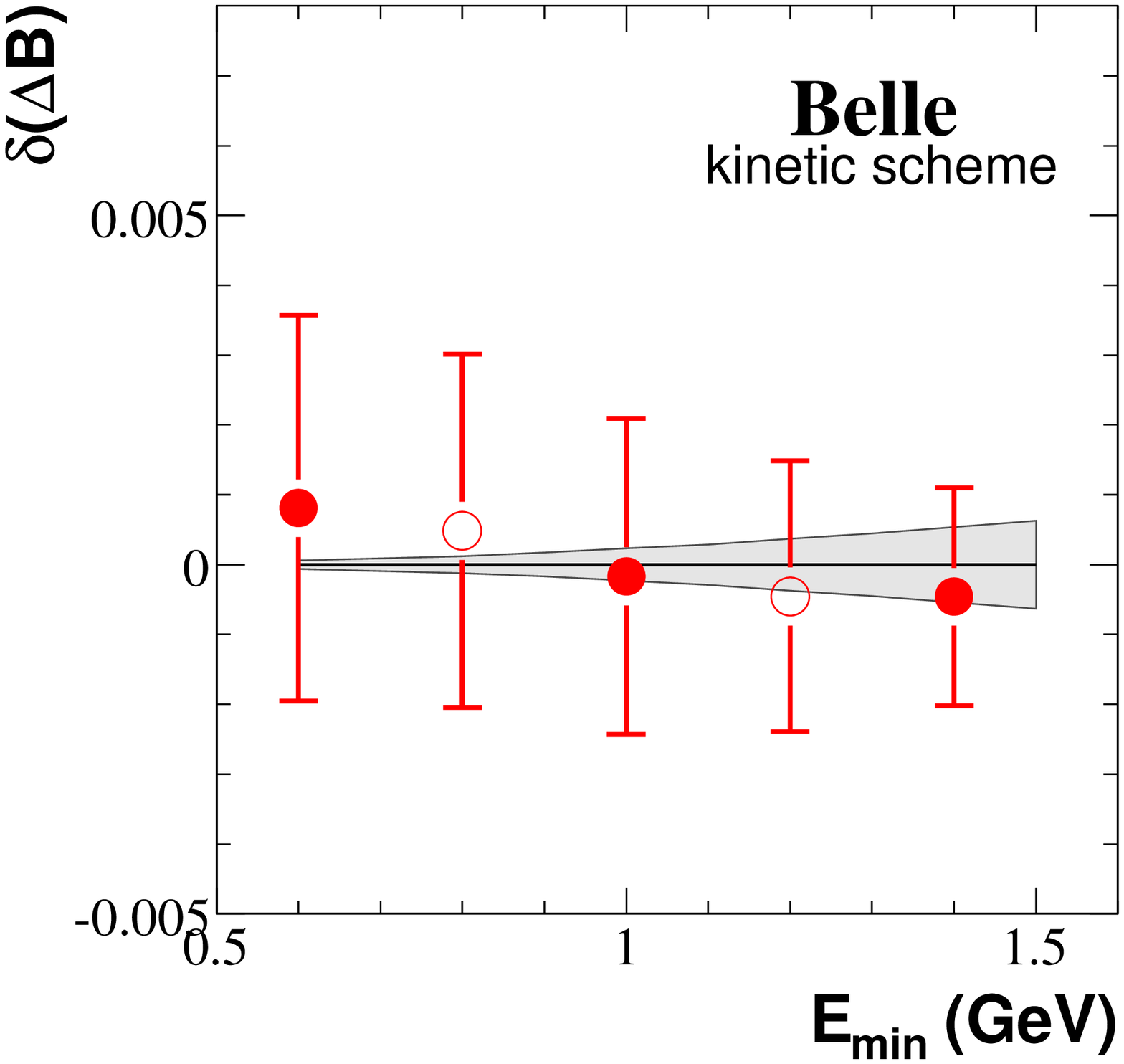}
    \includegraphics[width=0.24\columnwidth]{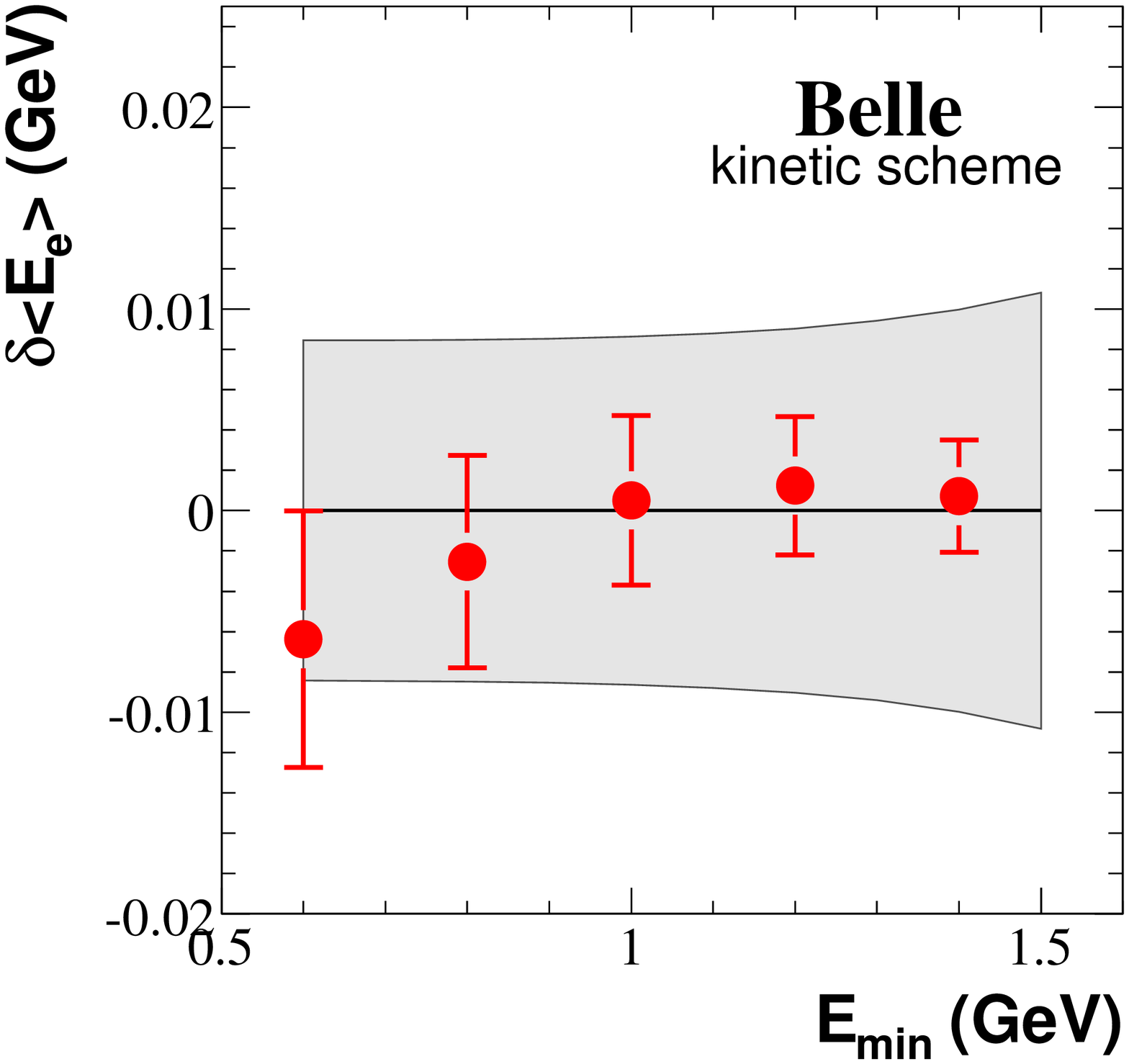}
    \includegraphics[width=0.24\columnwidth]{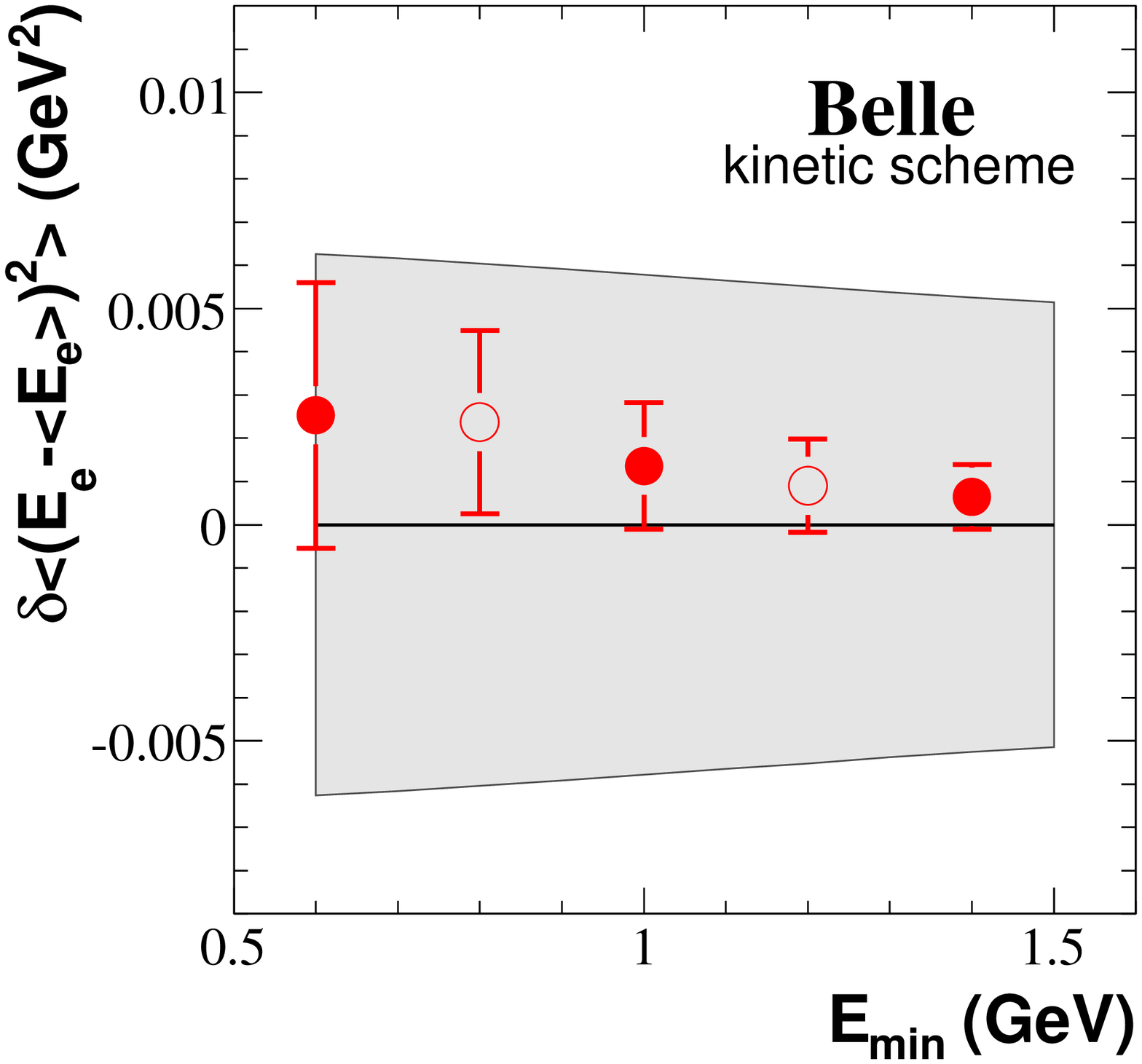}
    \includegraphics[width=0.24\columnwidth]{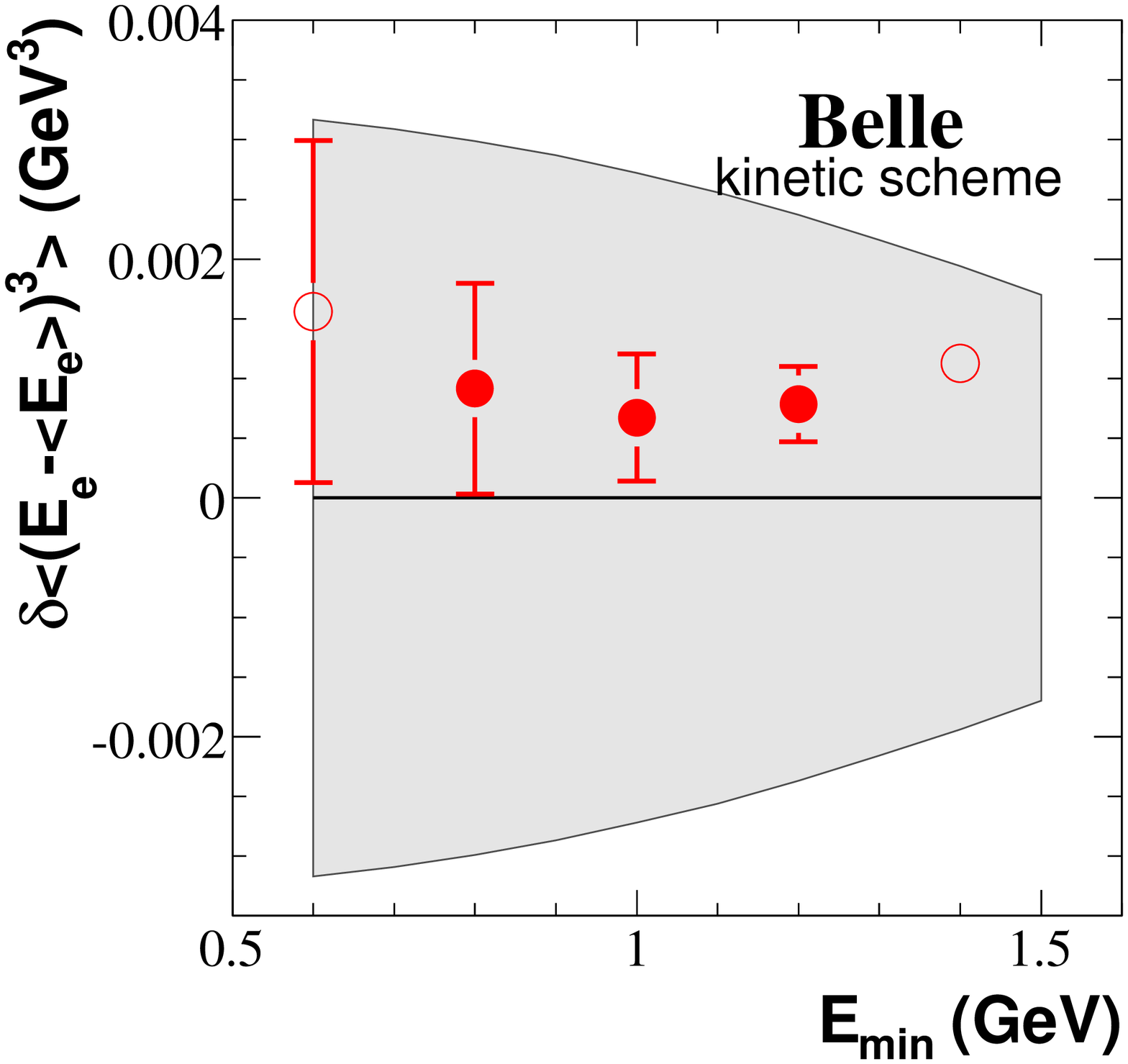}
  \end{center}
  \caption{Comparison of the measured electron energy moments and the
    kinetic scheme predictions (upper row), and difference between the
    measurements and the predictions (lower row). The error bars show
    the experimental uncertainties. The error bands represent the
    theory error. Filled circles are data points used in the fit, and
    open circles are unused measurements.} \label{fig:2_4}
\end{figure}

\begin{figure}
  \begin{center}
    \includegraphics[width=0.24\columnwidth]{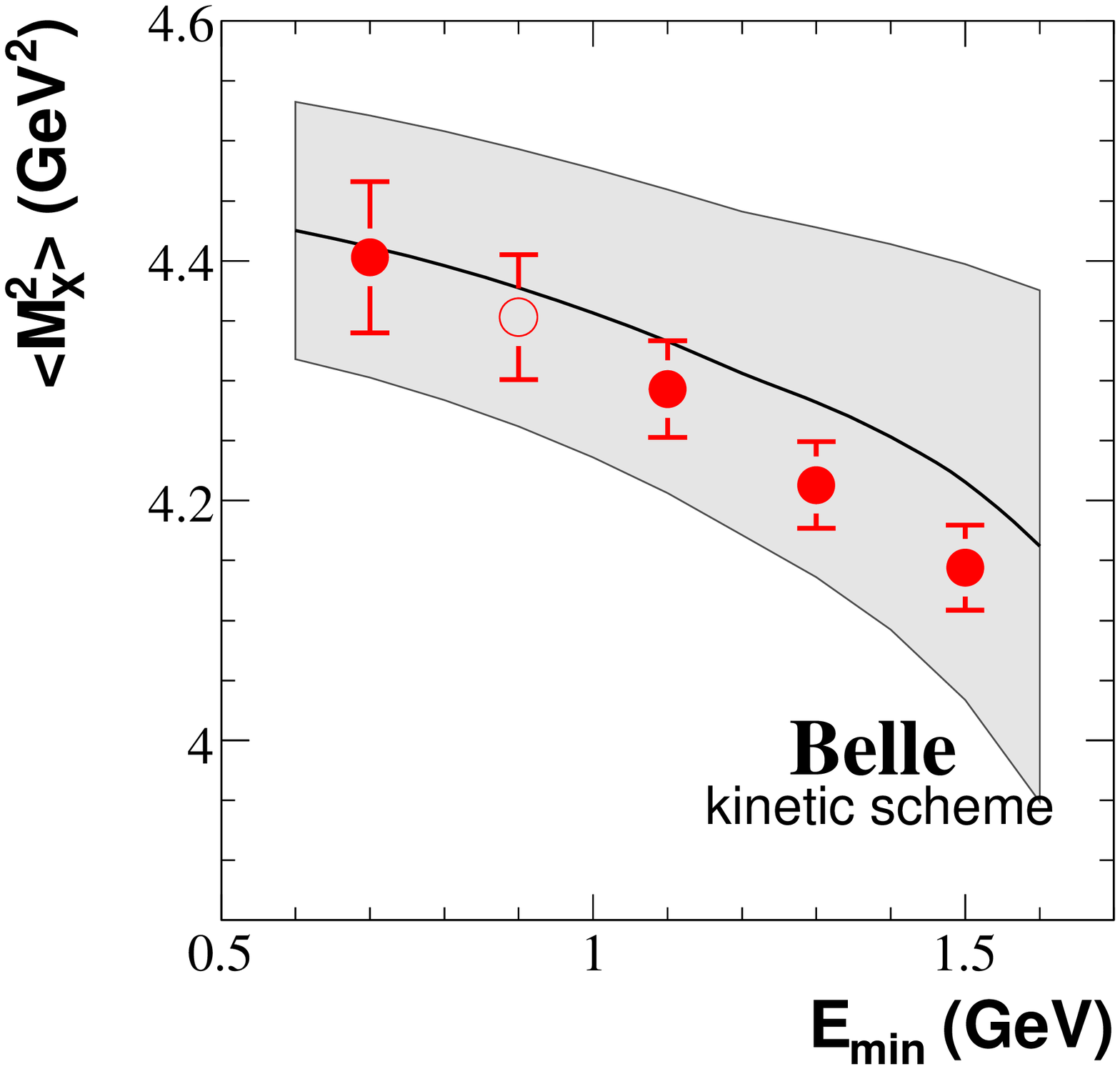}
    \includegraphics[width=0.24\columnwidth]{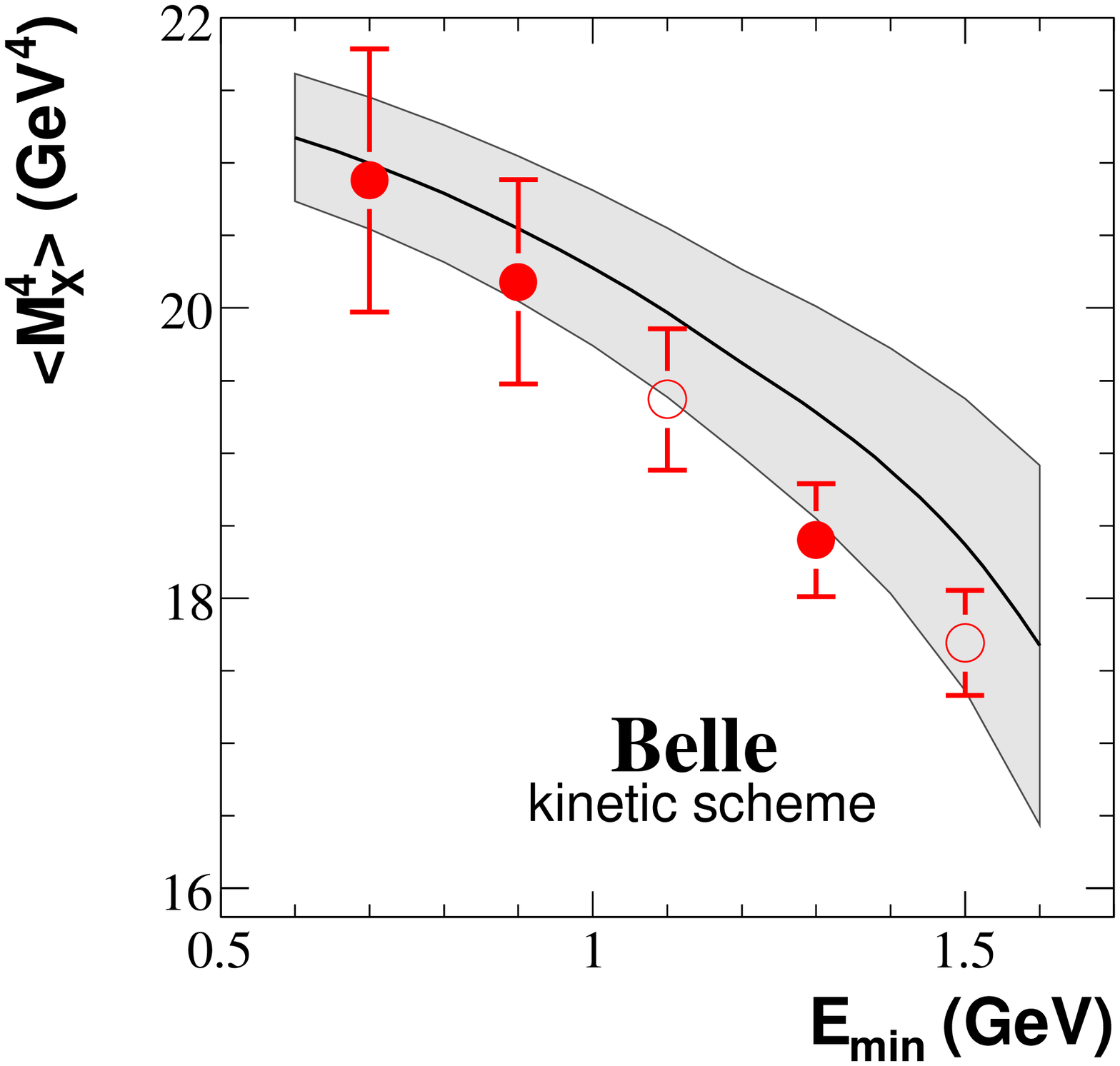}
    \includegraphics[width=0.24\columnwidth]{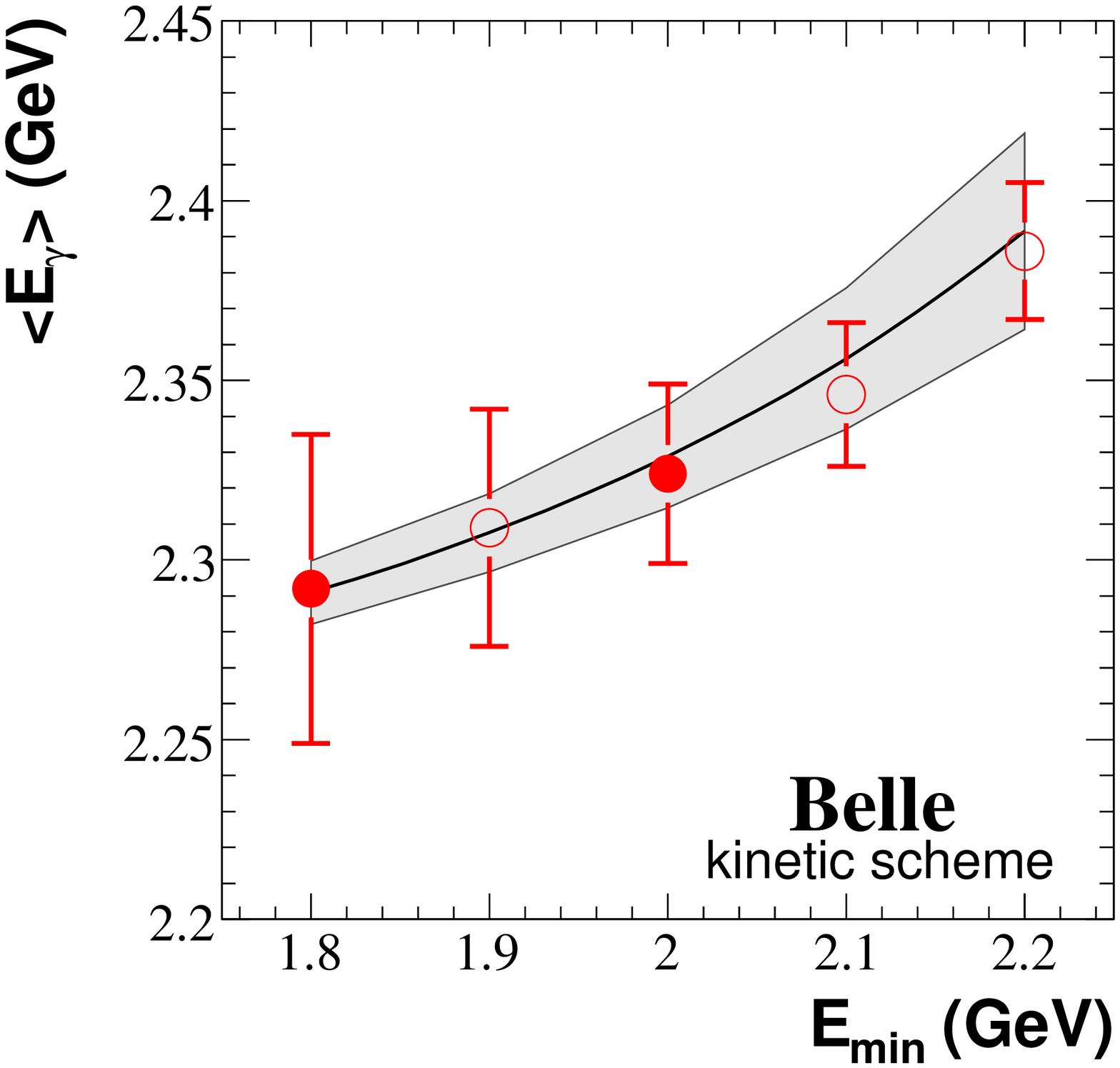}
    \includegraphics[width=0.24\columnwidth]{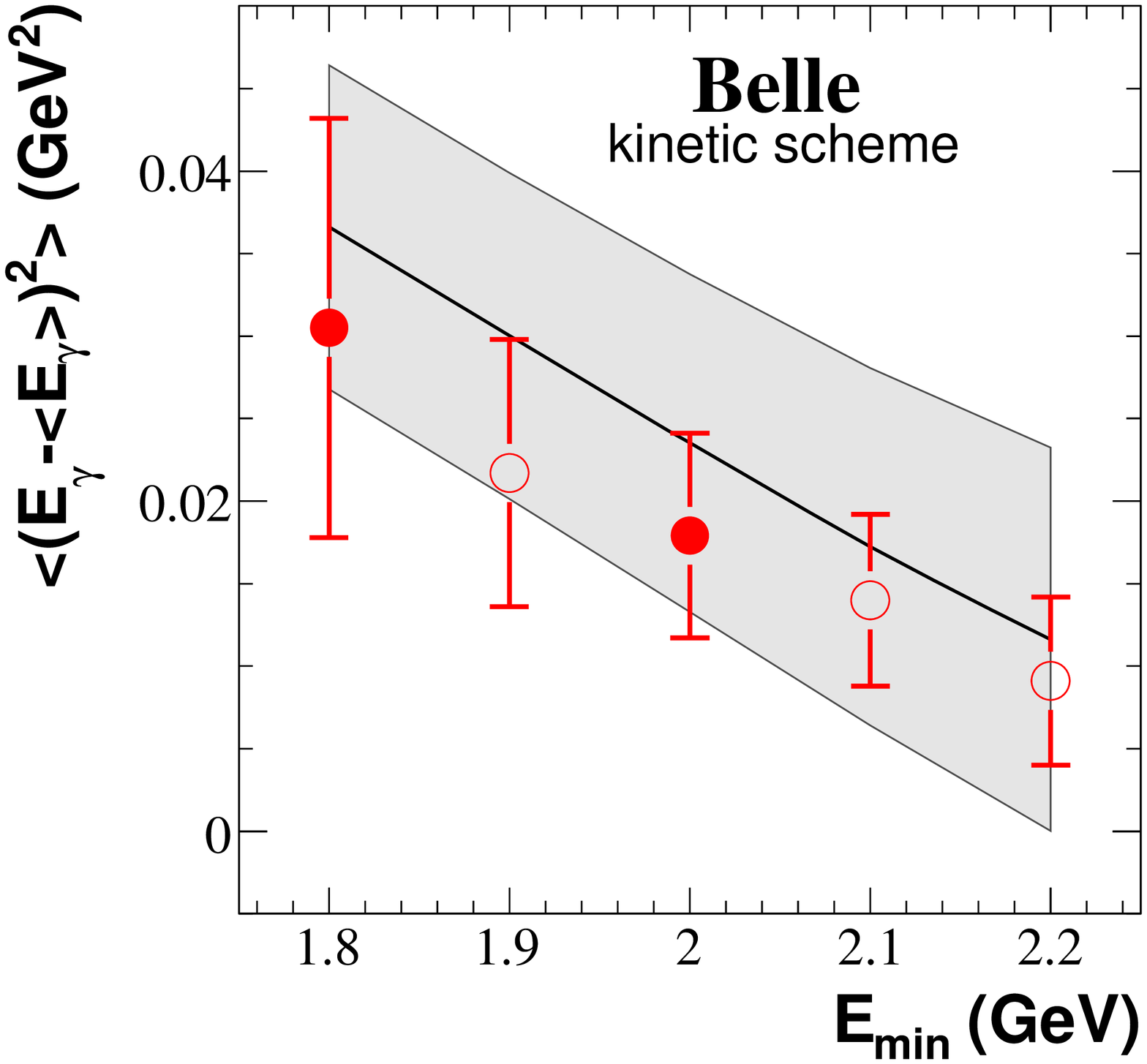}\\
    \includegraphics[width=0.24\columnwidth]{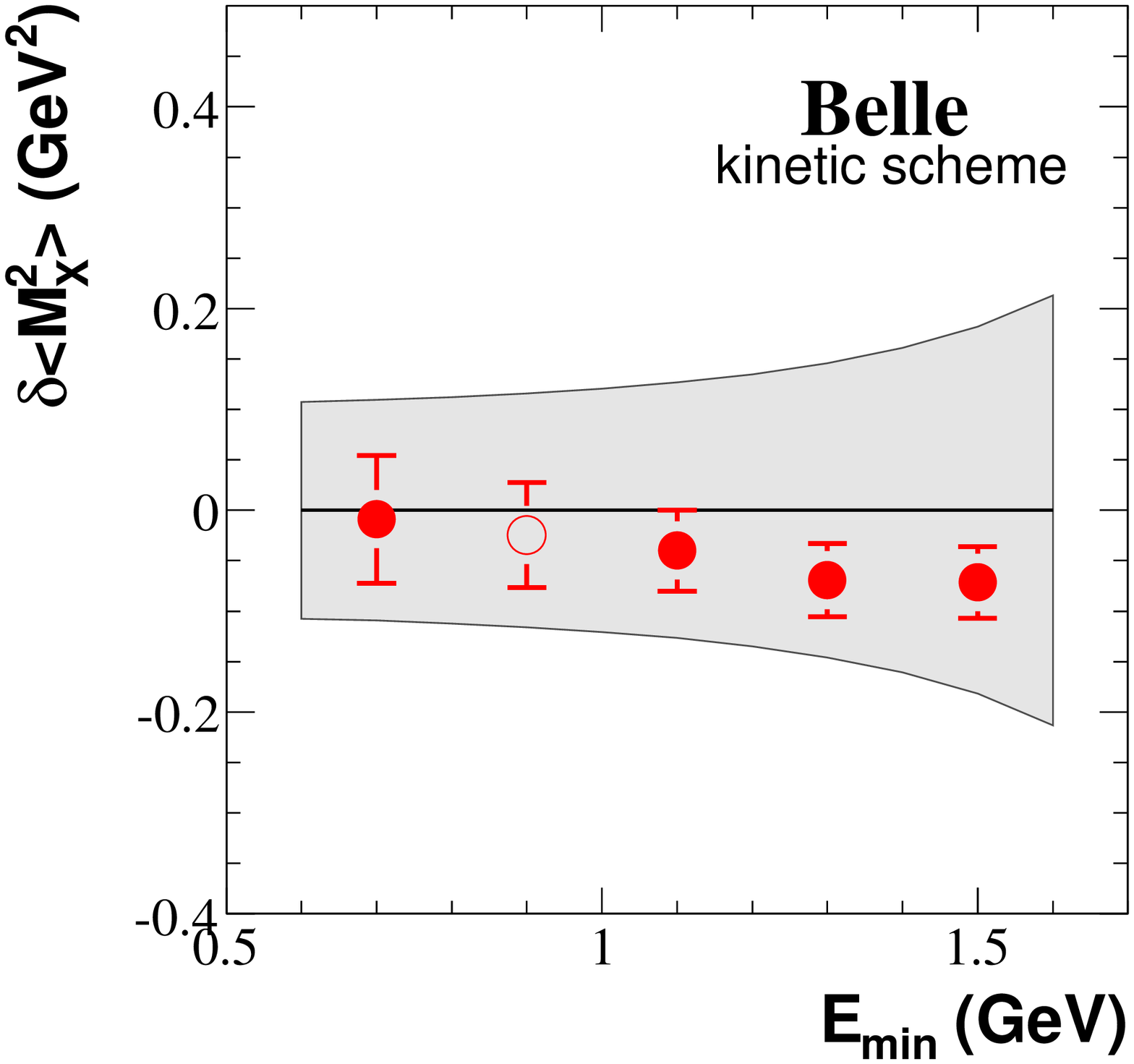}
    \includegraphics[width=0.24\columnwidth]{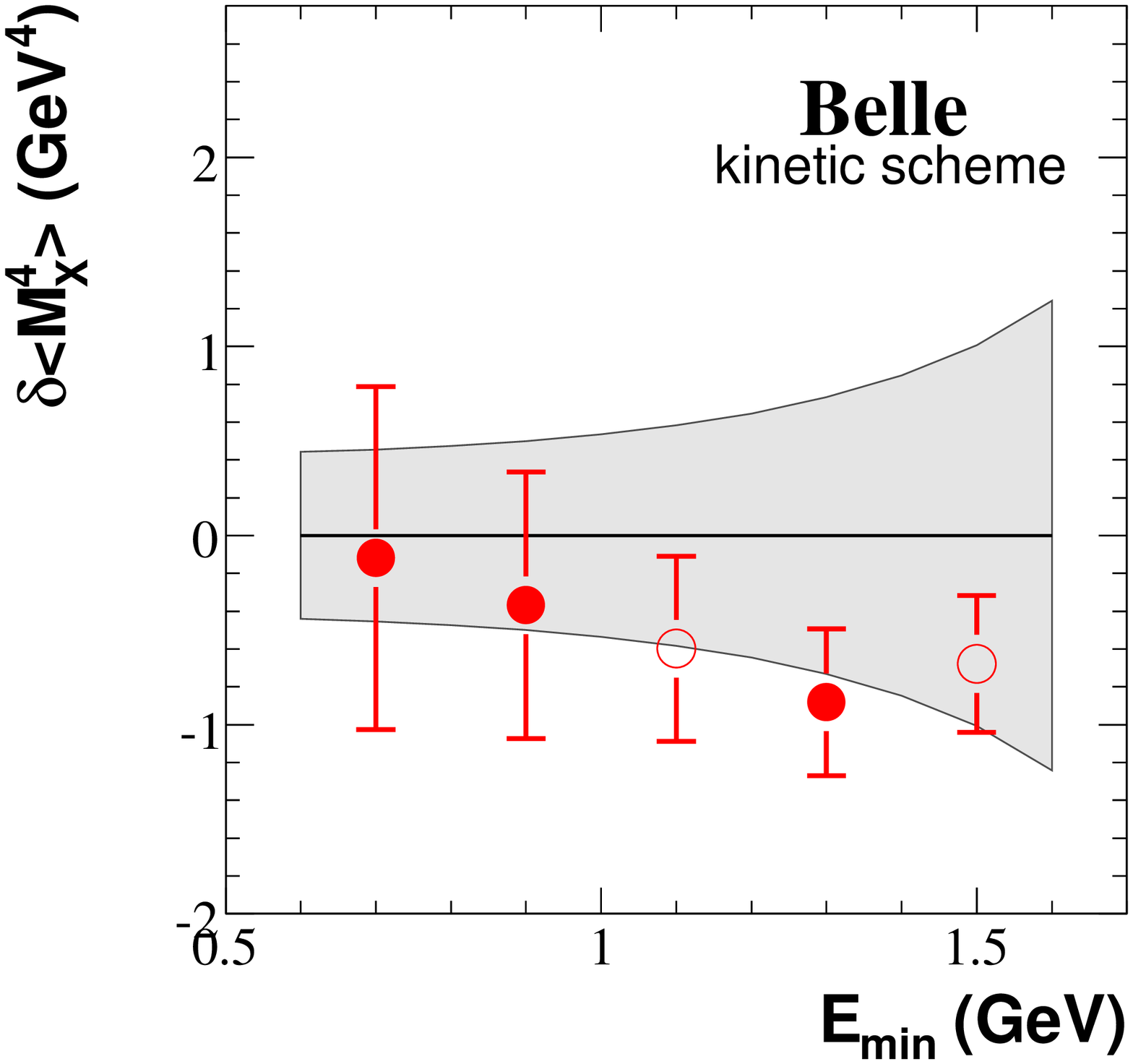}
    \includegraphics[width=0.24\columnwidth]{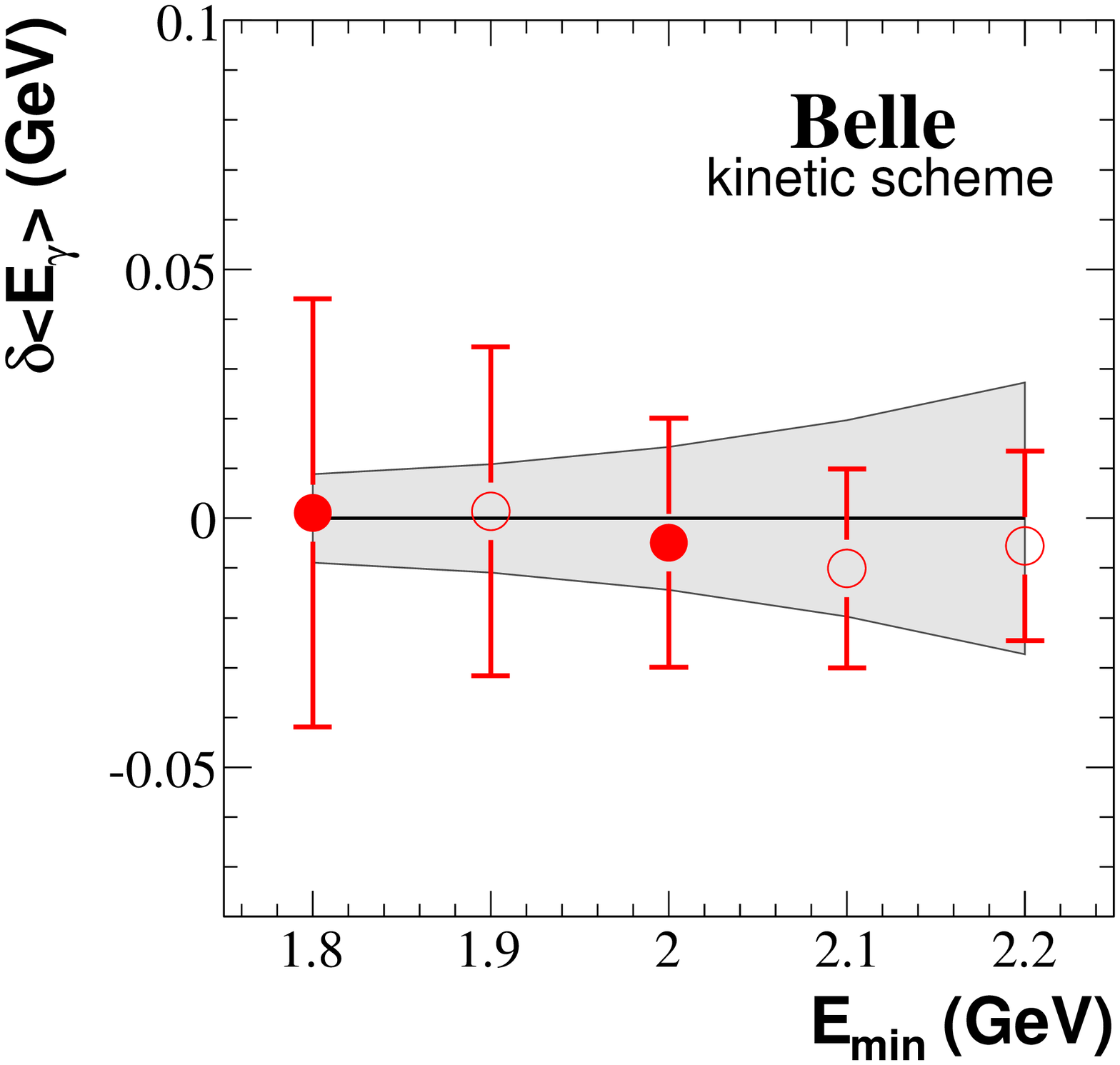}
    \includegraphics[width=0.24\columnwidth]{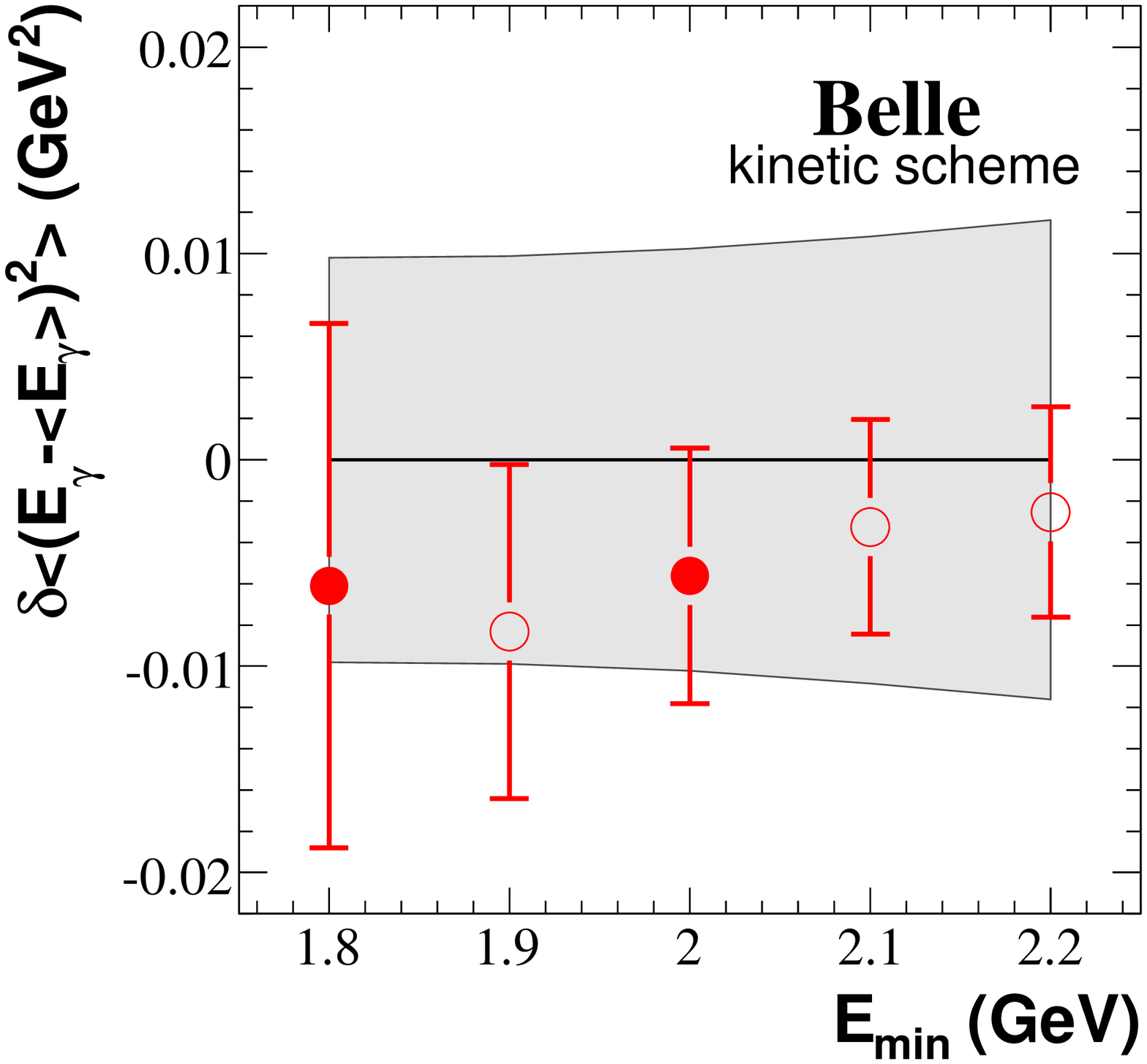}
  \end{center}
  \caption{Same as Fig.~\ref{fig:2_4} for the measured hadronic mass
    and photon energy moments and the kinetic scheme predictions.}
    \label{fig:2_5}
\end{figure}

The results of the fit in the kinetic mass scheme are shown in
Table~\ref{tab:2_4}. The value of the $\chi^2$~function at the minimum
is 4.7 for $25-7$ degrees of freedom. The semileptonic branching
fraction~$\mathcal{B}_{X_c\ell\nu}$ is found to be $(10.49\pm
0.23)\%$. The comparison of the measurements and the predictions in
the kinetic scheme is shown in Figs.~\ref{fig:2_4} and \ref{fig:2_5}.
\begin{table}
  \caption{Result of fit in the kinetic mass scheme. The
  $\sigma(\mathrm{fit})$~error contains the experimental and
  theoretical uncertainties in the moments. The $\sigma(\tau_B)$ and
  $\sigma$(th)~errors on $|V_{cb}|$ are due to the uncertainty
  in the average $B$~meson lifetime and the limited accuracy of
  Eq.~\ref{eq:2_2}, respectively. In the lower part of the table, the
  correlation matrix of the parameters is given.} \label{tab:2_4}
  \begin{center}
    \begin{tabular}{l|@{\extracolsep{.1cm}}ccccccc}
      \hline \hline
      & $|V_{cb}|$ (10$^{-3}$) & $m_b$ (GeV) & $m_c$ (GeV) &
      $\mu^2_\pi$ (GeV$^2$) & $\tilde\rho^3_D$ (GeV$^3$) & $\mu^2_G$
      (GeV$^2$) & $\rho^3_{LS}$ (GeV$^3$)\\
      \hline
      value & 41.58 & \phantom{$-$}4.543 & \phantom{$-$}1.055 & \phantom{$-$}0.539 & \phantom{$-$}0.166 & \phantom{$-$}0.362 & $-$0.153\\
      $\sigma$(fit) & 0.69 & \phantom{$-$}0.075 &
      \phantom{$-$}0.118 & \phantom{$-$}0.079 & \phantom{$-$}0.040 &
      \phantom{$-$}0.053 & \phantom{$-$}0.096\\
      $\sigma(\tau_B)$ & 0.08 & & & & & & \\
      $\sigma$(th) & 0.58 & & & & & & \\
      \hline
      $|V_{cb}|$ & 1.000 & $-$0.371 & $-$0.316 & \phantom{$-$}0.511 &
      \phantom{$-$}0.493 & $-$0.166 & \phantom{$-$}0.073\\
      $m_b$ & & \phantom{$-$}1.000 & \phantom{$-$}0.988 & $-$0.783 &
      $-$0.702 & $-$0.178 & $-$0.187\\
      $m_c$ & & & \phantom{$-$}1.000 & $-$0.771 & $-$0.715 & $-$0.262
      & $-$0.108\\
      $\mu^2_\pi$ & & & & \phantom{$-$}1.000 & \phantom{$-$}0.777 &
      \phantom{$-$}0.205 & \phantom{$-$}0.080\\
      $\tilde\rho^3_D$ & & & & & \phantom{$-$}1.000 &
      \phantom{$-$}0.108 & $-$0.158\\
      $\mu^2_G$ & & & & & & \phantom{$-$}1.000 & $-$0.103\\
      $\rho^3_{LS}$ & & & & & & & \phantom{$-$}1.000\\
      \hline \hline
    \end{tabular}
  \end{center}
\end{table}

\begin{figure}
  \begin{center}
    \includegraphics[width=0.49\columnwidth]{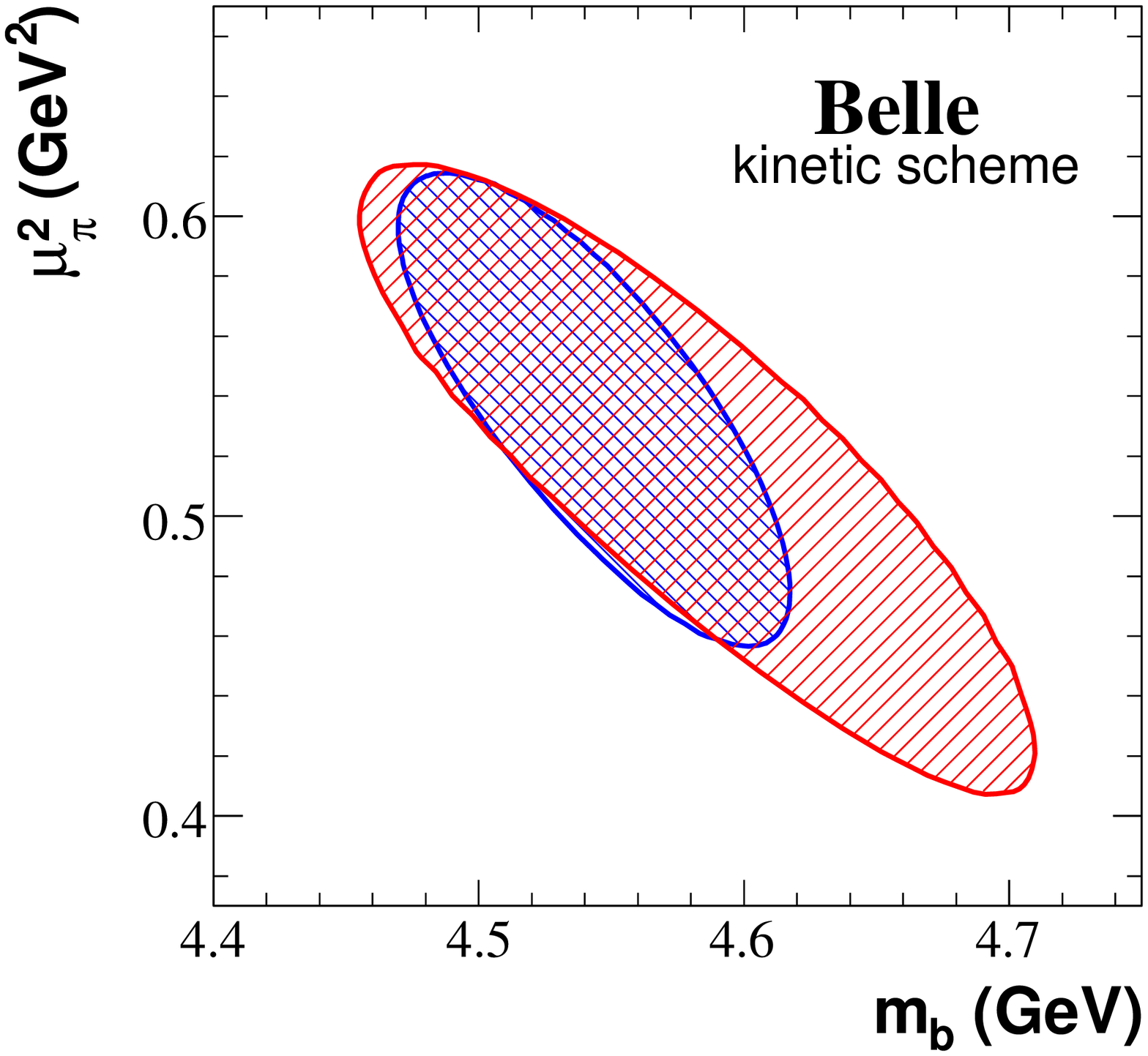}
    \includegraphics[width=0.49\columnwidth]{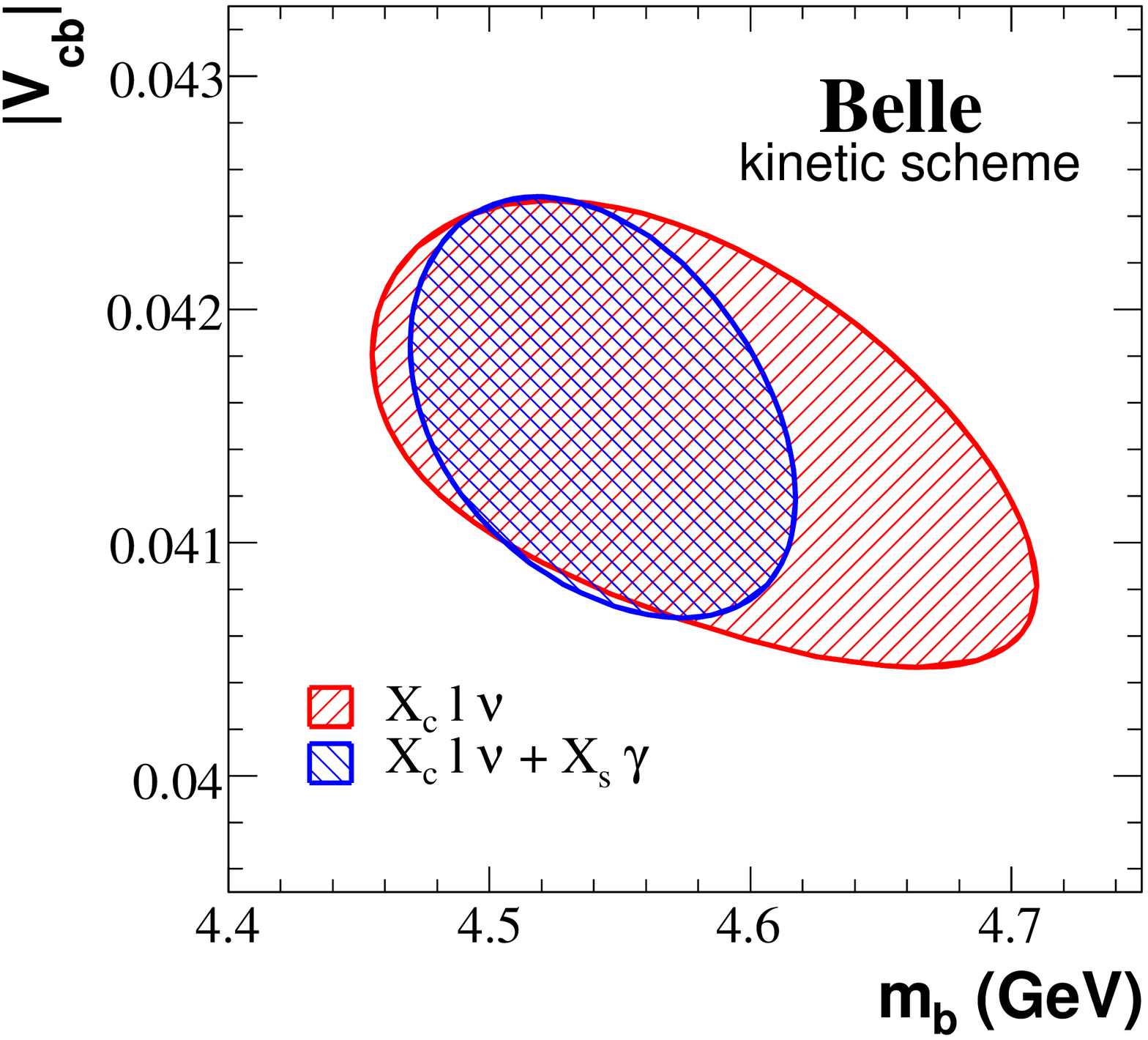}
  \end{center}
  \caption{$\Delta\chi^2=1$~contours for the fit to all moments and
  the fit to the $B\to X_c\ell\nu$~data only.} \label{fig:2_6}
\end{figure}

We have repeated the fit using the $B\to X_c\ell\nu$~moments only,
excluding $B\to X_s\gamma$~data
(Table~\ref{tab:2_5}). Figure~\ref{fig:2_6} shows the
$\Delta\chi^2=1$~contour plots for the fits corresponding to setups
(a) and (b) in Table~\ref{tab:2_5}.
\begin{table}
  \caption{Stability of the fit in the kinetic mass scheme. Setup (a)
    uses the $B\to X_c\ell\nu$~data only; setup (b) corresponds to the
    default fit.} \label{tab:2_5}
  \begin{center}
    \begin{tabular}{c|@{\extracolsep{.3cm}}cccc}
      \hline \hline
      Setup & $\chi^2/\mathrm{ndf.}$ & $|V_{cb}|$ (10$^{-3}$) &
      $m_b$ (GeV) & $\mu^2_\pi$ (GeV$^2$)\\
      \hline
      (a) & 4.2/14 & $41.51\pm 0.99$ & $4.573\pm 0.134$ & $0.523\pm 0.106$\\
      (b) & 4.7/18 & $41.58\pm 0.90$ & $4.543\pm 0.075$ & $0.539\pm 0.079$\\
      \hline\hline
    \end{tabular}
  \end{center}
\end{table}

\section{Summary}

We have determined the first and second moments of the photon energy
distribution in $B\to X_s\gamma$~decays, $\langle E_\gamma\rangle$ and
$\langle(E_\gamma-\langle E_\gamma\rangle)^2\rangle$, for minimum
photon energies in the $B$~meson rest frame ranging from 1.8 to
2.3~GeV using the measurement of this spectrum published in
Ref.~\cite{Koppenburg:2004fz}. The results are given in
Table~\ref{tab:1_4}. We have also evaluated the (statistical and
systematic) self- and cross-correlations between these measurements
(Tables~\ref{tab:1_5}$-$\ref{tab:1_7}).

In the second part of the present document, we have combined these
measurements with recent Belle data on the lepton energy and hadronic
mass moments in $B\to
X_c\ell\nu$~decays~\cite{Urquijo:2006wd,Schwanda:2006nf} to extract
$|V_{cb}|$, $m_b$ and other non-perturbative parameters using
theoretical expressions derived in the 1S~\cite{Bauer:2004ve} and
kinetic~\cite{Gambino:2004qm,Benson:2004sg} schemes.

The fits give consistent values of $|V_{cb}|$ in the two schemes. In
the 1S scheme analysis we find $|V_{cb}|=(41.56\pm
0.68(\mathrm{fit})\pm 0.08(\tau_B))\times 10^{-3}$ and
$m_b^\mathrm{1S}=(4.723\pm 0.055)$~GeV. In the kinetic scheme, we
obtain $|V_{cb}|=(41.58\pm 0.69(\mathrm{fit})\pm 0.08(\tau_B)\pm
0.58(\mathrm{th}))\times 10^{-3}$ and $m_b^\mathrm{kin}=(4.543\pm
0.075)$~GeV. Note that the $m_b$~values can only be compared after
scheme translation. The fit results using only the $B\to
X_c\ell\nu$~data are $|V_{cb}|=(41.55\pm 0.80(\mathrm{fit})\pm
0.08(\tau_B))\times 10^{-3}$ and $m_b^\mathrm{1S}=(4.718\pm
0.119)$~GeV in the 1S~scheme, and $|V_{cb}|=(41.51\pm
0.80(\mathrm{fit})\pm 0.08(\tau_B)\pm 0.58(\mathrm{th}))\times
10^{-3}$ and $m_b^\mathrm{kin}=(4.573\pm 0.134)$~GeV in the kinetic
scheme (see Tables~\ref{tab:2_3} and \ref{tab:2_5}).

The CKM magnitude $|V_{cb}|$ and the $b$-quark masses
$m_b^\mathrm{kin,1S}$ have been extracted with values that are
consistent with previous
determinations~\cite{Bauer:2004ve,Buchmuller:2005zv,Abdallah:2005cx,:2007yaa}.
In the 1S~scheme $|V_{cb}|$ has been measured with 1.6\%
precision. This is the most precise determination by any single
experiment so far~\cite{Abdallah:2005cx,:2007yaa}.

\section*{Acknowledgments}

We thank the theorists working on the 1S~scheme: C.W.\ Bauer, Z.\
Ligeti, M.\ Luke, A.V.\ Manohar and M.\ Trott, and those working on
the kinetic scheme: P.\ Gambino, N.\ Uraltsev and I.\ Bigi for
providing the Mathematica and Fortran codes that describe the
respective calculations.
We thank the KEKB group for the excellent operation of the
accelerator, the KEK cryogenics group for the efficient
operation of the solenoid, and the KEK computer group and
the National Institute of Informatics for valuable computing
and SINET3 network support. We acknowledge support from
the Ministry of Education, Culture, Sports, Science, and
Technology of Japan and the Japan Society for the Promotion
of Science; the Australian Research Council, the
Australian Department of Education, Science and Training, and the
David Hay Postgraduate Writing-Up Award;
the National Natural Science Foundation of China under
contract No.~10575109 and 10775142; the Department of
Science and Technology of India; 
the BK21 program of the Ministry of Education of Korea, 
the CHEP SRC program and Basic Research program 
(grant No.~R01-2005-000-10089-0) of the Korea Science and
Engineering Foundation, and the Pure Basic Research Group 
program of the Korea Research Foundation; 
the Polish State Committee for Scientific Research; 
the Ministry of Education and Science of the Russian
Federation and the Russian Federal Agency for Atomic Energy;
the Slovenian Research Agency;  the Swiss
National Science Foundation; the National Science Council
and the Ministry of Education of Taiwan; and the U.S.\
Department of Energy.


\begin{thebibliography}{99}
\bibitem{ref:1}
  The measurements of $<M_X^4>$ and $<(M_X^2 - <M_X^2>)^2>$ correspond
  to the same order in $M^2_X$ and are counted only once.
\bibitem{ref:3}
  In this analysis, all non-perturbative parameters in the kinetic
  scheme are defined at the scale $\mu=1$~GeV.
\end{thebibliography}
\end{document}